\documentclass[aps,prd,preprint,tightenlines,superscriptaddress,showpacs,byrevtex]{revtex4}
\usepackage{graphicx} 
\usepackage{dcolumn}  
\usepackage{bm}
\usepackage{pstricks}
\usepackage{pst-node}

\graphicspath{{ps}}

\usepackage{subfigure}
\usepackage{mathrsfs}
\usepackage{amssymb}
\usepackage{amsmath}
\usepackage{tabularx}
\usepackage{rotating}
\usepackage{multirow}

\def\bz{{B^0}}
\def\bzb{{\overline{B}{}^0}}

\def\dE{{\Delta E}}
\def\mb{{M_{\rm bc}}}
\def\mbc{{M_{\rm bc}}}

\newcommand{\fCP}{f_{CP}}

\newcommand{\ftag}{f_{\rm tag}}

\newcommand{\dm}{\Delta m_d}
\newcommand{\dmd}{\dm}

\newcommand*{\dwl}{\ensuremath{{\Delta w_l}}}

\def\pip{{\pi^+}}
\def\pim{{\pi^-}}
\def\piz{{\pi^0}}

\def\bbar{{\overline{B}}}

\def\lsig{{\cal L}_{\rm sig}}
\def\lbkg{{\cal L}_{\rm bkg}}
\def\rsigbkg{{\cal R}_{\rm s/b}}

\def\pbstar{p_B^{\rm cms}}

\def\nbb{449}
\def\nbbsvdone{152}
\def\nbbsvdtwo{297}
\def\lint{414}

\def\kakkoOverlineBeta{\raise1ex\hbox{\scriptsize ${}^($}
  \overline{\beta} \raise1ex\hbox{\scriptsize $\, {}^)$} \! \!}
\def\kakkoOverlineGamma{\raise0.5ex\hbox{\scriptsize ${}^($}
  \overline{\gamma} \raise0.5ex\hbox{\scriptsize $\, {}^)$} \! \!}
\def\kakkoOverlineF{\raise1ex\hbox{\scriptsize ${}^($}
  \overline{f} \raise1ex\hbox{\scriptsize $\, {}^{)}$} \! \!}

\begin{document}


\preprint{\vbox{ \hbox{   }
                 \hbox{BELLE-CONF-0650}
}}

\title{ \quad\\[0.5cm] \boldmath
$B^0 \to \pi^+ \pi^- \pi^0$ time-dependent Dalitz analysis from Belle}

\affiliation{Budker Institute of Nuclear Physics, Novosibirsk}
\affiliation{Chiba University, Chiba}
\affiliation{Chonnam National University, Kwangju}
\affiliation{University of Cincinnati, Cincinnati, Ohio 45221}
\affiliation{University of Frankfurt, Frankfurt}
\affiliation{The Graduate University for Advanced Studies, Hayama} 
\affiliation{Gyeongsang National University, Chinju}
\affiliation{University of Hawaii, Honolulu, Hawaii 96822}
\affiliation{High Energy Accelerator Research Organization (KEK), Tsukuba}
\affiliation{Hiroshima Institute of Technology, Hiroshima}
\affiliation{University of Illinois at Urbana-Champaign, Urbana, Illinois 61801}
\affiliation{Institute of High Energy Physics, Chinese Academy of Sciences, Beijing}
\affiliation{Institute of High Energy Physics, Vienna}
\affiliation{Institute of High Energy Physics, Protvino}
\affiliation{Institute for Theoretical and Experimental Physics, Moscow}
\affiliation{J. Stefan Institute, Ljubljana}
\affiliation{Kanagawa University, Yokohama}
\affiliation{Korea University, Seoul}
\affiliation{Kyoto University, Kyoto}
\affiliation{Kyungpook National University, Taegu}
\affiliation{Swiss Federal Institute of Technology of Lausanne, EPFL, Lausanne}
\affiliation{University of Ljubljana, Ljubljana}
\affiliation{University of Maribor, Maribor}
\affiliation{University of Melbourne, Victoria}
\affiliation{Nagoya University, Nagoya}
\affiliation{Nara Women's University, Nara}
\affiliation{National Central University, Chung-li}
\affiliation{National United University, Miao Li}
\affiliation{Department of Physics, National Taiwan University, Taipei}
\affiliation{H. Niewodniczanski Institute of Nuclear Physics, Krakow}
\affiliation{Nippon Dental University, Niigata}
\affiliation{Niigata University, Niigata}
\affiliation{University of Nova Gorica, Nova Gorica}
\affiliation{Osaka City University, Osaka}
\affiliation{Osaka University, Osaka}
\affiliation{Panjab University, Chandigarh}
\affiliation{Peking University, Beijing}
\affiliation{University of Pittsburgh, Pittsburgh, Pennsylvania 15260}
\affiliation{Princeton University, Princeton, New Jersey 08544}
\affiliation{RIKEN BNL Research Center, Upton, New York 11973}
\affiliation{Saga University, Saga}
\affiliation{University of Science and Technology of China, Hefei}
\affiliation{Seoul National University, Seoul}
\affiliation{Shinshu University, Nagano}
\affiliation{Sungkyunkwan University, Suwon}
\affiliation{University of Sydney, Sydney NSW}
\affiliation{Tata Institute of Fundamental Research, Bombay}
\affiliation{Toho University, Funabashi}
\affiliation{Tohoku Gakuin University, Tagajo}
\affiliation{Tohoku University, Sendai}
\affiliation{Department of Physics, University of Tokyo, Tokyo}
\affiliation{Tokyo Institute of Technology, Tokyo}
\affiliation{Tokyo Metropolitan University, Tokyo}
\affiliation{Tokyo University of Agriculture and Technology, Tokyo}
\affiliation{Toyama National College of Maritime Technology, Toyama}
\affiliation{University of Tsukuba, Tsukuba}
\affiliation{Virginia Polytechnic Institute and State University, Blacksburg, Virginia 24061}
\affiliation{Yonsei University, Seoul}
  \author{K.~Abe}\affiliation{High Energy Accelerator Research Organization (KEK), Tsukuba} 
  \author{K.~Abe}\affiliation{Tohoku Gakuin University, Tagajo} 
  \author{I.~Adachi}\affiliation{High Energy Accelerator Research Organization (KEK), Tsukuba} 
  \author{H.~Aihara}\affiliation{Department of Physics, University of Tokyo, Tokyo} 
  \author{D.~Anipko}\affiliation{Budker Institute of Nuclear Physics, Novosibirsk} 
  \author{K.~Aoki}\affiliation{Nagoya University, Nagoya} 
  \author{T.~Arakawa}\affiliation{Niigata University, Niigata} 
  \author{K.~Arinstein}\affiliation{Budker Institute of Nuclear Physics, Novosibirsk} 
  \author{Y.~Asano}\affiliation{University of Tsukuba, Tsukuba} 
  \author{T.~Aso}\affiliation{Toyama National College of Maritime Technology, Toyama} 
% ...
\collaboration{The Belle Collaboration}

\noaffiliation

\begin{abstract}

We present the results of a time-dependent Dalitz plot analysis of
$B^0 \to \pi^+ \pi^- \pi^0$ decays based on a $\lint {\rm fb}^{-1}$ data sample 
that contains $\nbb\times 10^6 B\bar{B}$ pairs collected 
on the $\Upsilon(4S)$ resonance
with the Belle detector at the KEKB asymmetric energy $e^+ e^-$
collider.
The direct $CP$ violating parameters of
 $B^0 \rightarrow \rho^\pm \pi^\mp$ decay mode are measured
to be
 $\mathcal{A}_{\rho \pi}^{+-}
 = +0.22\pm 0.08(\mathrm{stat.}) \pm 0.05(\mathrm{syst.})$
and
 $\mathcal{A}_{\rho \pi}^{-+}
 = +0.08\pm 0.17(\mathrm{stat.}) \pm 0.12(\mathrm{syst.})$.
We also measure the $CP$ violating parameters of the decay
mode $B^0 \rightarrow \rho^0 \pi^0$ as
$\mathcal{A}_{\rho^0 \pi^0}
 = -0.45\pm 0.35(\mathrm{stat.}) \pm 0.32(\mathrm{syst.})$
and
$\mathcal{S}_{\rho^0 \pi^0}
 = +0.15\pm 0.57(\mathrm{stat.}) \pm 0.43(\mathrm{syst.})$.
Combining our analysis with information on charged $B$ decay modes,
we perform a full Dalitz and isospin analysis
 and obtain a constraint on the CKM angle $\phi_2$,
$\phi_2 = (83^{+12}_{-23})^\circ$.
A large CKM-disfavored region
 ($\phi_2<8^\circ$ and $129^\circ <\phi_2$) also remains
at 68.3\% confidence level.
\end{abstract}

\pacs{11.30.Er, 12.15.Hh, 13.25.Hw}

\maketitle


{\renewcommand{\thefootnote}{\fnsymbol{footnote}}}
\setcounter{footnote}{0}

\section{Introduction}
\label{sec:introduction}
In the standard model (SM),
$CP$ violation arises from an irreducible phase in the 
Cabibbo-Kobayashi-Maskawa (CKM) matrix~\cite{Kobayashi:1973fv}.
The SM predicts
that measurement of a $CP$ asymmetry
 in time-dependent decay rates of $\bz$ and $\bzb$
gives access to the $CP$ violating phase
in the CKM matrix~\cite{Carter:1980hr,Carter:1980tk,Bigi:1981qs}.
The angle $\phi_2$ of the CKM unitarity triangle
can be measured via
 the tree diagram contribution
 in $b\rightarrow u \overline{u} d$ decay processes,
such as
$B^0 \rightarrow \pi^+ \pi^-$, 
$B^0 \rightarrow \rho^\pm\pi^\mp$, or
$B^0 \rightarrow \rho^+ \rho^-$~\cite{ChargeConjugate}.
In these decay processes,
however,
contributions from so-called penguin diagrams
could contaminate
the measurement of $\phi_2$.
Snyder and Quinn pointed out
 that a Dalitz plot analysis
of $B^0 \rightarrow \rho \pi$,
which includes $B^0 \rightarrow \rho^+\pi^-$,
$B^0 \rightarrow \rho^-\pi^+$, and
$B^0 \rightarrow \rho^0\pi^0$,
offers a unique way to determine $\phi_2$
without ambiguity.
The Dalitz plot analysis takes into account
a possible
contamination from a $b\rightarrow d$ penguin contribution~\cite{Snyder:1993mx}.
In addition, an isospin analysis~\cite{Lipkin:1991st,Gronau:1991dq}
involving the charged decay modes,
$B^+\rightarrow \rho^+\pi^0$ and $B^+\rightarrow \rho^0\pi^+$,
provides further improvement
of the $\phi_2$ determination.

\subsection{KEKB and Belle Detector}
KEKB~\cite{KEKB} operates at the $\Upsilon(4S)$ resonance 
($\sqrt{s}=10.58$~GeV) with a peak luminosity that exceeds
$1.6\times 10^{34}~{\rm cm}^{-2}{\rm s}^{-1}$.

At KEKB, the $\Upsilon(4S)$ is produced
with a Lorentz boost of $\beta\gamma=0.425$ nearly along
the electron beamline ($z$).
Since the $B^0$ and $\bzb$ mesons are approximately at 
rest in the $\Upsilon(4S)$ center-of-mass system (cms),
$\Delta t$ can be determined from the displacement in the $z$ direction,
 $\Delta z$, between the vertices of the two $B$ mesons:
$\Delta t \simeq  \Delta z/\beta\gamma c$.

The Belle detector is a large-solid-angle magnetic
spectrometer that
consists of a silicon vertex detector (SVD),
a 50-layer central drift chamber (CDC), an array of
aerogel threshold \v{C}erenkov counters (ACC), 
a barrel-like arrangement of time-of-flight
scintillation counters (TOF), and an electromagnetic calorimeter
comprised of CsI(Tl) crystals (ECL) located inside 
a super-conducting solenoid coil that provides a 1.5~T
magnetic field.  An iron flux-return located outside of
the coil is instrumented to detect $K_L^0$ mesons and to identify
muons (KLM).  The detector
is described in detail elsewhere~\cite{Belle}.
Two inner detector configurations were used. A 2.0 cm beampipe
and a 3-layer silicon vertex detector was used for the first data sample
of $\nbbsvdone\times 10^6 B\bar{B}$ pairs (DS-I),
while a 1.5 cm beampipe, a 4-layer
silicon detector and a small-cell inner drift chamber were used to record
the remaining $\nbbsvdtwo\times 10^6 B\bar{B}$ pairs
 (DS-II)~\cite{Ushiroda:2003bu}.

\subsection{Analysis flow}
The analysis proceeds in the following steps.
First, we extract the signal
 fraction (Sec.~\ref{sec:selection_and_reconstruction}).
We then determine the sizes
and phases of the contributions from radial excitations
(Sec.~\ref{sec:lineshape_determine}).
Using the parameters determined in the steps above,
we perform a time-dependent Dalitz plot analysis
 (Sec.~\ref{sec:time_depend_analysis_pdf}-\ref{sec:systematic_errors}).
The fit results are interpreted as
 quasi-two-body $CP$ violation parameters
 (Sec.~\ref{sec:quasi_two_body})
and are used to constrain the CKM angle $\phi_2$ (Sec.~\ref{sec:constraint_on_ph2}).

\subsection{Differential decay width of time-dependent Dalitz plot}
We measure the decay process $B^0 \rightarrow \pi^+ \pi^- \pi^0$,
where we denote the four-momenta of the $\pi^+$, $\pi^-$, and
 $\pi^0$ by $p_+$, $p_-$, and $p_0$, respectively.
The invariant-mass squared of their combinations
\begin{equation}
 s_+ = (p_+ + p_0)^2, \quad
  s_- = (p_- + p_0)^2, \quad
  s_0 = (p_+ + p_-)^2
\end{equation}
satisfies the following equation
\begin{equation}
 s_+ + s_- + s_0 = m_{B^0}^2 + 2 m_{\pi^+}^2 + m_{\pi^0}^2
  \label{equ:dalitz_var_add_relation}
\end{equation}
by energy and momentum conservation.
The differential (time-integrated) decay width with respect to the
variables above (Dalitz plot) is
\begin{equation}
 d\Gamma = \frac{1}{(2\pi)^3}
  \frac{|{\overset{(-)}{A}}_{3\pi}|^2}{8 m_{B^0}^2} ds_+ ds_- \;,
  \label{equ:amplitude_dalitz_width}
\end{equation}
where ${\overset{(-)}{A}}_{3\pi}$ is the Lorentz-invariant
amplitude of the $B^0 (\bzb) \rightarrow \pi^+ \pi^- \pi^0$ decay.

In the decay chain
 $\Upsilon(4S) \rightarrow \bz \bzb \rightarrow f_1 f_2$,
where one of the $B$'s decays into final state $f_1$ at time $t_1$
 and the other decays into another final state $f_2$ at time $t_2$,
the time dependent amplitude is
\begin{eqnarray}
 A(t_1, t_2) & \sim &
  e^{-(\Gamma/2 + iM)(t_1 + t_2)}
  \Bigl\{
  \cos[\dmd (t_1-t_2)/2] (A_1 \overline{A}_2 - \overline{A}_1 A_2)
  \nonumber \\
  & &
   \quad
  - i \sin[\dmd (t_1-t_2)/2]
  \left(\frac{p}{q} A_1 A_2 - \frac{q}{p} \overline{A}_1 \overline{A}_2
       \right)
  \Bigr\} \;.
\end{eqnarray}
Here, $p$ and $q$ define the mass eigenstates of neutral $B$ mesons
as $p \bz \pm q \bzb$, with average mass $M$ and width $\Gamma$,
and mass difference $\Delta m_d$.
The width difference is assumed to be zero.

The decay amplitudes are defined as follows,
\begin{eqnarray}
 A_1 & \equiv & A(B^0 \rightarrow f_1) \;, \\
 \overline{A}_1 & \equiv & A(\bzb \rightarrow f_1) \;, \\
 A_2 & \equiv & A(B^0 \rightarrow f_2) \;, \\
 \overline{A}_2 & \equiv & A(\bzb \rightarrow f_2) \;.
\end{eqnarray}
In this analysis,
we take $A_{3\pi}$ as $A_1$
and choose $f_2$ to be
 a flavor eigenstate,
i.e., $A_2 = 0$ or $\overline{A}_2 = 0$.
Here we call the $B$ decaying into $f_1 = \fCP = \pi^+ \pi^- \pi^0$
the $CP$ side $B$ while the other $B$ is the tag side $B$,
 $f_2 = f_\mathrm{tag}(\overline{f}_\mathrm{tag})$.
The differential decay width dependence on time difference
$\Delta t \equiv t_{CP} - t_\mathrm{tag}$ is then,
\begin{equation}
 \begin{array}{l}
  d \Gamma  \sim 
   e^{-\Gamma |\Delta t|}  \Bigl\{
   \left(|A_{3\pi}|^2 + |\overline{A}_{3\pi}|^2\right)
   \\
  \hspace{15mm} - q_\mathrm{tag} \cdot
    \left(|A_{3\pi}|^2 - |\overline{A}_{3\pi}|^2\right)
    \cos (\dmd \Delta t)
    +
    q_\mathrm{tag}\cdot
     2 \mathrm{Im}\left[
   		     \frac{q}{p}
   		     A_{3\pi}^*
   		     \overline{A}_{3\pi}
   		    \right]
    \sin (\dmd \Delta t)
    \Bigr\}
    d \Delta t \;,
    \\
 \end{array}
 \label{equ:amplitude_dt_width}
\end{equation}
where we assume $|q/p| = 1$ ($CP$ and $CPT$ conservation in the mixing)
and
 $|A(B^0\rightarrow f_\mathrm{tag})| = |A(\bzb\rightarrow \overline{f}_\mathrm{tag})|$,
and integrate
over $t_\mathrm{sum} = t_{CP} + t_\mathrm{tag}$.
Here $q_\mathrm{tag}$ is the $b$-flavor charge
and $q_\mathrm{tag} = +1(-1)$  when the tag side
$B$ decays into a $\bz$ ($\bzb$) flavor eigenstate.

Combining the Dalitz plot decay width (\ref{equ:amplitude_dalitz_width})
and the time dependent decay width
(\ref{equ:amplitude_dt_width}),
we obtain the time dependent Dalitz plot decay width
\begin{equation}
 d\Gamma \sim |A(\Delta t; s_+, s_-)|^2 \: d\Delta t \: ds_+ \: ds_- \;,
\end{equation}
where
\begin{equation}
 \begin{array}{l}
 |A(\Delta t; s_+, s_-)|^2
  = 
  e^{-\Gamma |\Delta t|}
  \Bigl\{
   \left(|A_{3\pi}|^2 + |\overline{A}_{3\pi}|^2\right)
   \\
  \hspace{15mm} - q_\mathrm{tag} \cdot 
   \left(|A_{3\pi}|^2 - |\overline{A}_{3\pi}|^2\right)
   \cos (\dmd \Delta t)
   + q_\mathrm{tag} \cdot 
    2 \mathrm{Im}\left[
		     \frac{q}{p}
		     A_{3\pi}^*
		     \overline{A}_{3\pi}
		    \right]
   \sin (\dmd \Delta t)
  \Bigr\} \;,
  \label{equ:amplitude_time_dep_three_pi}
 \end{array}
\end{equation}
\begin{equation}
 A_{3\pi} = A_{3\pi}(s_+, s_-) \;, \quad
  \overline{A}_{3\pi} = \overline{A}_{3\pi}(s_+, s_-) \;.
\end{equation}

Although there could exist
 contributions from $B^0$ decays into 
non-$\rho\pi$ $\pi^+\pi^-\pi^0$ final states, such as
$f_0(980)\pi^0$, $f_0(600) \pi^0$, $\omega \pi^0$, and
non-resonant $\pi^+\pi^-\pi^0$,
we confirm that these contributions are small and their effects are taken
into account as systematic
 uncertainties (Sec.~\ref{sec:syst_other_pipipi}).
We therefore assume that the $B^0 \to \pi^+\pi^-\pi^0$ decay is dominated by
the $B^0 \rightarrow (\rho \pi)^0$ amplitudes:
$B^0 \rightarrow \rho^+ \pi^-$, $B^0 \rightarrow \rho^- \pi^+$,
and $B^0 \rightarrow \rho^0 \pi^0$,
where $\rho$ can be $\rho(770)$, $\rho(1450)$,
or $\rho(1700)$.
The Dalitz plot amplitude $A_{3\pi}(s_+, s_-)$
can then be written as
\begin{eqnarray}
 A_{3\pi}(s_+, s_-) & = &
  f_+(s_+, s_-) A^+ + f_-(s_+, s_-) A^- + f_0(s_+, s_-) A^0 \;,
  \label{equ:amplitude_q2b_b}
 \\
 \frac{q}{p}
  \overline{A}_{3\pi}(s_+, s_-)
  & = & \overline{f}_+(s_+, s_-) \overline{A}{}^+
  + \overline{f}_-(s_+, s_-) \overline{A}{}^-
  + \overline{f}_0(s_+, s_-) \overline{A}{}^0 \;,
  \label{equ:amplitude_q2b_bbar}
\end{eqnarray}
where the functions $f_\kappa(s_+, s_-)$
 (with $\rho$ charge $\kappa = {+,-,0}$)
incorporate the kinematic and dynamical properties 
of the $B^0$ decay into a vector $\rho$ and a pseudoscalar
$\pi$.
The goal of this analysis is to measure
the complex coefficients $A^+$, $A^-$, $A^0$,
$\overline{A}{}^+$, $\overline{A}{}^-$, and $\overline{A}{}^0$;
and constrain the CKM angle $\phi_2$ using them.

\subsection{\boldmath Kinematics of
  $B^0 \rightarrow (\rho \pi)^0$
 \label{sec:square_dalitz_kinema_angular}
}
The function $f_\kappa(s_+, s_-)$
can be factorized into two parts as
\begin{equation}
 {\mathop{f}^{(-)}}_\kappa(s_+, s_-) = T_J^\kappa F_{\pi}(s_\kappa)
  \quad (\kappa = +, -, 0) \;,
  \label{equ:f_kappa_definition_with_unique_lineshape_assumption}
\end{equation}
where $F_{\pi}(s_\kappa)$
and $T_J^\kappa$
correspond to the lineshape of $\rho$
and the helicity distribution of the $\rho$,
 respectively.
Here we assume that
 a single unique functional form for
the lineshape $F_{\pi}(s)$ can be used
for all six $\kakkoOverlineF_\kappa$\footnote{
Strictly speaking, the difference between the $\pi^\pm$ and $\pi^0$ masses
is taken into acccount and thus
$F_{\pi}(s_\pm)$ and $F_{\pi}(s_0)$
are slightly different.
The essential point here is, however,
that one unique set of $(\beta, \gamma)$ is used
for all six $\kakkoOverlineF_\kappa$.
}.
Since this assumption has no good theoretical or experimental
foundation,
we check the goodness of the assumption with data
and assign systematic errors.
\par
The lineshape is parameterized with Breit-Wigner functions
corresponding to the $\rho(770)$, $\rho(1450)$, and $\rho(1700)$
resonances:
\begin{equation}
 F_{\pi}(s) =
  BW_{\rho(770)} + \beta \cdot BW_{\rho(1450)}
  + \gamma \cdot BW_{\rho(1700)} \;,
  \label{equ:fpi_lineshape_beta_gamma}
\end{equation}
where the amplitudes $\beta$ and $\gamma$ (denoting the relative size of two
resonances)
are complex numbers.
We use the Gounaris-Sakurai(GS)
model~\cite{Gounaris:1968mw} for the Breit-Wigner shape.
\par
\begin{figure}[htbp]
 \begin{center}
  \includegraphics[scale=0.3]{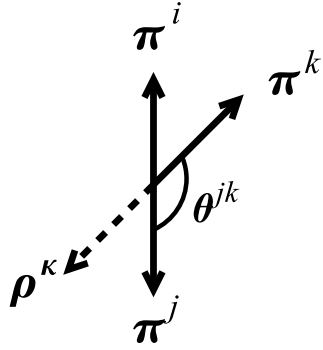}
  \caption{
  \label{fig:amplitude_rho_pi_kinema_in_rho_rest}
  The relation between three pions in the rest frame of $\rho^\kappa$.
  }
 \end{center}
\end{figure}

In the case of pseudoscalar-vector ($J=1$) decay,
$T_J^\kappa$ is given by 
\begin{equation}
 T_1^\kappa = - 4 |\vec{p}_j| |\vec{p}_k| \cos \theta^{jk} \;,
\end{equation}
\begin{equation}
 \left(
  \begin{array}{c}
   T_1^+ = - 4 |\vec{p}_+| |\vec{p}_-| \cos \theta^{+-} \;, \\
   T_1^- = - 4 |\vec{p}_0| |\vec{p}_+| \cos \theta^{0+} \;, \\
   T_1^0 = - 4 |\vec{p}_-| |\vec{p}_0| \cos \theta^{-0} \;, \\
  \end{array}
 \right)
\end{equation}
where $\vec{p}_j, \vec{p}_k$ are the three momenta
of the $\pi^j$ and $\pi^k$ in the rest frame of $\rho^\kappa$
 (or the $\pi^i\pi^j$ system),
and the $\theta^{jk} (\equiv \theta_\kappa)$
is the angle between $\vec{p}_j$ and $\vec{p}_k$
(see Fig.~\ref{fig:amplitude_rho_pi_kinema_in_rho_rest}).

\subsection{Fitting parameters
\label{sec:square_dlz_fitting_parameters}}
Inserting
(\ref{equ:f_kappa_definition_with_unique_lineshape_assumption}),
into expressions 
(\ref{equ:amplitude_q2b_b}) and (\ref{equ:amplitude_q2b_bbar}),
the coefficients of equation
 (\ref{equ:amplitude_time_dep_three_pi})
become
\begin{equation}
 |A_{3\pi}|^2 \pm |\overline{A}_{3\pi}|^2
  = \sum_{\kappa \in \{+,-,0\}} |f_\kappa|^2 U^\pm_\kappa
  + 2 \sum_{\kappa<\sigma \in \{+,-,0\}}
  \left(
   \mathrm{Re}[f_\kappa f^*_\sigma] U^{\pm,\mathrm{Re}}_{\kappa\sigma}
   - \mathrm{Im}[f_\kappa f^*_\sigma] U^{\pm,\mathrm{Im}}_{\kappa\sigma}
  \right)
  \;,
\end{equation}
\begin{equation}
 \mathrm{Im}\left(\frac{q}{p} A_{3\pi}^* \overline{A}_{3\pi} \right)
  = \sum_{\kappa \in \{+,-,0\}} |f_\kappa|^2 I_\kappa
  + \sum_{\kappa<\sigma \in \{+,-,0\}}
  \left(
   \mathrm{Re}[f_\kappa f^*_\sigma] I^\mathrm{Im}_{\kappa\sigma}
   + \mathrm{Im}[f_\kappa f^*_\sigma] I^\mathrm{Re}_{\kappa\sigma}
  \right)
  \;,
\end{equation}
with
\begin{eqnarray}
 U^\pm_{\kappa}
  & = & (|A^\kappa|^2 \pm |\overline{A}{}^\kappa|^2) / N \;,
  \label{equ:fit_params_first}
  \\
 U^{\pm,\mathrm{Re}(\mathrm{Im})}_{\kappa\sigma}
  & = & \mathrm{Re}(\mathrm{Im})
  \left[ A^\kappa A^{\sigma *} \pm \overline{A}{}^\kappa 
    \overline{A}{}^{\sigma *}
  \right] / N \;,
  \\
 I_\kappa & = & \mathrm{Im}
  \left[
   \overline{A}{}^{\kappa} A^{\kappa *}
  \right] / N \;,
  \\
 I^\mathrm{Re}_{\kappa \sigma} & = & \mathrm{Re}
  \left[
   \overline{A}{}^\kappa A^{\sigma *}
   - \overline{A}{}^\sigma A^{\kappa *}
  \right] / N \;,
 \\
 I^\mathrm{Im}_{\kappa \sigma} & = & \mathrm{Im}
  \left[
   \overline{A}{}^\kappa A^{\sigma *}
   + \overline{A}{}^\sigma A^{\kappa *}
  \right] / N \;,
  \label{equ:fit_params_last}
\end{eqnarray}
where $N$ is a normalization factor.
The 27 coefficients
(\ref{equ:fit_params_first})-(\ref{equ:fit_params_last})
are the parameters determined by the fit.
This parameterization allows us
to describe the differential decay width as a linear combination
of independent functions,
whose coefficients are the fit parameters
in a well behaved fit.
We fix the overall normalization by requiring $U^+_{+} = 1$,
i.e., we use $N = |A^+|^2 + |\overline{A}{}^+|^2$.
Thus, 26 of the 27 coefficients are free parameters in the fit.

\subsection{Square Dalitz plot (SDP)}
\begin{figure}[p]
 \begin{center}
  \subfigure[Usual Dalitz plot]{
  \includegraphics[scale=0.3]{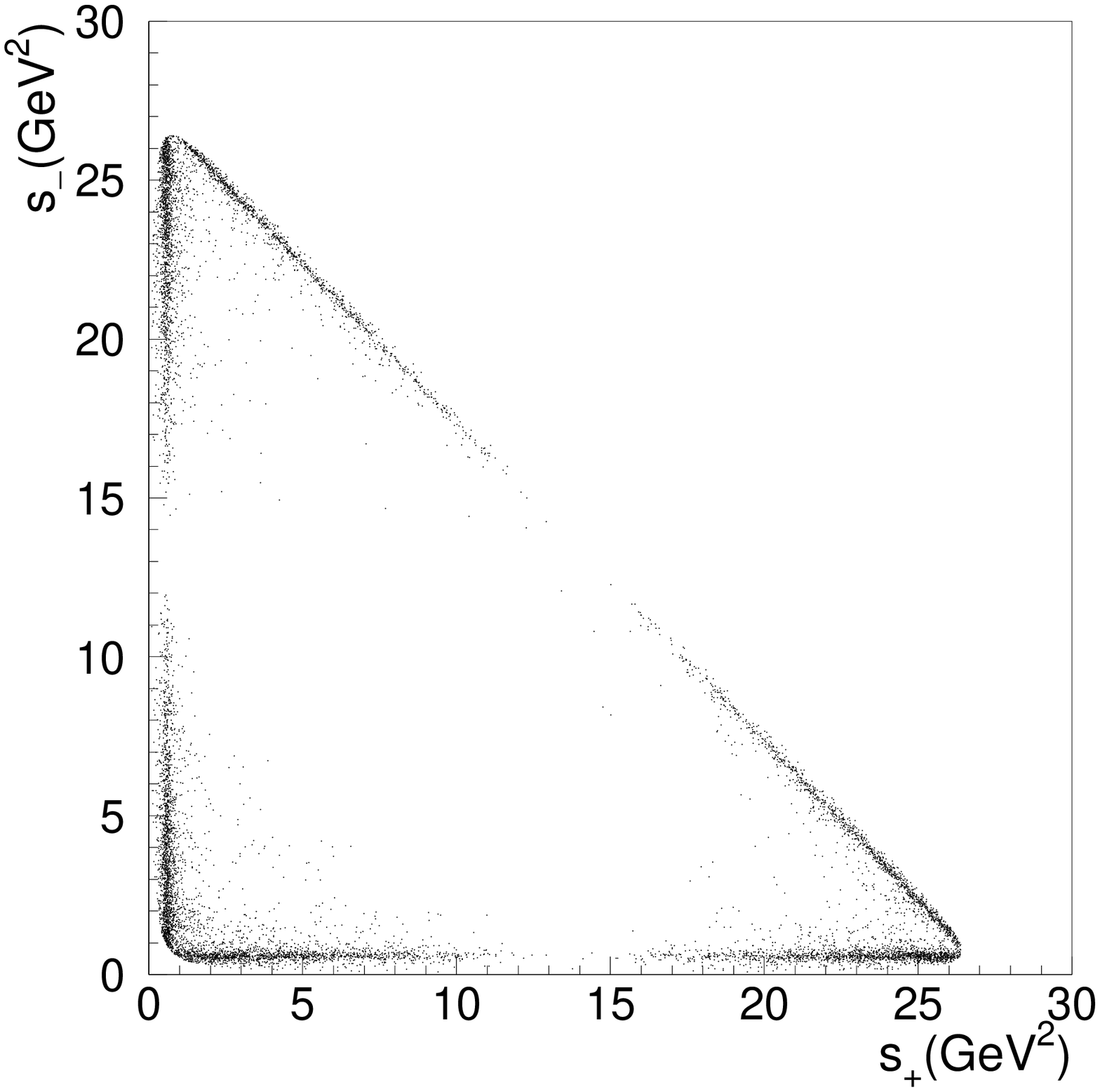}
  }
  \subfigure[Square Dalitz plot]{
  \includegraphics[scale=0.3]{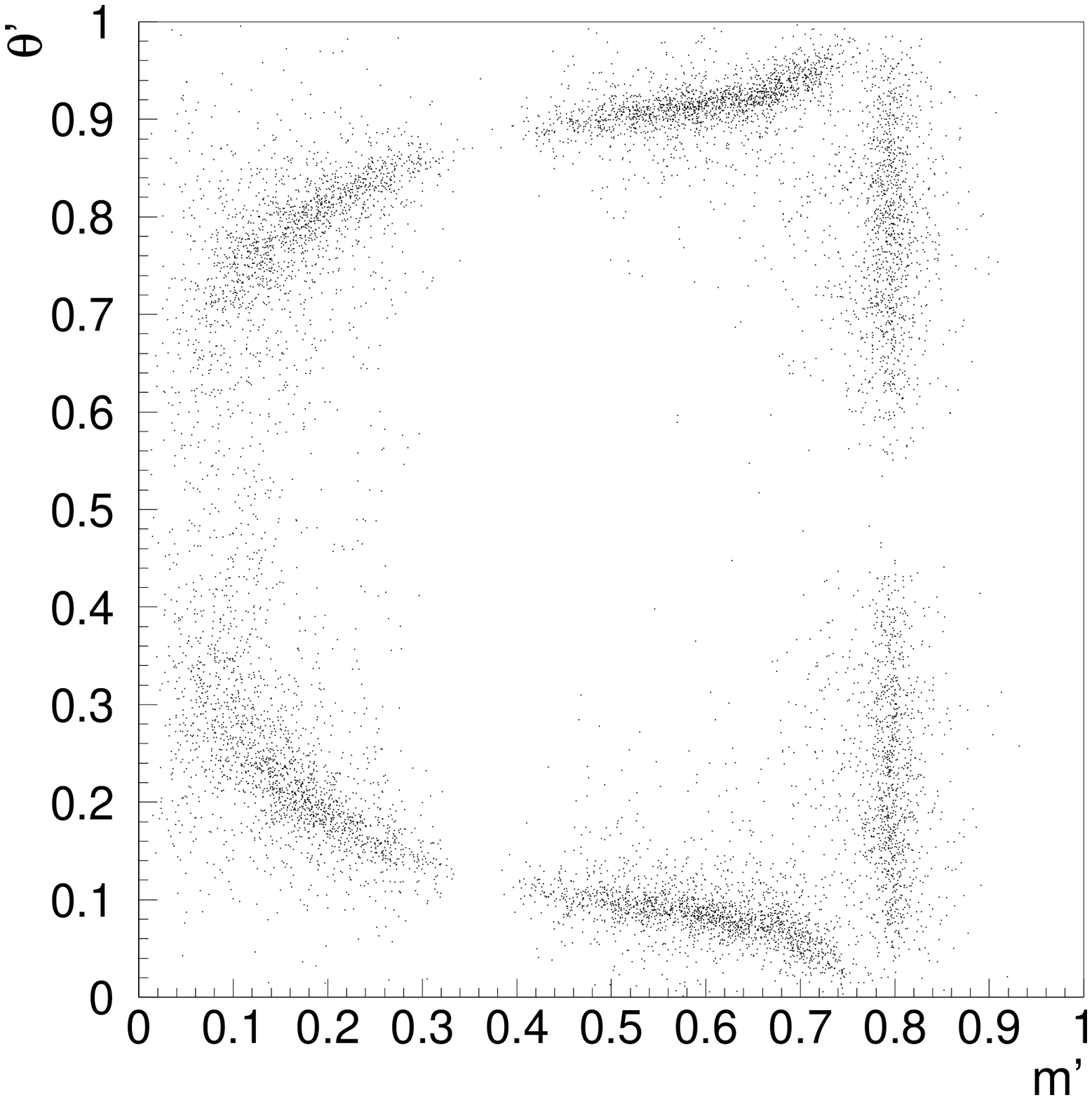}
  }
  \caption{
  \label{fig:sdp_sig_distr}
  Distribution of signal Monte Carlo (without detector efficiency and
  smearing)
  in the Dalitz plot.
  }
 \end{center}
\end{figure}

\begin{figure}[p]
 \begin{center}
  \subfigure[Usual Dalitz plot]{
  \includegraphics[scale=0.3]{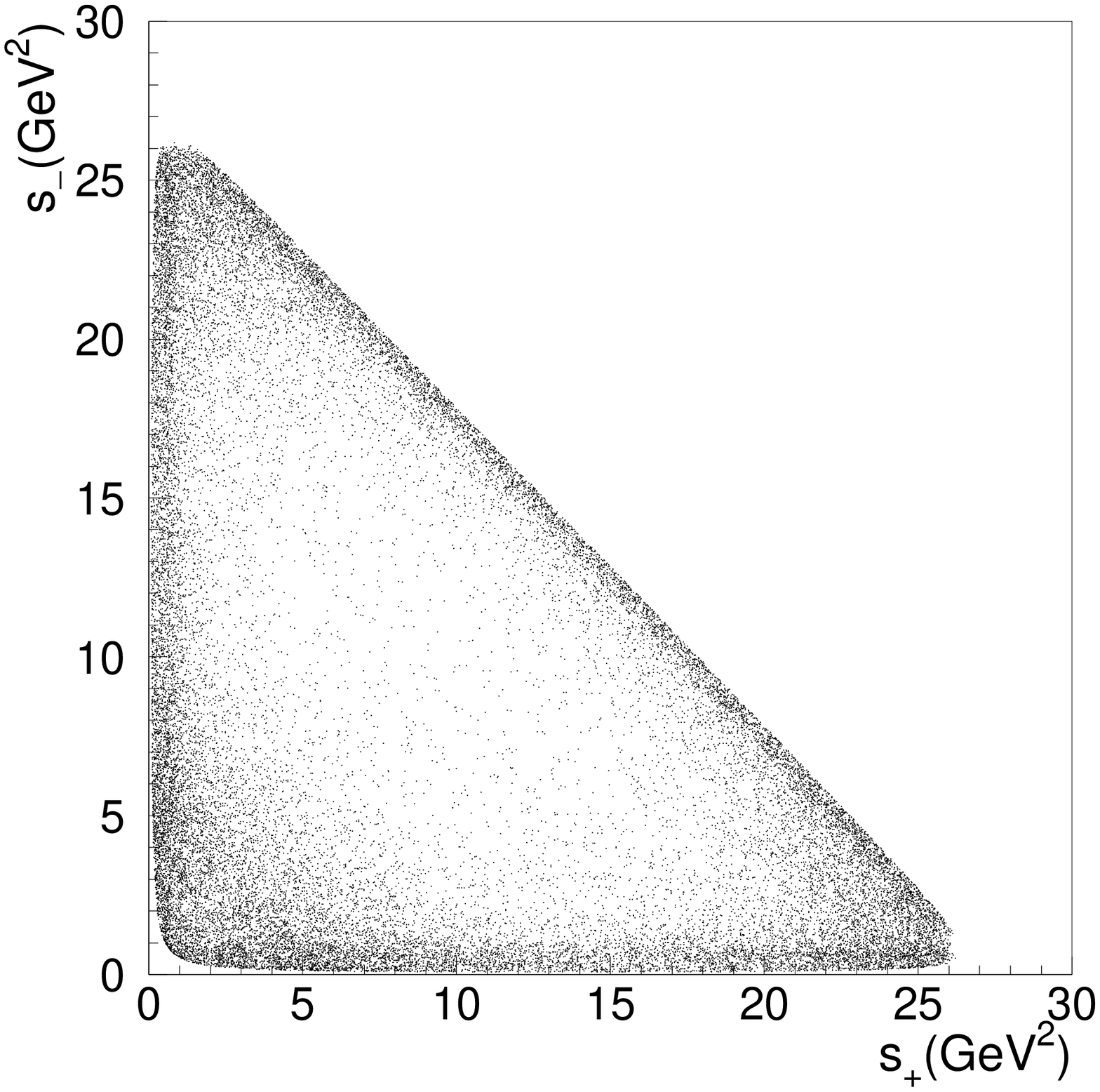}
  }
  \subfigure[Square Dalitz plot]{
  \includegraphics[scale=0.3]{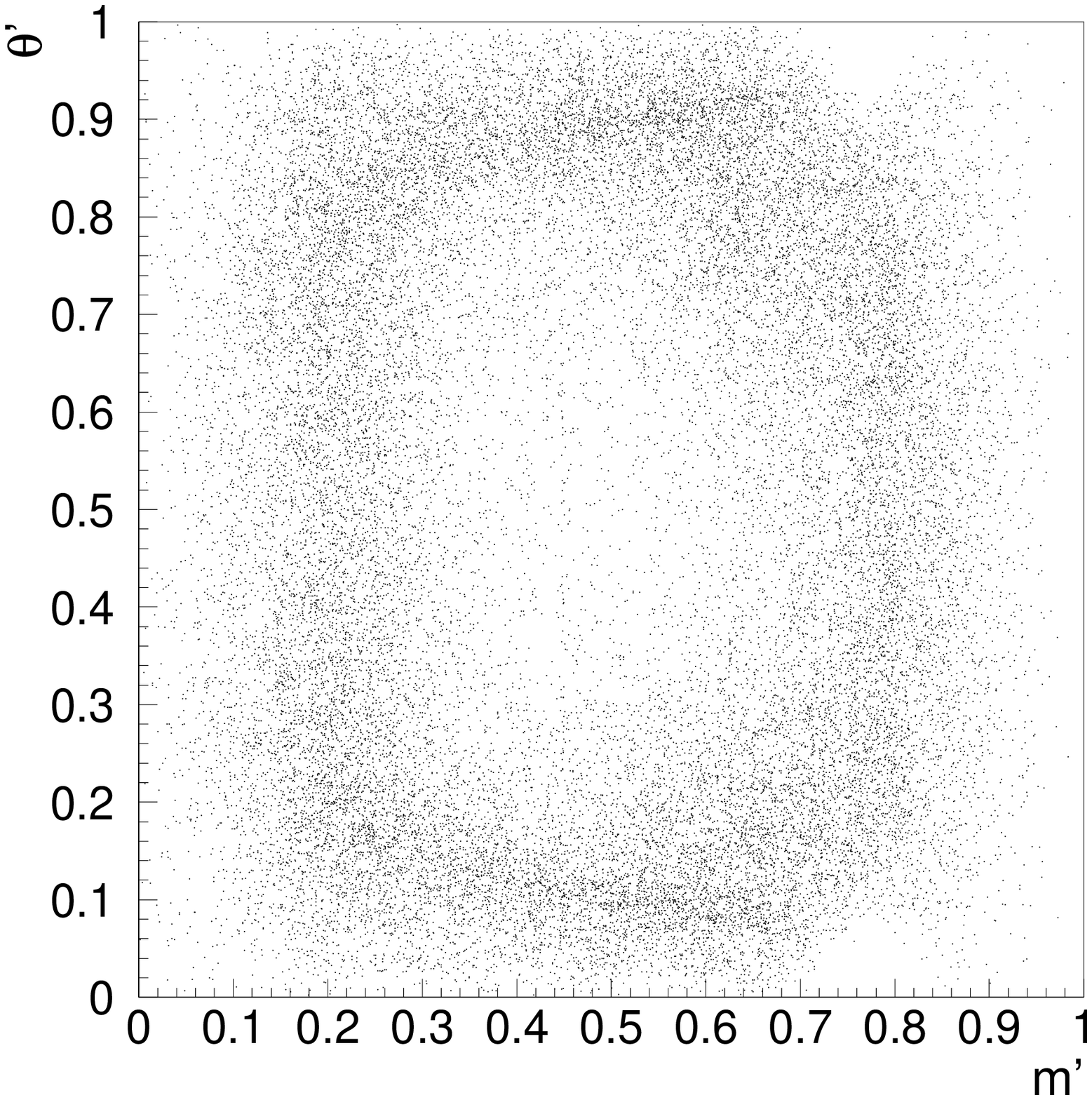}
  }
  \caption{
  \label{fig:sdp_qq_distr}
  Distribution of 
  $q\overline{q}$ background (from the data $\mbc$ sideband)
  in the Dalitz plot.
  }
 \end{center}
\end{figure}

The signal and the continuum background
 $e^+e^- \rightarrow q\overline{q} (q=u, d, s, c)$,
which is the dominant background in this analysis,
 populate the kinematic boundaries of the usual Dalitz plot
as shown in Figs.
 \ref{fig:sdp_sig_distr} and  \ref{fig:sdp_qq_distr}.
Since we model part of the Dalitz plot probability density function (PDF)
 with a binned histogram,
the distribution concentrated in a narrow region
is not easy to treat.
We therefore apply the transformation
\begin{equation}
 ds_+ ds_- \rightarrow |\det \boldsymbol{J}| dm' d\theta' \;,
\end{equation}
which defines the square Dalitz plot (SDP)~\cite{Aubert:2004iu}.
The new coordinates are
\begin{equation}
 \label{equ:square_dalitz_mprim_definition}
 m' \equiv \frac{1}{\pi} \arccos
  \left(
   2 \frac{m_0 - m_0^\mathrm{min}}{m_0^\mathrm{max} - m_0^\mathrm{min}}
   - 1
  \right) \;,
\end{equation}
\begin{equation}
 \theta' \equiv \frac{1}{\pi} \theta_0
  \quad
  \left( = \frac{1}{\pi} \theta^{-0} \right) \;,
\end{equation}
where $m_0 = \sqrt{s_0}$,
$m_0^\mathrm{max} = m_{B^0} - m_{\pi^0}$
and  $m_0^\mathrm{min} = 2m_{\pi^+}$
are the kinematic limits of $m_0$,
and $\boldsymbol{J}$ is the Jacobian of the transformation.
The determinant of the Jacobian is given by
\begin{equation}
 |\det \boldsymbol{J}| = 4|\vec{p}_+| |\vec{p}_0| m_0
  \cdot \frac{m_0^\mathrm{max} - m_0^\mathrm{min}}{2} \pi \sin (\pi m')
  \cdot \pi \sin(\pi \theta') \;,
\end{equation}
where $\vec{p}_+$ and $\vec{p}_0$
are the three momenta of $\pi^+$ and $\pi^0$
in the $\pi^+\pi^-$ rest frame.

\clearpage
\newpage

\section{Event Selection and Reconstruction
 \label{sec:selection_and_reconstruction}
}
To reconstruct candidate $\bz\to\pip\pim\piz$ decays,
charged tracks reconstructed with the CDC and SVD are required to
originate from the interaction point (IP) and to have
transverse momenta greater than 0.1 GeV/$c$.
We distinguish charged kaons from pions based on
a kaon (pion) likelihood $\mathcal{L}_{K(\pi)}$
derived from the TOF, ACC and $dE/dx$ measurements in the CDC.
Tracks that are positively identified as electrons are rejected.

Photons are identified as isolated ECL clusters
that are not matched to any charged track.
We reconstruct $\piz$ candidates from pairs of photons
detected in the barrel (end-cap) ECL with $E_\gamma > 0.05$ (0.1) GeV,
where $E_\gamma$ is the photon energy measured with the ECL. 
Photon pairs with momenta greater than 0.1 GeV/$c$ in the
laboratory frame and
with an invariant mass between 0.1178 GeV/$c^2$ to 0.1502 GeV/$c^2$,
roughly corresponding to $\pm 3\sigma$ of the mass resolution,
are used as $\piz$ candidates.

We identify $B$ meson decays 
using the energy difference $\dE\equiv E_B^{\rm cms}-E_{\rm beam}^{\rm cms}$ and
the beam-energy constrained mass $\mb\equiv\sqrt{(E_{\rm beam}^{\rm cms})^2-
(\pbstar)^2}$, where $E_{\rm beam}^{\rm cms}$ is
the beam energy in the cms, and
$E_B^{\rm cms}$ and $\pbstar$ are the cms energy and momentum of the
reconstructed $B$ candidate, respectively.

We select candidates in a fit region that is defined as
$-0.2\,\mathrm{GeV} < \dE < 0.2\,\mathrm{GeV}$
 and $5.2\,\mathrm{GeV}/c^2 < \mb < 5.3\,\mathrm{GeV}/c^2$.
The fit region consists of the signal region that is
defined as $-0.1\,\mathrm{GeV} < \dE < 0.08\,\mathrm{GeV}$
 and $\mb >5.27\,\mathrm{GeV}/c^2$,
and the complement, called the sideband region, which is
dominated by background events.

The vertex position for the $\bz\to\pip\pim\piz$ decay 
is reconstructed using charged tracks that have
enough SVD hits~\cite{Tajima:2003bu}.
The $\ftag$ vertex is obtained with well 
reconstructed tracks that are not assigned to $\fCP$.
A constraint on the interaction-region profile 
in the plane perpendicular to the beam axis
is also used with the selected tracks.

The $b$-flavor of the accompanying $B$ meson is identified
from inclusive properties of particles
that are not associated with the reconstructed $\bz \to \fCP$ 
decay. 
We use two parameters, the $b$-flavor charge $q_\mathrm{tag}$ and $r$,
to represent the tagging information~\cite{Kakuno:2004cf}.
The parameter $r$ is an event-by-event,
Monte Carlo (MC) determined flavor-tagging dilution factor
that ranges from $r=0$ for no flavor
discrimination to $r=1$ for unambiguous flavor assignment.
It is used only to sort data into six $r$ intervals.
The wrong tag fractions for the six $r$ intervals, 
$w_l~(l=1,6)$, and differences 
between $\bz$ and $\bzb$ decays, $\dwl$,
are determined 
using a high-statistics control sample of semileptonic and 
hadronic $b\to c$ decays~\cite{Kakuno:2004cf,Abe:2004mz,Chen:2005dr}.

The dominant background for the $\bz \to \pi^+\pi^-\pi^0$ signal comes from
continuum events. 
To distinguish these jet-like events from the spherical $B$ decay signal 
events, we combine a set of variables that characterize the event topology
into a signal (background) likelihood variable $\cal L_{\rm sig(bkg)}$,
and impose requirements on the likelihood ratio
$\rsigbkg \equiv \lsig/(\lsig+\lbkg)$ that depend on
the quality of flavor tagging.

When more than one candidate in the same event is found
in the fit region, we select the best candidate
based on the reconstructed $\piz$ mass and $\rsigbkg$.

After the best candidate selection, we reconstruct
the Dalitz variables $s_+$, $s_0$ and $s_-$
from 1) the four momenta of the $\pi^+$ and $\pi^-$,
2) the helicity angle of the $\rho^0$
(i.e., the helicity angle of the $\pi^+ \pi^-$ system),
and 3) the relation:
${m_{\bz}}^2 + 2{m_{\pip}}^2 + {m_{\piz}}^2 = s_+ + s_- + s_0$.
Note that the energy of $\pi^0$ is not explicitly used here,
which improves the resolution of the Dalitz plot variables.
We reject candidates that are located in
one of the following regions in the Dalitz plot:
$\sqrt{s_0} > 0.95$ GeV/$c^2$ and $\sqrt{s_+}>1.0$ GeV$/c^2$ and $\sqrt{s_-} > 1.0$ GeV/$c^2$;
$\sqrt{s_0} < 0.55$ GeV/$c^2$ or $\sqrt{s_+}<0.55$ GeV$/c^2$ or $\sqrt{s_-} < 0.55$ GeV/$c^2$.
In these regions, the fraction of $\bz\to\rho\pi$ signal is small.
However,
 radial excitations
($\rho(1450)$ and $\rho(1700)$)
are the dominant $\bz \rightarrow \pi^+\pi^-\pi^0$ contributions
 in the region with $\sqrt{s} > 1.0$ GeV$/c^2$.
Since the amplitudes
of the radial excitations
 are in general independent of
the amplitude of the $\rho(770)$,
they are considered to be background in our analysis;
 vetoing the high mass region considerably reduces
the systematic uncertainties due to their contributions.

\begin{figure}[htb]
 \includegraphics[width=0.4\textwidth]{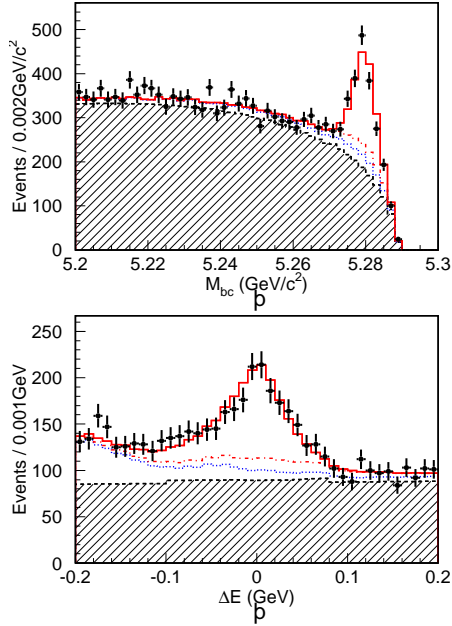}
 \caption{The $\mbc$(upper) and $\dE$(lower) distributions
 within the $\dE$ and $\mbc$ signal regions.
 The histograms are cumulative. Solid, dot-dashed, dotted and dashed
 hatched histograms correspond to correctly reconstructed signal,
 SCF, $B\bbar$, and continuum PDFs, respectively.
 }
 \label{fig:mbc_and_de_plots}
\end{figure}

Figure \ref{fig:mbc_and_de_plots} shows the $\mbc$
and $\dE$ distributions for the reconstructed
$B^0 \rightarrow \pi^+\pi^-\pi^0$ candidates
within the $\dE$ and $\mbc$ signal regions, respectively.
The signal yield is determined from an unbinned four-dimensional
extended-maximum-likelihood fit to the $\dE$-$\mbc$
and Dalitz plot distribution in the fit region defined as
$\mbc > 5.2\,\mathrm{GeV}/c^2$ and $-0.2\,\mathrm{GeV} < \dE < 0.2
\,\mathrm{GeV}$,
where the Dalitz plot distribution is only used for the events
inside the $\dE$-$\mbc$ signal region.
The $\dE$-$\mbc$ distribution of signal is modeled with binned histograms
obtained from MC,
where the correlation between $\dE$ and $\mbc$,
 the dependence on $p_{\pi^0}$,
and the difference between data and MC are taken into account.
We also take into account incorrectly reconstructed signal events,
 which we call self cross feed (SCF)
 and which amounts to $\sim 20\%$ of the signal.
In a SCF event, one of the three pions in $\fCP$ is swapped
 with a pion in $f_\mathrm{tag}$ or the $\pi^0$ in $\fCP$ is misreconstructed.
We give the details of the $\dE$-$\mb$ and Dalitz plot PDFs of the SCF events
 in appendix \ref{sec:appendix_pdf_definitions}.
For continuum,
we use the ARGUS parameterization~\cite{Albrecht:1990am}
for $\mbc$
and a linear function for $\dE$.
The $\dE$-$\mbc$ distribution of $B\overline{B}$ background
is modeled by binned histograms based on MC.
The Dalitz plot distributions for all components
are modeled in the same way as the time-dependent fit
described later,
but integrated over the dimensions of the proper time difference,
$\Delta t$, and the flavor of the tag side $B$, $q_\mathrm{tag}$.
The fit yields $987 \pm 42$
$B^0 \rightarrow \pi^+\pi^-\pi^0$ events in the signal region,
where the errors are statistical only.

\section{Determination of the contributions from radial excitations
 \label{sec:lineshape_determine}
}
\begin{figure}[htb]
 \includegraphics[width=0.32\textwidth]{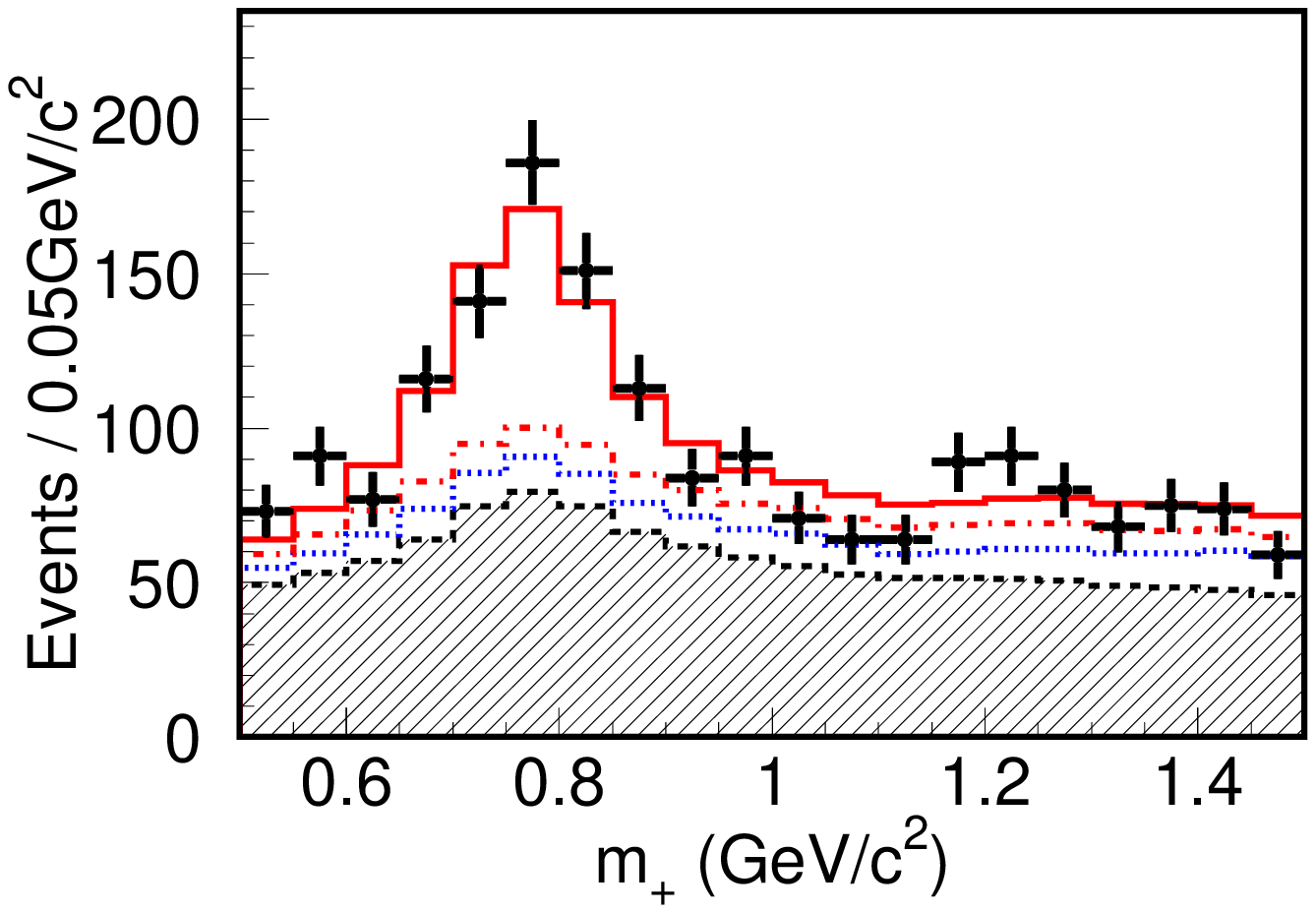}
 \includegraphics[width=0.32\textwidth]{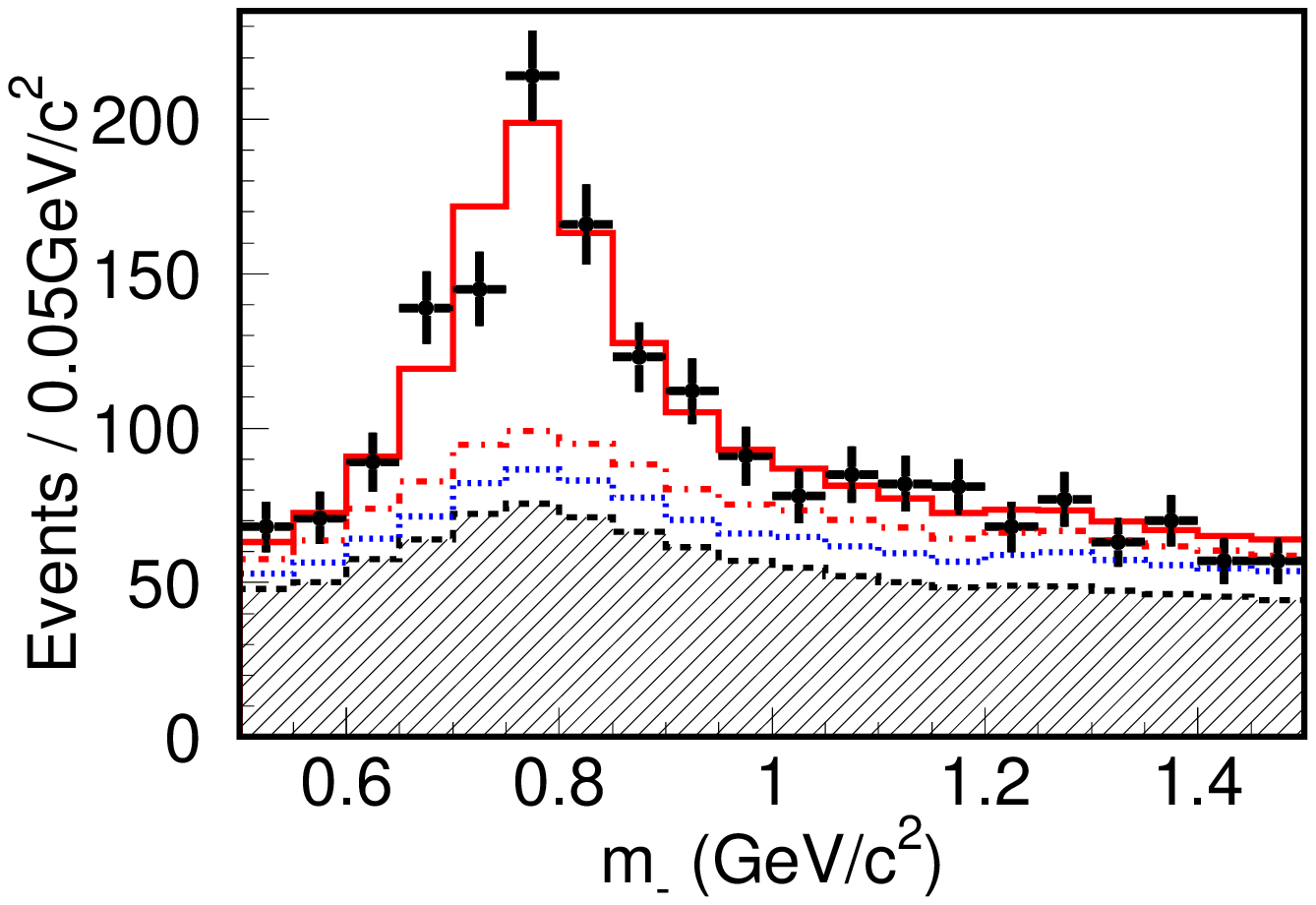}
 \includegraphics[width=0.32\textwidth]{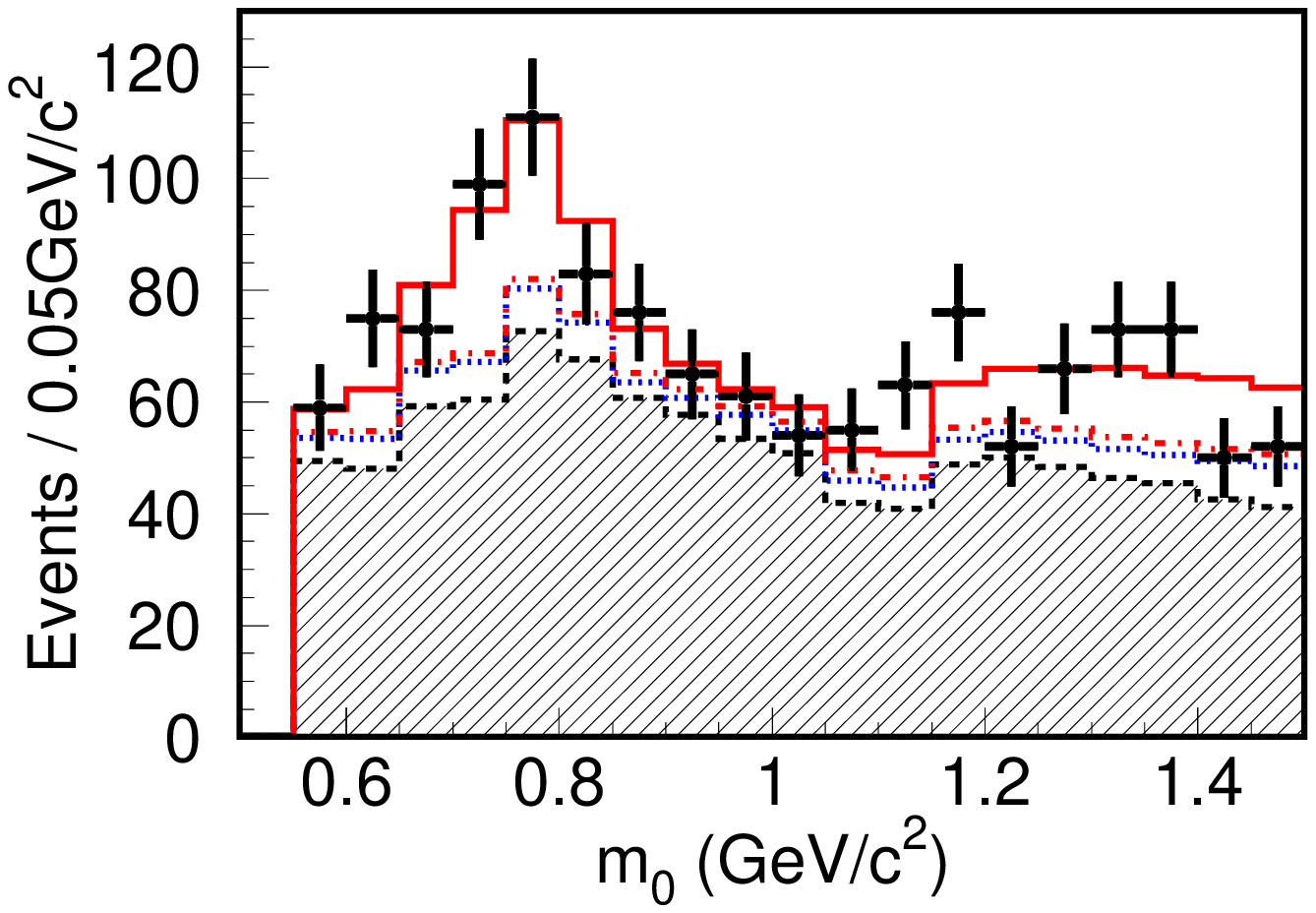}
 \caption{
 Mass distributions and fitted lineshapes
 in $\rho^+\pi^-$ (left), $\rho^-\pi^+$ (middle),
 and $\rho^0\pi^0$ (right) enhanced regions.
 The histograms are cumulative. Solid, dot-dashed, dotted and dashed
 hatched histograms correspond to correctly reconstructed signal,
 SCF, $B\bbar$, and continuum PDFs, respectively.
 Note
 that there are feed-downs from
 other quasi-two-body components than
 those of interest,
 especially in the high-mass regions.
 For example,
 the
 high-mass region ($m_0 \gtrsim 1.0\, \mathrm{GeV}/c^2$)
 of the $\rho^0\pi^0$ enhanced region (right)
 includes large contributions from
 $\rho^\pm \pi^\mp$.
 }
 \label{fig:mass_plot_upto_high_mass}
\end{figure}

\begin{figure}[htb]
 \includegraphics[width=0.6\textwidth]{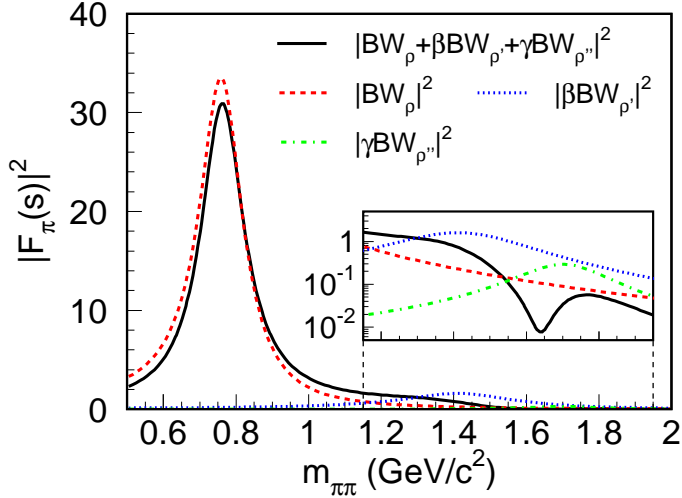}
 \caption{
 A schematic figure of the fit result of the lineshape
 and the contributions from radial excitations.
 Note that 
 our definition of
 $F_{\pi}(s)$ does not include the factor $1/(1+\beta+\gamma)$
 as in equation (\ref{equ:fpi_lineshape_beta_gamma}).
 One can see that the $\rho(770)$ and $\rho(1450)$
 destructively interfere with each other
 near $\sqrt{s} \equiv m_{\pi\pi} = 1.4 (\mathrm{GeV}/c^2)$,
 which means that
 the $\rho(1450)$ has a large impact on
 the phase of $F_{\pi}(s)$
 although the absolute value of $|F_{\pi}(s)|$ is not much affected.
 }
 \label{fig:lineshape_fitted_schematic}
\end{figure}

Using the same data sample as described above
but performing a time-integrated Dalitz plot fit
 with a wider Dalitz plot acceptance,
$0.55\,\mathrm{GeV}/c^2 < \sqrt{s_0} < 1.5\,\mathrm{GeV}/c^2$
 or $\sqrt{s_+}<1.5\,\mathrm{GeV}/c^2$
 or $\sqrt{s_-} < 1.5\,\mathrm{GeV}/c^2$,
we determine the $\rho$ lineshape,
i.e. the phases and amplitudes of the coefficients $\beta$
and $\gamma$ in equation (\ref{equ:fpi_lineshape_beta_gamma}),
which we use for all of the decay amplitudes.
In this fit, we use the PDG values~\cite{Eidelman:2004wy}
for the masses and widths of the $\rho(1450)$ and $\rho(1700)$.
The fit yields
\begin{equation}
 \label{equ:nominal_beta_gamma}
 |\beta| = 0.30^{+0.06}_{-0.05}\;, \quad
  \arg \beta = (213^{+15}_{-19})^\circ\;, \quad
  |\gamma| = 0.07\pm 0.03\;, \quad
  \arg \gamma = (91^{+27}_{-32})^\circ \; .
\end{equation}
The mass distributions and fit results are shown in
 Fig.~\ref{fig:mass_plot_upto_high_mass}.
Figure \ref{fig:lineshape_fitted_schematic} schematically shows
how the radial excitations
contribute to our fit result.
Note that the values given above for
 $\beta$ and $\gamma$ and their errors 
are not meaningful measurements of physics parameters
but rather are quantities needed for the time-dependent Dalitz fit.
This is because these parameters are determined from
the interference region,
the interference between $\rho^+\pi^-$ and $\rho^-\pi^+$, etc.,
and depend on the unfounded common lineshape assumption
of the equation
(\ref{equ:f_kappa_definition_with_unique_lineshape_assumption}).
However, because statistics are still low,
the time-dependent Dalitz analysis would not be possible if
 we were to discard the common lineshape assumption.

Thus, it is important to
 determine the common or {\it average} lineshape as well as
 obtain an upper limit on the deviation
from the average lineshape for each of
the 6 decay amplitudes.
%
For this purpose,
we put constraints on
additional amplitudes that describe
1) the excess in the high mass region, $\sqrt{s} > 0.9\,\mathrm{GeV}/c^2$,
and 2) the interferences between radial excitations
and the lowest resonance, the $\rho(770)$:
interferences between $\rho(770)^+\pi^-$
and $\rho(1450)^-\pi^+$, etc.
The nominal fit is performed
with the average lineshape determined above,
fixing all of the additional amplitudes to zero.
When floating the additional amplitudes for the other resonances,
we obtain results consistent with zero for all of the additional amplitudes
but with large uncertainties compared to the
errors for the average lineshape parameters above.
We use the fit result with the additional lineshape parameters floating
including their uncertainties in the systematic error study.


\section{Time-dependent Dalitz plot analysis
 \label{sec:time_depend_analysis_pdf}
}
To determine the 26 Dalitz plot parameters,
we define the following event-by-event PDF:
\begin{eqnarray}
 \label{equ:dembc_dalitz_simultaneous-total_pdf}
 p(\vec{x})
  &\equiv&
  \nonumber
  f_\mathrm{sig} \;
  p^l_\mathrm{sig}(\Delta E, \mbc; m', \theta';\Delta t, q_\mathrm{tag}; p_{\pi^0})\\ 
  \nonumber
  & & + f_{BB} \;
  p^l_{BB}(\Delta E, \mbc; m', \theta';\Delta t, q_\mathrm{tag})\\
  & &
  + f_{qq} \;
  p^l_{qq}(\Delta E, \mbc; m', \theta';\Delta t, q_\mathrm{tag})
  \;,
\end{eqnarray}
where
$p^l_\mathrm{sig}$, $p^l_{BB}$ and $p^l_{qq}$ are PDF's 
for signal, $B\bbar$ background and continuum background
in flavor tagging region $l$, respectively,
and
$f_\mathrm{sig}$, $f_{BB}$ and $f_{qq}$ are
the corresponding fractions that satisfy
\begin{equation}
 f_\mathrm{sig} + f_{BB} + f_{qq} = 1 \;.
\end{equation}
The vector $\vec{x}$,
the argument of $p$,
 corresponds to a set of event-by-event variables:
\begin{equation}
 \vec{x} \equiv
  (\Delta E, \mbc; m', \theta';\Delta t, q_\mathrm{tag}, l; p_{\pi^0}) \;.
\end{equation}
The PDF for signal events consists of
a PDF for the correctly reconstructed events
 $p_\mathrm{true}^l$ and
PDFs for SCF(Self Cross Feed) events $p^l_{i}$ ($i = \mathrm{NR,CR}$):
\begin{equation}
 \label{equ:signal_pdf_definition}
p^l_\mathrm{sig} = 
 \frac{p_\mathrm{true}^l(\dE, \mbc; m',\theta';\Delta t, q_\mathrm{tag}; p_{\pi^0})
 +\sum_{i=\mathrm{NR,CR}} p^l_{i}(\dE, \mbc; m', \theta';\Delta t, q_\mathrm{tag})
 }
 {
 n_\mathrm{true} \quad + \quad \sum_{i = \mathrm{NR,CR}} n_i
 }
  \;, 
\end{equation}
where NR and CR represent the $\pi^0$ (neutral)
 replaced and $\pi^\pm$ (charged) replaced SCF, respectively,
and $n$ are the integrals of the PDFs.
A detailed description of the PDF for each component can be found
in appendix \ref{sec:appendix_pdf_definitions}.

With the PDF defined above,
we form the likelihood function
\begin{equation}
 \mathcal{L} \equiv \prod_i p(\vec{x}_i) \;,
\end{equation}
where $i$ is an index over events.
We perform an unbinned-maximum-likelihood fit
and determine the 26 Dalitz plot parameters
using the likelihood function
with the signal fraction and the lineshape parameters
 obtained in
 Sec.~\ref{sec:selection_and_reconstruction}
and Sec.~\ref{sec:lineshape_determine}, respectively.

%

\section{Fit result}
A fit to the 2,824 events in the signal region yields
the result listed in Table \ref{tbl:dt_all_data}.
The correlation matrix of the 26 parameters
after combining statistical and systematic errors
is shown in the appendix \ref{sec:appendix_correlation}.
Figure \ref{fig:dalitz_plot}
shows the projections of the square Dalitz plot in data
with the fit result superimposed.
We also show the mass and helicity distribution
for each $\rho \pi$ enhanced region
along with projections of the fit (Fig.~\ref{fig:mass_helicity_plot}).
Figure \ref{fig:dt_plot_result}
shows the $\Delta t$ distributions and background 
subtracted asymmetries.
We define the asymmetry in each $\Delta t$ bin
by
 $(N_{q_\mathrm{tag}=+1} - N_{q_\mathrm{tag}=-1}) / (N_{q_\mathrm{tag}=+1} + N_{q_\mathrm{tag}=-1})$,
where $N_{q_\mathrm{tag}=+1(-1)}$ corresponds the
background subtracted number of events with $q_\mathrm{tag}=+1(-1)$.
The $\rho^- \pi^+$ enhanced region shows a significant asymmetry,
corresponding to a non-zero value of $U^-_-$.

\subsection{Treatment of statistical errors}
With a Toy MC study, we check the pull distribution,
where the pull is defined
 as the residual divided by the MINOS error.
Here, the MINOS error,
which corresponds to the deviation from the best fit parameter
when $-2\ln(\mathcal{L}/\mathcal{L}_\mathrm{max})$ is changed by one,
is an estimate of the statistical error.
Although the pull is expected to follow
a Gaussian distribution with unit width,
we find that the width of the pull distribution
 tends to be significantly larger than one for the interference terms
due to small statistics.
To restore the pull width to unity,
we multiply the MINOS errors of the interference terms by a factor of 1.19,
which is the average pull width for the interference terms
obtained above,
and quote the results as the statistical errors.
For the non-interfering terms,
we quote the MINOS errors without the correction factor.

%
%
%
\begin{figure}[htb]
\includegraphics[width=0.6\textwidth]{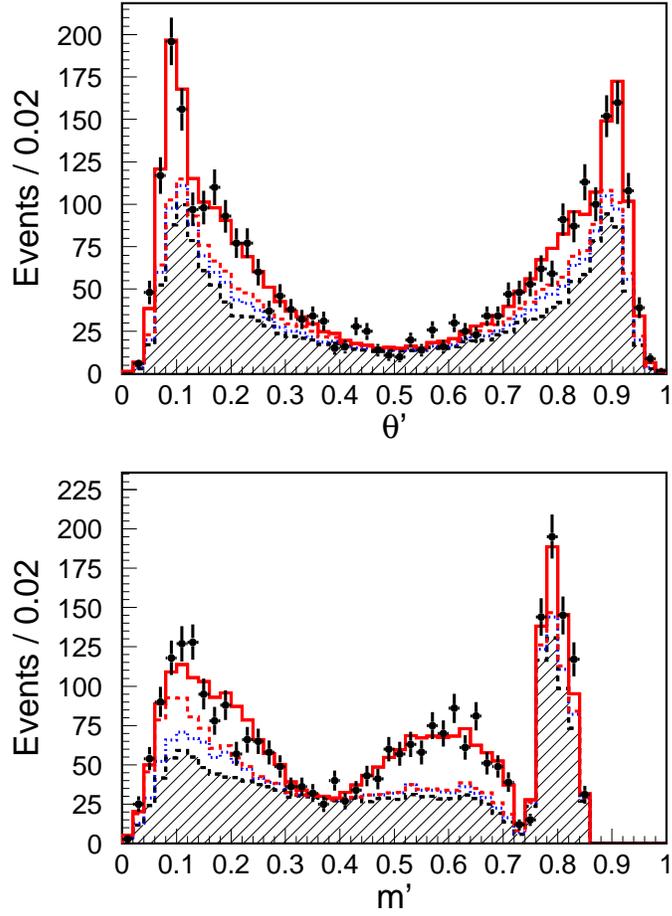}
 \caption{Distributions of $\theta'$ (upper) and $m'$ (lower) with
 fit results.
 The histograms are cumulative. Solid, dot-dashed, dotted and dashed
 hatched histograms correspond to correctly reconstructed signal,
 SCF, $B\bbar$, and continuum PDFs, respectively.
 }
\label{fig:dalitz_plot}
\end{figure}
\begin{figure}[htb]
 \includegraphics[width=0.32\textwidth]{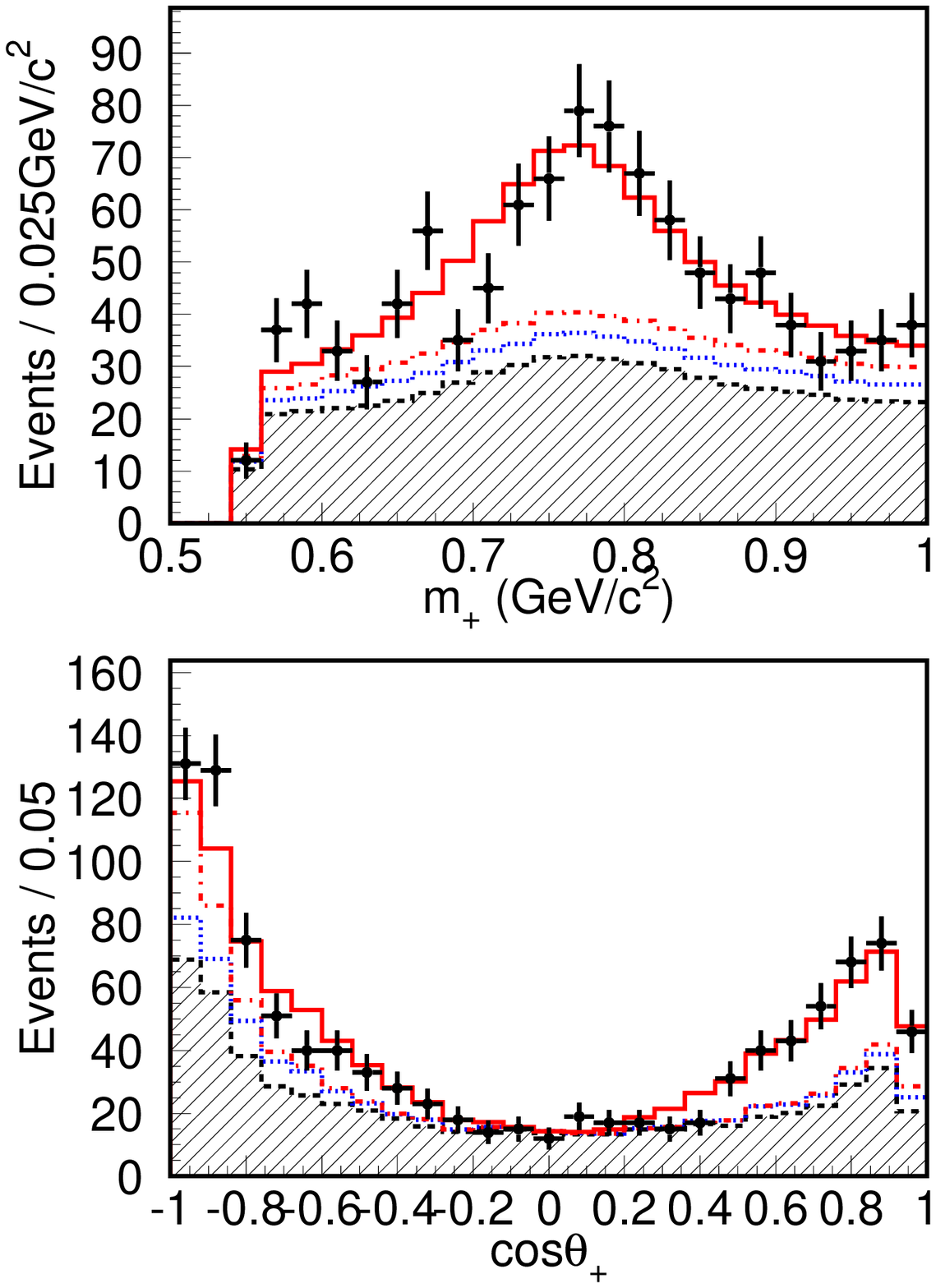}
 \includegraphics[width=0.32\textwidth]{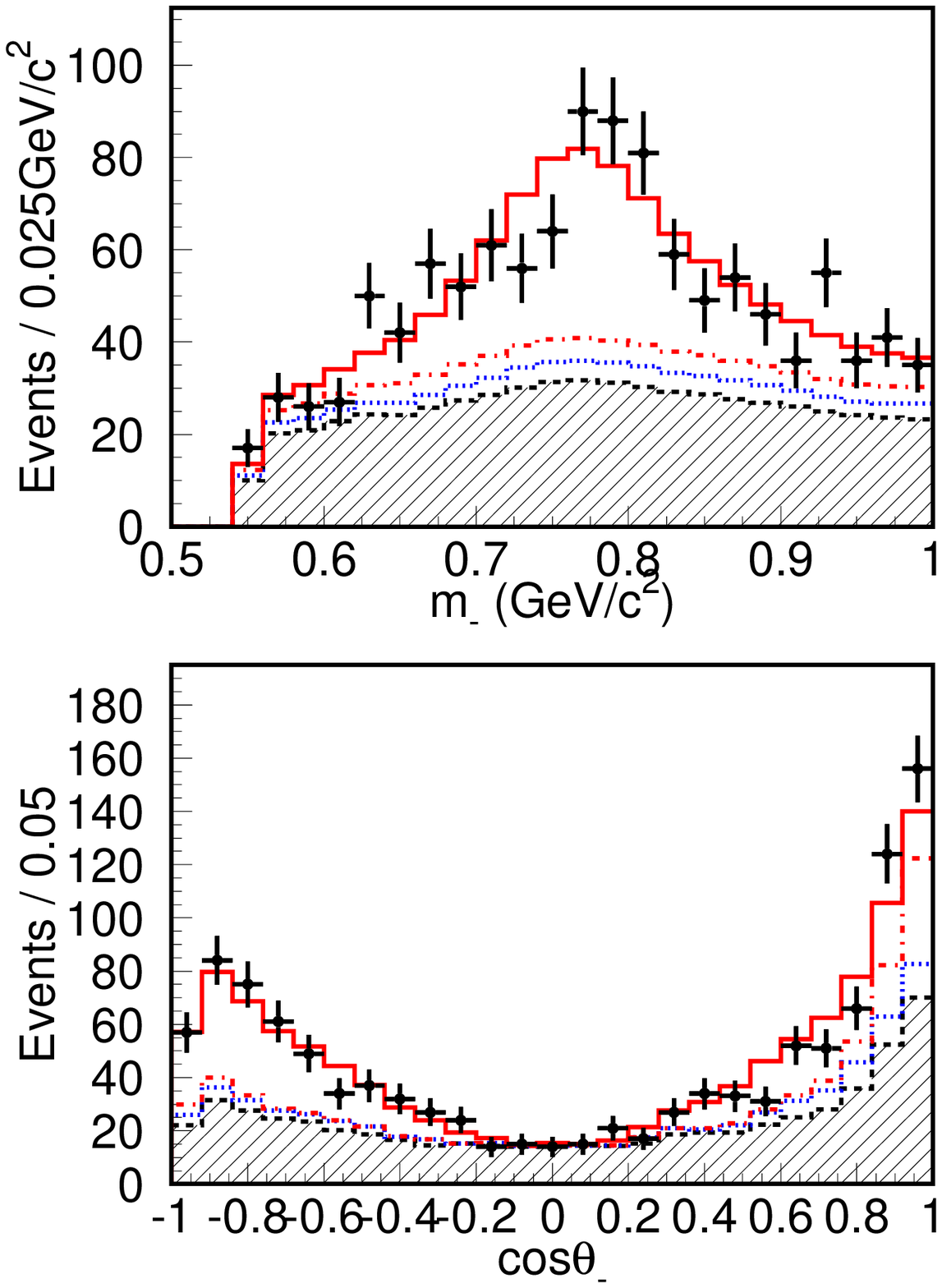}
 \includegraphics[width=0.32\textwidth]{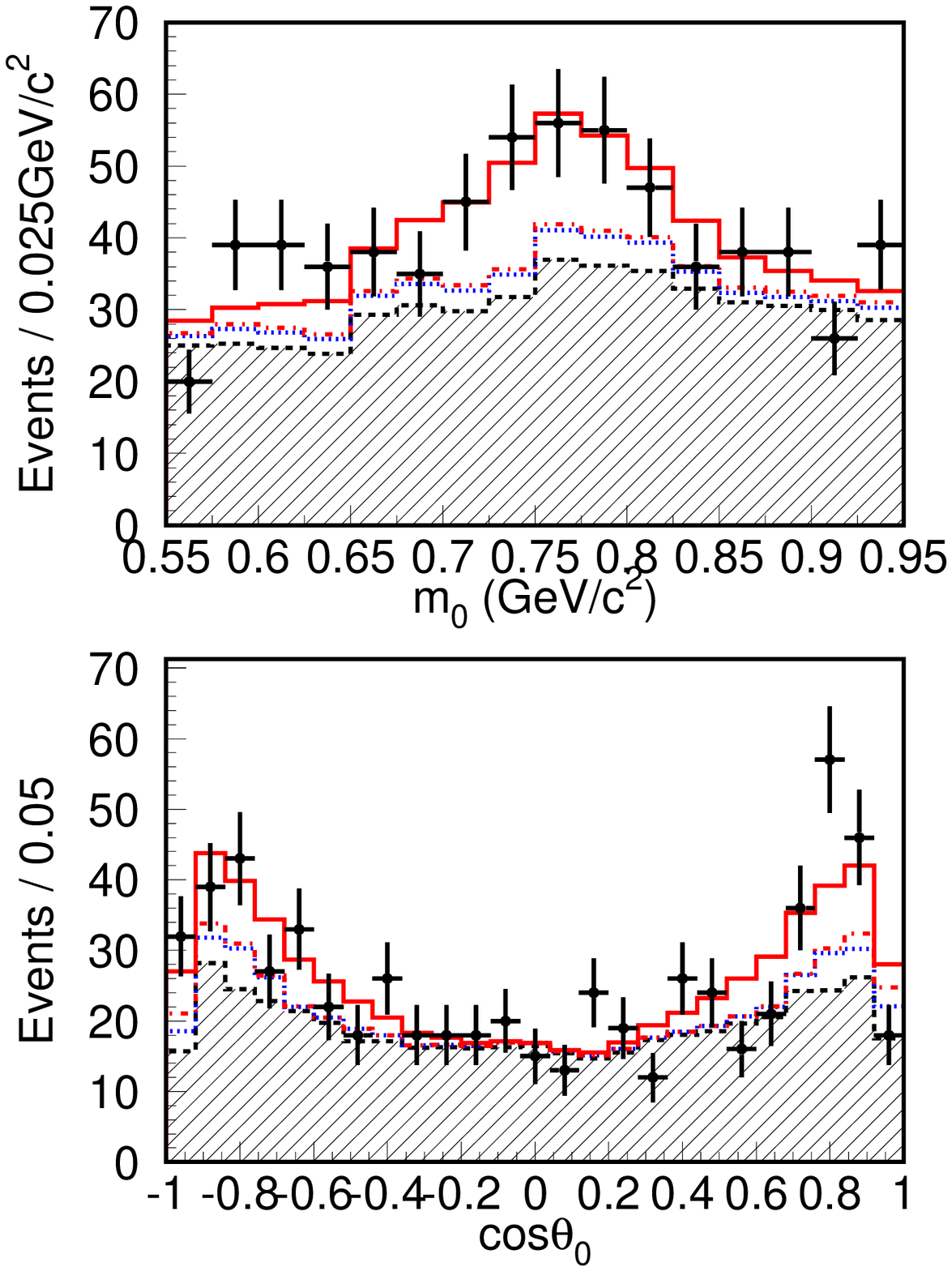}
 \caption{
 Mass (upper) and helicity (lower) distribution of
 $\rho^+ \pi^-$ (left), $\rho^- \pi^+$ (middle), 
 and $\rho^0 \pi^0$ (right) enhanced regions.
 The histograms are cumulative. Solid, dot-dashed, dotted and dashed
 hatched histograms correspond to correctly reconstructed signal,
 SCF, $B\bbar$, and continuum PDFs, respectively.
 }
 \label{fig:mass_helicity_plot}
\end{figure}
\begin{figure}[htbp]
 \includegraphics[width=0.32\textwidth]{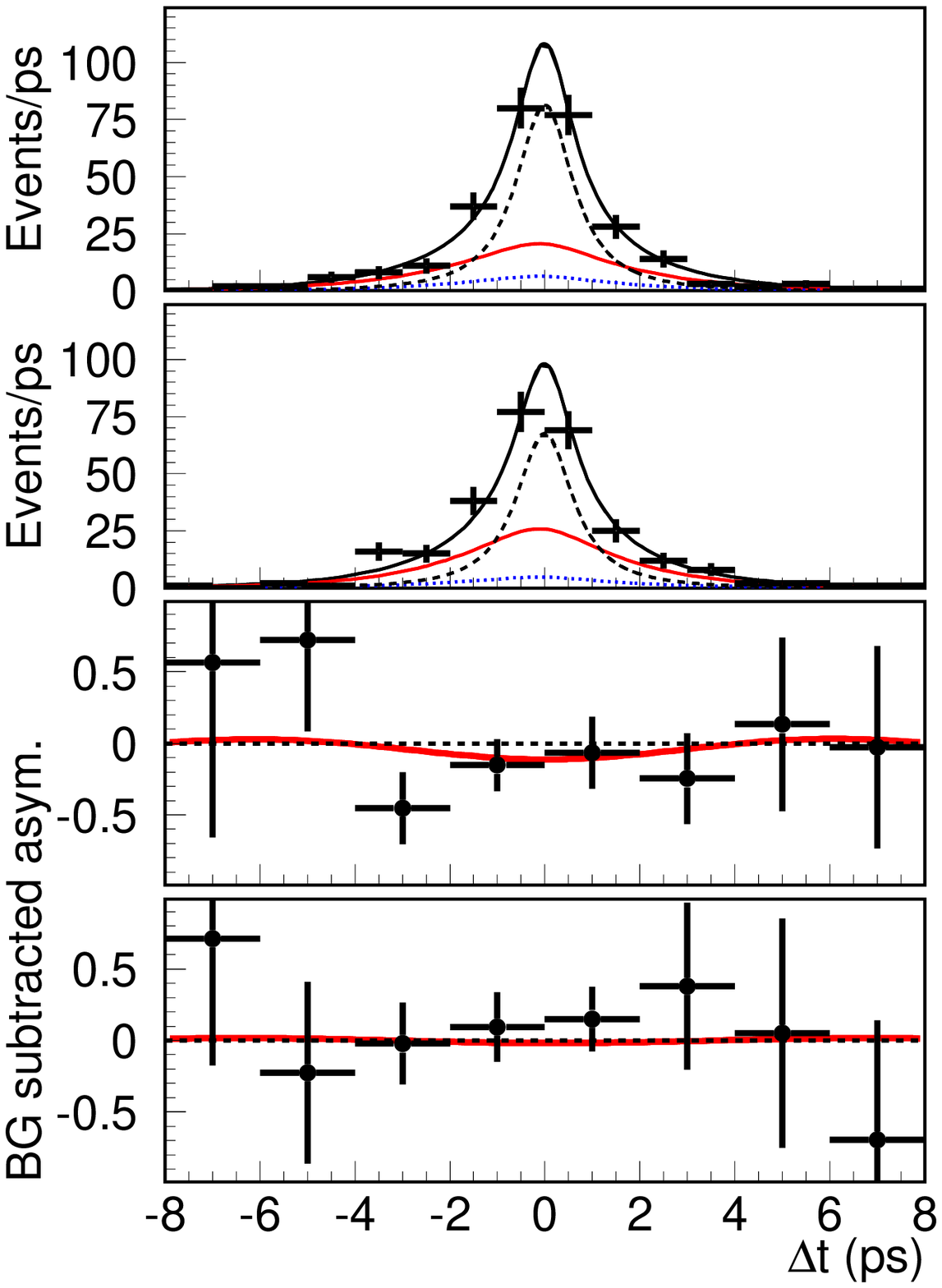}
 \includegraphics[width=0.32\textwidth]{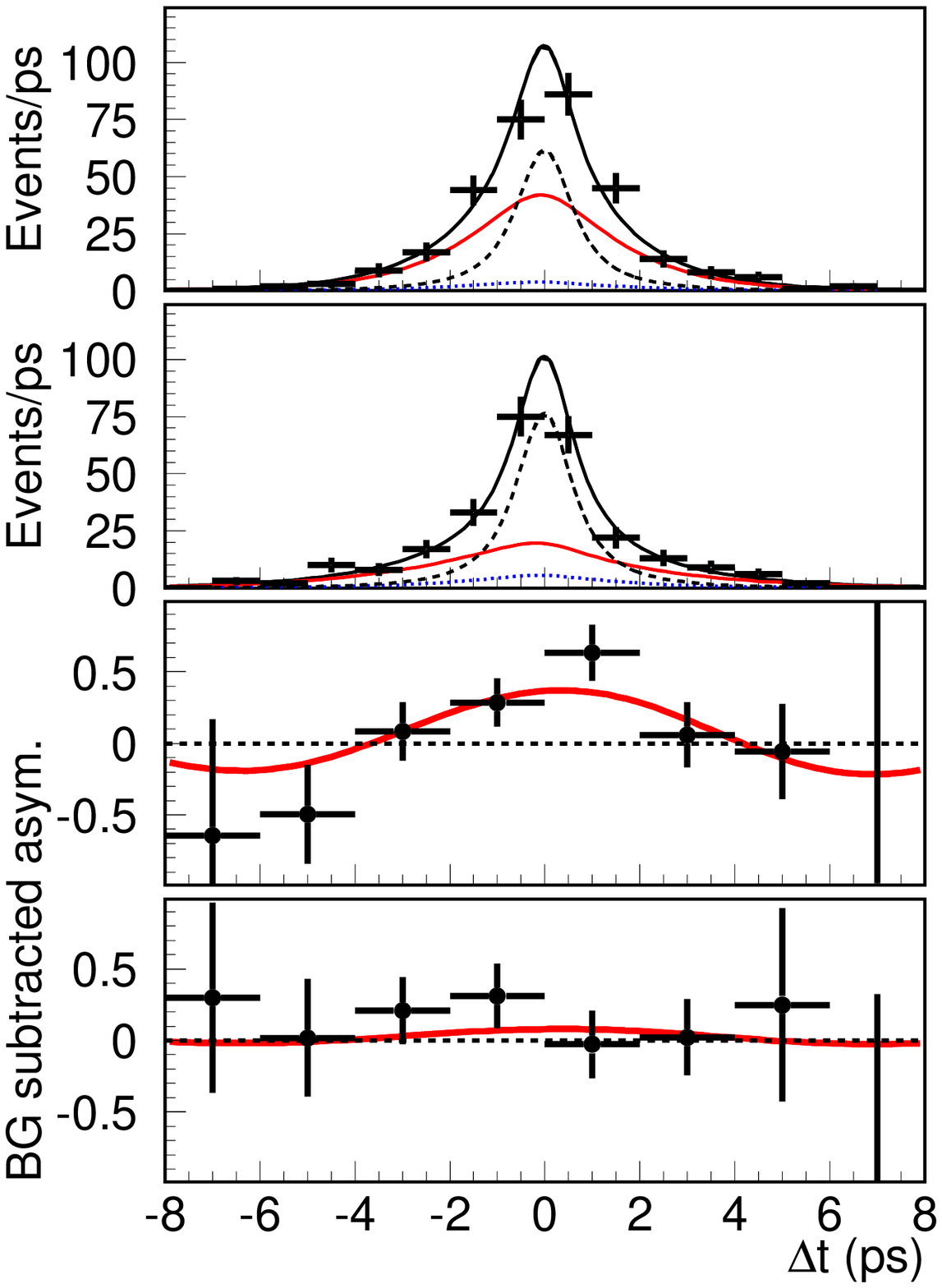}
 \includegraphics[width=0.32\textwidth]{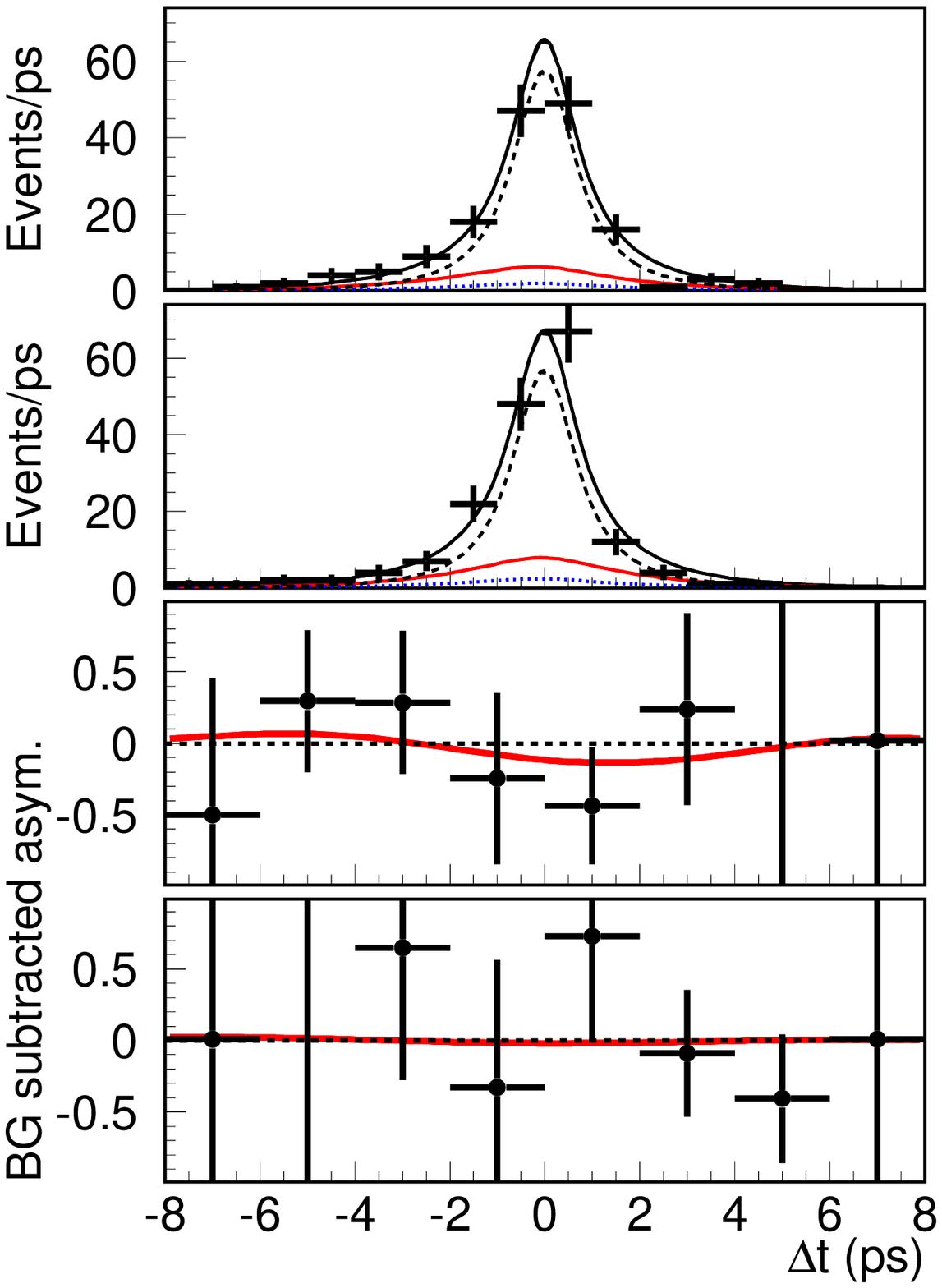}
 \caption{
 Proper time distributions of good tag ($r>0.5$) regions
 for $f_\mathrm{tag} = \bz$ (upper) and
 $f_\mathrm{tag} = \bzb$ (middle upper),
 in $\rho^+\pi^-$ (left), $\rho^- \pi^+$ (middle),
 $\rho^0\pi^0$ (right) enhanced regions,
 where solid (red), dotted, and dashed curves
 correspond to signal, continuum, and $B\bbar$ PDFs.
 The middle lower and lower plots
 show the background subtracted asymmetries
 in the good tag ($r>0.5$) and poor tag ($r<0.5$) regions, respectively.
 The significant asymmetry in the $\rho^- \pi^+$ enhanced region
 (middle)
 corresponds to a non-zero value of $U^-_-$.
 \label{fig:dt_plot_result}
 }
\end{figure}
\begin{table}[htb]
\caption{ Results of the time-dependent Dalitz fit.}
\label{tbl:dt_all_data}
\begin{tabular}
{@{\hspace{0.5cm}}l@{\hspace{0.5cm}}|@{\hspace{0.5cm}}c@{\hspace{0.5cm}}}
\hline \hline
                       & Fit Result                       \\
\hline
$U^+_+$                & $+1 (\rm{fixed})$                  \\
$U^+_-$                & $+1.28\pm 0.13(\rm{stat.})\pm 0.08 (\rm{syst.})$	 \\
$U^+_0$                & $+0.30\pm 0.06(\rm{stat.})\pm 0.05 (\rm{syst.})$	 \\
$U^{+, \mathrm{Re}}_{+-}$ & $+0.62\pm 0.80(\rm{stat.})\pm 0.57 (\rm{syst.})$	 \\
$U^{+, \mathrm{Re}}_{+0}$ & $+0.41\pm 0.52(\rm{stat.})\pm 0.46 (\rm{syst.})$	 \\
$U^{+, \mathrm{Re}}_{-0}$ & $+0.49\pm 0.65(\rm{stat.})\pm 0.44 (\rm{syst.})$	 \\
$U^{+, \mathrm{Im}}_{+-}$ & $+0.86\pm 0.83(\rm{stat.})\pm 0.49 (\rm{syst.})$	 \\
$U^{+, \mathrm{Im}}_{+0}$ & $-0.53\pm 0.39(\rm{stat.})\pm 0.47 (\rm{syst.})$	 \\
$U^{+, \mathrm{Im}}_{-0}$ & $-1.72\pm 0.69(\rm{stat.})\pm 0.53 (\rm{syst.})$	 \\
\hline
$U^-_+$                & $+0.22\pm 0.15(\rm{stat.})\pm 0.10 (\rm{syst.})$	 \\
$U^-_-$                & $-0.62\pm 0.17(\rm{stat.})\pm 0.09 (\rm{syst.})$	 \\
$U^-_0$                & $+0.14\pm 0.11(\rm{stat.})\pm 0.09 (\rm{syst.})$	 \\
$U^{-, \mathrm{Re}}_{+-}$ & $-1.70\pm 1.59(\rm{stat.})\pm 0.77 (\rm{syst.})$	 \\
$U^{-, \mathrm{Re}}_{+0}$ & $-2.46\pm 1.39(\rm{stat.})\pm 0.86 (\rm{syst.})$	 \\
$U^{-, \mathrm{Re}}_{-0}$ & $-0.70\pm 1.59(\rm{stat.})\pm 0.86 (\rm{syst.})$	 \\
$U^{-, \mathrm{Im}}_{+-}$ & $-2.21\pm 1.71(\rm{stat.})\pm 1.03 (\rm{syst.})$	 \\
$U^{-, \mathrm{Im}}_{+0}$ & $-0.83\pm 0.98(\rm{stat.})\pm 0.65 (\rm{syst.})$	 \\
$U^{-, \mathrm{Im}}_{-0}$ & $-0.79\pm 1.59(\rm{stat.})\pm 1.05 (\rm{syst.})$	 \\
\hline
$I_+$                  & $-0.03\pm 0.11(\rm{stat.})\pm 0.06 (\rm{syst.})$	 \\
$I_-$                  & $+0.11\pm 0.11(\rm{stat.})\pm 0.05 (\rm{syst.})$	 \\
$I_0$                  & $+0.02\pm 0.09(\rm{stat.})\pm 0.06 (\rm{syst.})$	 \\
$I^{\mathrm{Re}}_{+-}$ & $+1.62\pm 2.65(\rm{stat.})\pm 1.23 (\rm{syst.})$	 \\
$I^{\mathrm{Re}}_{+0}$ & $+1.45\pm 2.41(\rm{stat.})\pm 1.12 (\rm{syst.})$	 \\
$I^{\mathrm{Re}}_{-0}$ & $-0.65\pm 1.63(\rm{stat.})\pm 1.49 (\rm{syst.})$	 \\
$I^{\mathrm{Im}}_{+-}$ & $-1.76\pm 2.42(\rm{stat.})\pm 1.31 (\rm{syst.})$	\\
$I^{\mathrm{Im}}_{+0}$ & $+0.00\pm 2.06(\rm{stat.})\pm 1.15 (\rm{syst.})$	\\
$I^{\mathrm{Im}}_{-0}$ & $-2.58\pm 1.72(\rm{stat.})\pm 1.33 (\rm{syst.})$	\\
 \hline
\end{tabular}
\end{table}

\clearpage

\section{Systematic Uncertainties
 \label{sec:systematic_errors}
}
Tables \ref{tbl:systematics1}-\ref{tbl:systematics3}
list the systematic errors for the 26 time-dependent
Dalitz plot parameters.
The total systematic error is obtained by adding each
source of systematic uncertainty in quadrature.

\subsection{\boldmath Radial excitations ($\rho'$ and $\rho''$)}
The largest contribution for the interference terms
tends to come from radial excitations.
The systematic error related to the radial excitations
($\rho(1450)$ and $\rho(1700)$, or $\rho'$ and $\rho''$)
 can be categorized into
three classes:
1) uncertainties coming from the lineshape variation,
i.e., the lineshape difference between
each decay amplitude,
2) uncertainties in
external parameters,
 $m_{\rho(1450)}$, $\Gamma_{\rho(1450)}$, 
 $m_{\rho(1700)}$, $\Gamma_{\rho(1700)}$, and
3) uncertainties in the common lineshape parameters
$\beta$ and $\gamma$ used for the nominal fit.

In our nominal fit, we assume all of 6 decay amplitudes
have the same contribution from $\rho(1450)$ and $\rho(1700)$,
i.e.,
we assume equation
 (\ref{equ:f_kappa_definition_with_unique_lineshape_assumption}).
This assumption, however, is not well grounded.
In general, the contributions from $\rho(1450)$ and $\rho(1700)$
can be different for each of the decay amplitudes and thus
the systematic uncertainty from this assumption
must be addressed.
Without the assumption about the higher resonances,
equation
(\ref{equ:f_kappa_definition_with_unique_lineshape_assumption}) becomes
\begin{equation}
 {\mathop{f}^{(-)}}_\kappa
 = T^\kappa_1 {\mathop{F_{\pi}}^{(-)}}{}^\kappa(s_\kappa) \;,
\end{equation}
where
\begin{equation}
 {\mathop{F_{\pi}}^{(-)}}{}^\kappa(s)
  \equiv BW^{GS}_{\rho(770)}(s)
  + (\beta + \Delta {\mathop{\beta}^{(-)}}_\kappa) BW^{GS}_{\rho(1450)}(s)
  + (\gamma + \Delta {\mathop{\gamma}^{(-)}}_\kappa)
  BW^{GS}_{\rho(1700)}(s) \;.
\end{equation}
The variation of the contributions from radial excitations
is described by non-zero
$\Delta \kakkoOverlineBeta {}_\kappa$
and $\Delta \kakkoOverlineGamma {}_\kappa$,
which are 12 complex variables.
We generate various Toy MC samples,
where the input $A^\kappa$ and $\overline{A}{}^\kappa$ are fixed
but the values of
$\Delta \kakkoOverlineBeta {}_\kappa$
and
$\Delta \kakkoOverlineGamma {}_\kappa$
are randomly varied
 according to the constraints
 on $\Delta \kakkoOverlineBeta {}_\kappa$
and
$\Delta \kakkoOverlineGamma {}_\kappa$;
 these constraints are obtained from the
 results in Sec.~\ref{sec:lineshape_determine}, and are
 combined with the isospin relation~\cite{Lipkin:1991st,Gronau:1991dq},
 which improves the constraints.
The statistics for each pseudo-experiment are set to be
large enough so that the statistical uncertainty is negligible.
We assign the variations and the biases
of the fit results
due to the 
$\Delta \kakkoOverlineBeta {}_\kappa$
and
$\Delta \kakkoOverlineGamma {}_\kappa$
variation as systematic errors.

For the masses and widths
of the $\rho(1450)$ and $\rho(1700)$,
 we quote the values from
the PDG~\cite{Eidelman:2004wy}. 
To estimate the systematic error coming from uncertainties
in their parameters,
we generate Toy MC varying the input masses and widths
and fit them with the masses and widths of the nominal fit.
Here, we vary the masses by two times the PDG error
 ($\pm 50 \, \mathrm{MeV}/c^2$ for $\rho(1450)$
and $\pm 40 \, \mathrm{MeV}/c^2$ for $\rho(1700)$)
since the variations between
 independent experiments are much larger than the $1\sigma$ PDG errors,
while we vary the widths by the $\pm 1\sigma$ PDG errors.
We quote the mean shift of the Toy MC ensemble as the systematic errors.
We also take into account the systematic errors
from the uncertainties in $\beta$ and $\gamma$
for the nominal fit (equation (\ref{equ:nominal_beta_gamma}))
 in the same way.

\subsection{SCF}
Systematic errors due to SCF
are dominated by the uncertainty
in the difference between data and MC;
these errors are determined from
the $B \rightarrow D^{(*)}\rho$ control samples
that contain a single $\pi^0$ in the final state.
We vary the amount of SCF by its $1\sigma$ error, which is
$\pm 100\%$ for the CR SCF and ${}^{+30(60)}_{-30}\%$ for the NR SCF
in the DS-I(DS-II);
we quote the difference from the nominal
 fit as the systematic error.
The event fraction for each $r$ region (for CR and NR),
the wrong tag fractions (for CR) 
and lifetime used in the $\Delta t$ PDF (for CR), which are obtained from MC,
are also varied and the difference in the fit results is assigned as systematic
error.

\subsection{Signal Dalitz PDF}
Systematic errors due to the Dalitz PDF for signal
is mainly from the Dalitz plot dependent efficiency.
We take account of
MC statistics in the efficiency
and uncertainty in the $\pi^0$ momentum
dependent efficiency correction, $\epsilon'(p_{\pi^0})$,
obtained from the control samples
of the decay modes
$\bzb\to\rho^-D^{(*)+}$, $\bzb\to\pi^-D^{(*)+}$, $B^-\to\rho^-D^{(*)0}$
and $B^-\to\pi^-D^{(*)0}$.
The Dalitz plot efficiency
obtained from MC is
found to have a small charge asymmetry ($\sim 3\%$ at most).
We use this asymmetric efficiency for our nominal fit.
To estimate the systematic error from the asymmetry,
we fit the data using a symmetric
 efficiency and quote twice the difference
between symmetric and asymmetric efficiencies
as the systematic error.
The Dalitz plot efficiency is $r$ region dependent
and obtained as a product with the event fraction
in the corresponding region,
$\mathcal{F}^l_\mathrm{sig} \cdot \epsilon^l(m', \theta')$,
using MC.
The difference in the fraction
for data and MC is
estimated to be $\sim 10\%$
using the $B^0 \rightarrow D^{*-} \pi^+$ decay mode
as a control sample.
The fractions are varied by $\pm 10\%$
to estimate the systematic error.

\subsection{Background Dalitz PDF}
The Dalitz plot for continuum background
has an uncertainty due to the limited statistics
of the sideband events,
which we use to model the PDF.
We estimate the uncertainty
by performing a Toy MC study of sideband events.
With each pseudo-experiment,
we model the PDF in the same way as we do for real data;
with the PDF, we fit the data in the signal region;
and we quote the variation of fit results as the systematic error.
The flavor-asymmetry parameters of the continuum background,
which are fitted from sideband events,
are varied by their uncertainties.
Systematic uncertainty from
the statistics of the $B\bbar$ MC,
which is used to model the $B\bbar$ Dalitz plot PDF,
is also taken into account.

\subsection{\boldmath 
$B^0 \rightarrow \pi^+\pi^-\pi^0$ processes other than $B^0
  \rightarrow (\rho\pi)^0$
 \label{sec:syst_other_pipipi}
}
The primary contribution to the systematic errors
of the non-interfering parameters
tends to come from the $B^0 \rightarrow \pi^+\pi^-\pi^0$
decay processes that are not $B^0 \rightarrow (\rho\pi)^0$.
We take account of the contributions from
$B^0 \rightarrow f_0(980) \pi^0$,
$B^0 \rightarrow f_0(600) \pi^0$,
$B^0 \rightarrow \omega \pi^0$,
and 
non-resonant $B^0 \rightarrow \pi^+ \pi^- \pi^0$.
Upper limits on their contributions
 are determined from data,
except for $B^0 \rightarrow \omega \pi^0$,
for which we use the world average of
 $\mathcal{B}(B^0 \rightarrow \omega \pi^0)$~\cite{unknown:2006bi}
and $\mathcal{B}(\omega \rightarrow \pi^+\pi^-)$~\cite{Eidelman:2004wy}.
For the mass and width parameters of the $f_0(600)$ resonance,
we use recent measurements by
BES~\cite{Ablikim:2004qn},
CLEO~\cite{Muramatsu:2002jp},
and E791~\cite{Aitala:2000xu}
and take the largest variation.
We find no significant signals for any of the above decay modes.
Using the $1\sigma$ upper limits as input,
we generate Toy MC
for each mode
with the interference between
the $B^0 \rightarrow (\rho \pi)^0$
and the other
$B^0 \rightarrow \pi^+ \pi^- \pi^0$ mode taken into account.
We obtain the systematic error by
fitting the Toy MC assuming
$B^0 \rightarrow (\rho \pi)^0$ only in the PDF.
Within physically allowed regions,
we vary the $CP$ violation parameters
of the other $B^0 \rightarrow \pi^+ \pi^- \pi^0$ modes and
 the relative phase difference
between $B^0 \rightarrow (\rho \pi)^0$
and the other $B^0 \rightarrow \pi^+ \pi^- \pi^0$
as the input parameters,
and use the largest deviation as the systematic error for each decay mode.

\subsection{Background fraction}
Systematic errors
due to the event-by-event $\dE$-$\mbc$ background fractions
are studied
by varying the PDF shape parameters and the fraction of continuum background,
and the correction factor to the signal PDF shape
by $\pm 1\sigma$.
We also vary the fractions of the $B\bbar$ background,
which are estimated with MC,
by $\pm 50\%$ ($\pm 20 \%$) for the
$b\rightarrow c$ ($b\rightarrow u$) transition.

\subsection{Physics parameters}
We use the world average~\cite{Eidelman:2004wy,unknown:2006bi}
 for the following physics parameters:
$\tau_{B^0}$ and $\dmd$ (used for signal and $B\bbar$ background $\Delta t$),
the CKM angles of $\phi_1$ and $\phi_2$ (used in $B\bbar$ background),
and the branching fractions of $b\rightarrow u$ decay modes (used in
 $B\bbar$ background).
The systematic error is assigned by varying them by $\pm 1\sigma$.
The charge asymmetry of $B^0 \rightarrow a_1^\pm \pi^\mp$,
for which there is no measurement
and we use zero in the nominal fit,
is varied in the physically allowed region,
i.e., $\pm 1$.

\subsection{\boldmath Background $\Delta t$ PDF}
Systematic errors
from uncertainties in the background $\Delta t$ shapes
for both continuum and $B\bbar$ backgrounds
are estimated by varying each parameter by $\pm 1\sigma$.

\subsection{Vertex reconstruction}
To determine the systematic error
that arises from uncertainties in the vertex reconstruction,
the track and vertex selection criteria are varied to search
for possible systematic biases.
Systematic error due to the IP constraint
in the vertex reconstruction is estimated by varying
the smearing used to account for the $B$ flight length by $\pm 10 \, \mu m$.

\subsection{\boldmath Resolution function for the $\Delta t$ PDF}
Systematic errors due to uncertainties in the resolution function are
estimated by varying each resolution parameter obtained from data (MC)
by $\pm 1\sigma$ ($\pm 2\sigma$).
Systematic errors due to uncertainties in the wrong tag fractions
are also studied by varying the wrong tag fraction individually
for each $r$ region.

\subsection{Fit bias}
We observed fit bias due to small statistics for some of the fitted parameters.
Since this bias is much smaller than the statistical error,
we take it into account in the systematic errors.
We estimate the size of the fit bias by Toy MC study
and quote the bias as the systematic errors.
We also confirm that the bias
is consistent between Toy MC and full detector MC simulation.

\subsection{Tag-side interference}
Finally, we investigate the effects of tag-side interference (TSI),
which is the interference between
CKM-favored and CKM-suppressed $B\rightarrow D$ transitions
in the $f_\mathrm{tag}$ final state~\cite{Long:2003wq}.
A small correction to the PDF for the signal distribution arises
from the interference.
We estimate the size of the correction using the
 $\bz \rightarrow D^{*-}\ell^+ \nu$ sample.
We then generate MC pseudo-experiments and make an 
ensemble test to obtain the systematic biases.

\begin{table}[htbp]
 \caption{
 Table of systematic errors (1).
 The notation ``$<0.01$'' means that the value is small and less than
 0.01, and thus invisible in the number of significant digits shown
 here. We calculate the total systematic error including these small
 contributions.
 \label{tbl:systematics1}
 }
\newcolumntype{R}{>{\raggedleft\arraybackslash}X}
 \begin{tabularx}{150mm}{l|*{8}{R@{\hspace{4mm}}}}
   \hline
   \hline
  & \multicolumn{1}{c}{$U^+_-$}  & \multicolumn{1}{c}{$U^+_0$}  & \multicolumn{1}{c}{$U^{+,\mathrm{Re}}_{+-}$}  & \multicolumn{1}{c}{$U^{+,\mathrm{Re}}_{+0}$}  & \multicolumn{1}{c}{$U^{+,\mathrm{Re}}_{-0}$}  & \multicolumn{1}{c}{$U^{+,\mathrm{Im}}_{+-}$}  & \multicolumn{1}{c}{$U^{+,\mathrm{Im}}_{+0}$}  & \multicolumn{1}{c}{$U^{+,\mathrm{Im}}_{-0}$}  \\
   \hline
 $\rho'$ and $\rho''$  & 0.01 & 0.01 & 0.29 & 0.19 & 0.26 & 0.32 & 0.37 & 0.29  \\
 SCF  & 0.01 & 0.02 & 0.31 & 0.14 & 0.17 & 0.03 & 0.03 & 0.10  \\
 Signal Dalitz  & 0.02 & 0.01 & 0.24 & 0.15 & 0.19 & 0.13 & 0.06 & 0.13  \\
 BG Dalitz  & 0.02 & 0.01 & 0.16 & 0.12 & 0.14 & 0.14 & 0.12 & 0.22  \\
 Other $\pi\pi\pi$  & 0.06 & 0.03 & 0.10 & 0.08 & 0.10 & 0.15 & 0.10 & 0.08  \\
 BG fraction  & 0.03 & 0.02 & 0.14 & 0.19 & 0.13 & 0.23 & 0.07 & 0.22  \\
 Physics  & 0.02 & $<0.01$ & 0.01 & 0.02 & 0.02 & 0.01 & 0.01 & 0.02  \\
 BG $\Delta t$  & $<0.01$ & $<0.01$ & 0.03 & 0.01 & 0.02 & 0.02 & 0.01 & 0.02  \\
 Vertexing  & 0.02 & 0.02 & 0.02 & 0.16 & 0.11 & 0.08 & 0.08 & 0.09  \\
 Resolution  & $<0.01$ & $<0.01$ & 0.04 & 0.07 & 0.03 & 0.04 & 0.03 & 0.02  \\
 Flavor tagging  & $<0.01$ & $<0.01$ & $<0.01$ & $<0.01$ & 0.01 & $<0.01$ & $<0.01$ & 0.01  \\
 Fit bias  & 0.01 & 0.01 & 0.16 & 0.22 & 0.07 & 0.09 & 0.22 & 0.24  \\
 TSI  & $<0.01$ & $<0.01$ & 0.01 & 0.01 & 0.01 & 0.03 & 0.01 & 0.01  \\
   \hline
 Total  & 0.08 & 0.05 & 0.57 & 0.46 & 0.44 & 0.49 & 0.47 & 0.53  \\
   \hline
 \end{tabularx}
\end{table}
\begin{table}
 \caption{
 Table of systematic errors (2).
 The notation ``$<0.01$'' means that the value is small and less than
 0.01, and thus invisible in the number of significant digits shown
 here. We calculate the total systematic error including these small
 contributions.
 \label{tbl:systematics2}
 }
\newcolumntype{R}{>{\raggedleft\arraybackslash}X}
 \begin{tabularx}{164mm}{l|*{9}{R@{\hspace{4mm}}}}
   \hline
   \hline
   & \multicolumn{1}{c}{$U^-_+$}  & \multicolumn{1}{c}{$U^-_-$}  & \multicolumn{1}{c}{$U^-_0$}  & \multicolumn{1}{c}{$U^{-,\mathrm{Re}}_{+-}$}  & \multicolumn{1}{c}{$U^{-,\mathrm{Re}}_{+0}$}  & \multicolumn{1}{c}{$U^{-,\mathrm{Re}}_{-0}$}  & \multicolumn{1}{c}{$U^{-,\mathrm{Im}}_{+-}$}  & \multicolumn{1}{c}{$U^{-,\mathrm{Im}}_{+0}$}  & \multicolumn{1}{c}{$U^{-,\mathrm{Im}}_{-0}$}  \\
   \hline
$\rho'$ and $\rho''$  & 0.02 & 0.02 & 0.05 & 0.42 & 0.31 & 0.41 & 0.77 & 0.45 & 0.36  \\
SCF  & 0.02 & 0.03 & 0.03 & 0.29 & 0.27 & 0.32 & 0.09 & 0.25 & 0.17  \\
Signal Dalitz  & 0.01 & 0.02 & 0.02 & 0.28 & 0.32 & 0.32 & 0.38 & 0.15 & 0.53  \\
BG Dalitz  & 0.04 & 0.03 & 0.02 & 0.29 & 0.36 & 0.30 & 0.31 & 0.22 & 0.41  \\
Other $\pi\pi\pi$  & 0.05 & 0.05 & 0.03 & 0.12 & 0.11 & 0.14 & 0.15 & 0.11 & 0.13  \\
BG fraction  & 0.03 & 0.04 & 0.02 & 0.31 & 0.30 & 0.32 & 0.38 & 0.22 & 0.49  \\
Physics  & 0.01 & 0.01 & $<0.01$ & 0.04 & 0.03 & 0.04 & 0.04 & 0.02 & 0.06  \\
BG $\Delta t$  & $<0.01$ & $<0.01$ & $<0.01$ & 0.03 & 0.04 & 0.02 & 0.04 & 0.02 & 0.05  \\
Vertexing  & 0.04 & 0.02 & 0.05 & 0.17 & 0.45 & 0.16 & 0.08 & 0.10 & 0.27  \\
Resolution  & 0.01 & 0.01 & 0.01 & 0.16 & 0.17 & 0.32 & 0.11 & 0.10 & 0.29  \\
Flavor tagging  & 0.01 & 0.01 & $<0.01$ & 0.03 & 0.04 & 0.04 & 0.05 & 0.03 & 0.03  \\
Fit bias  & 0.01 & 0.03 & $<0.01$ & 0.05 & 0.07 & 0.12 & 0.18 & 0.02 & 0.23  \\
TSI  & 0.04 & 0.04 & 0.01 & 0.05 & 0.07 & 0.03 & 0.02 & 0.06 & 0.01  \\
   \hline
Total  & 0.10 & 0.09 & 0.09 & 0.77 & 0.86 & 0.86 & 1.03 & 0.65 & 1.05  \\
   \hline
   \end{tabularx}
\end{table}
\begin{table}
 \caption{
 Table of systematic errors (3).
 The notation ``$<0.01$'' means that the value is small and less than
 0.01, and thus invisible in the number of significant digits shown
 here. We calculate the total systematic error including these small
 contributions.
 \label{tbl:systematics3}
 }
\newcolumntype{R}{>{\raggedleft\arraybackslash}X}
 \begin{tabularx}{164mm}{l|*{9}{R@{\hspace{4mm}}}}
   \hline
   \hline
 & \multicolumn{1}{c}{$I_+$}  & \multicolumn{1}{c}{$I_-$}  & \multicolumn{1}{c}{$I_0$}  & \multicolumn{1}{c}{$I^{\mathrm{Re}}_{+-}$}  & \multicolumn{1}{c}{$I^{\mathrm{Re}}_{+0}$}  & \multicolumn{1}{c}{$I^{\mathrm{Re}}_{-0}$}  & \multicolumn{1}{c}{$I^{\mathrm{Im}}_{+-}$}  & \multicolumn{1}{c}{$I^{\mathrm{Im}}_{+0}$}  & \multicolumn{1}{c}{$I^{\mathrm{Im}}_{-0}$}  \\
   \hline
$\rho'$ and $\rho''$  & 0.03 & 0.02 & 0.04 & 0.95 & 0.59 & 1.32 & 0.89 & 0.84 & 0.89  \\
SCF  & 0.01 & 0.01 & 0.01 & 0.09 & 0.64 & 0.07 & 0.50 & 0.08 & 0.65  \\
Signal Dalitz  & 0.01 & 0.01 & 0.01 & 0.33 & 0.29 & 0.30 & 0.31 & 0.35 & 0.31  \\
BG Dalitz  & 0.01 & 0.01 & 0.01 & 0.34 & 0.38 & 0.30 & 0.32 & 0.34 & 0.33  \\
Other $\pi\pi\pi$  & 0.03 & 0.03 & 0.02 & 0.17 & 0.15 & 0.18 & 0.22 & 0.15 & 0.20  \\
BG frac.  & 0.02 & 0.01 & 0.01 & 0.44 & 0.34 & 0.33 & 0.32 & 0.37 & 0.29  \\
Physics  & 0.01 & 0.01 & $<0.01$ & 0.05 & 0.06 & 0.03 & 0.05 & 0.05 & 0.05  \\
BG $\Delta t$  & $<0.01$ & $<0.01$ & $<0.01$ & 0.05 & 0.04 & 0.04 & 0.05 & 0.04 & 0.11  \\
Vertexing  & 0.02 & 0.02 & 0.04 & 0.16 & 0.28 & 0.14 & 0.42 & 0.37 & 0.28  \\
Resolution  & 0.01 & 0.01 & 0.01 & 0.30 & 0.21 & 0.18 & 0.35 & 0.25 & 0.28  \\
Flavor tagging  & $<0.01$ & $<0.01$ & $<0.01$ & 0.04 & 0.07 & 0.04 & 0.03 & 0.07 & 0.03  \\
Fit bias  & $<0.01$ & 0.01 & $<0.01$ & 0.12 & 0.01 & 0.27 & 0.09 & 0.09 & 0.22  \\
TSI  & 0.01 & $<0.01$ & 0.01 & 0.09 & 0.07 & 0.04 & 0.04 & 0.03 & 0.07  \\
   \hline
Total  & 0.06 & 0.05 & 0.06 & 1.23 & 1.12 & 1.49 & 1.31 & 1.15 & 1.33  \\
   \hline
  \end{tabularx}
\end{table}

\clearpage

\section{Quasi-two-body parameters
 \label{sec:quasi_two_body}
}
We calculate quasi-two-body $CP$ violation parameters
as
\begin{equation}
 \mathcal{C}^+ = \frac{U^-_+}{U^+_+}\; , \quad
 \mathcal{C}^- = \frac{U^-_-}{U^+_-}\; , \quad
 \mathcal{S}^+ = \frac{2 I_+}{U^+_+}\; , \quad
 \mathcal{S}^- = \frac{2 I_-}{U^+_-}\; , \quad
 \mathcal{A}_{\rho \pi}^{CP} = \frac{U^+_+ - U^+_-}{U^+_+ + U^+_-} \;,
\end{equation}
and
\begin{equation}
 \mathcal{C} \equiv \frac{\mathcal{C}^+ + \mathcal{C}^-}{2} \;, \quad
 \Delta \mathcal{C} \equiv \frac{\mathcal{C}^+ - \mathcal{C}^-}{2} \;, \quad
 \mathcal{S} \equiv \frac{\mathcal{S}^+ + \mathcal{S}^-}{2} \;, \quad
 \Delta \mathcal{S} \equiv \frac{\mathcal{S}^+ - \mathcal{S}^-}{2} \;.
\end{equation}
We obtain
\begin{eqnarray}
   \mathcal{A}_{\rho \pi}^{CP}  & = & -0.12 \pm 0.05 \pm 0.03 \; ,\\
   \mathcal{C}                  & = & -0.13 \pm 0.09 \pm 0.06 \; ,\\
   \Delta \mathcal{C}           & = & +0.35 \pm 0.10 \pm 0.06 \; ,\\
   \mathcal{S}                  & = & +0.06 \pm 0.13 \pm 0.07 \; ,\\
   \Delta \mathcal{S}           & = & -0.12 \pm 0.14 \pm 0.07 \; ,
\end{eqnarray}
where first errors and second errors are statistical and systematic, respectively.
The correlation matrix is shown in Table \ref{tab:q2b_correlation_matrix}.
\begin{table}
 \caption{
 Correlation matrix of the quasi-two-body parameters,
 with statistical and systematic error combined.
 \label{tab:q2b_correlation_matrix}
 }
  \begin{tabular*}{8cm}{@{\extracolsep{\fill}}@{\hspace{3mm}}c@{\hspace{3mm}}|ccccc}
   \hline
   \hline
   &
   $\mathcal{A}_{\rho \pi}^{CP}$ &
   $\mathcal{C}$                 &
   $\Delta \mathcal{C}$          &
   $\mathcal{S}$                 &
   $\Delta \mathcal{S}$          \\
   \hline
   $\mathcal{A}_{\rho \pi}^{CP}$  & $+1.00$ \\
   $\mathcal{C}$                  & $-0.15$ & $+1.00$ \\
   $\Delta \mathcal{C}$           & $+0.08$ & $+0.19$ & $+1.00$ \\
   $\mathcal{S}$                  & $+0.02$ & $-0.02$ & $-0.00$ & $+1.00$ \\
   $\Delta \mathcal{S}$           & $-0.02$ & $-0.00$ & $-0.01$ & $+0.30$ & $+1.00$ \\
   \hline
  \end{tabular*}
\end{table}

\begin{figure}[htbp]
 \begin{center}
  \includegraphics[scale=1.0]{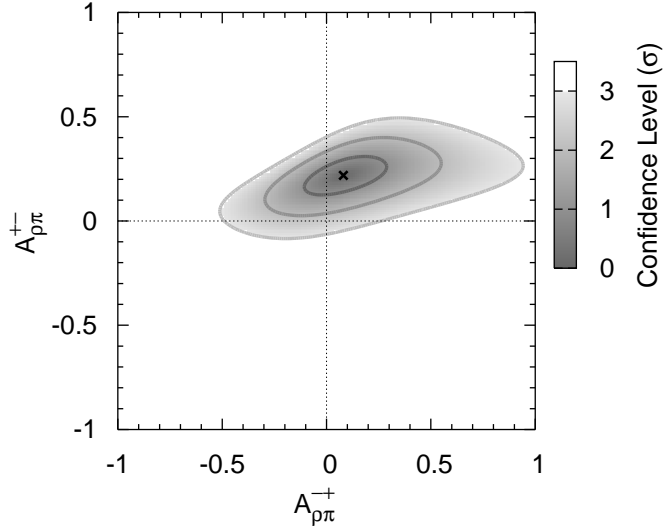}
  \caption{
  \label{fig:apm_amp_contour}
  Contour plot of the confidence level for the direct $CP$ violation
  parameters $\mathcal{A}_{\rho \pi}^{+-}$ vs.
  $\mathcal{A}_{\rho \pi}^{-+}$.
  }
 \end{center}
\end{figure}

One can transform the parameters into
the direct $CP$ violation parameters
 $\mathcal{A}_{\rho \pi}^{+-}$
and 
 $\mathcal{A}_{\rho \pi}^{-+}$
defined as
\begin{equation}
 \mathcal{A}_{\rho \pi}^{+-}
  \equiv
  - \frac{\mathcal{A}_{\rho \pi}^{CP} + \mathcal{C} + \mathcal{A}_{\rho
  \pi}^{CP} \Delta \mathcal{C}}
  {1 + \Delta \mathcal{C} + \mathcal{A}_{\rho \pi}^{CP} \: \mathcal{C}} \;,
\end{equation}
\begin{equation}
 \mathcal{A}_{\rho \pi}^{-+}
  \equiv
   \frac{\mathcal{A}_{\rho \pi}^{CP} - \mathcal{C} - \mathcal{A}_{\rho
  \pi}^{CP} \Delta \mathcal{C}}
  {1 - \Delta \mathcal{C} - \mathcal{A}_{\rho \pi}^{CP} \: \mathcal{C}} \;,
\end{equation}
which can be more convenient for interpretation.
We obtain
\begin{eqnarray}
 \mathcal{A}_{\rho \pi}^{+-} & = & +0.22 \pm 0.08 \pm 0.05 \;, \\
 \mathcal{A}_{\rho \pi}^{-+} & = & +0.08 \pm 0.17 \pm 0.12 \;,
\end{eqnarray}
with a correlation coefficient of $+0.53$.
Our result differs from the case with no direct $CP$
asymmetry ($\mathcal{A}_{\rho \pi}^{+-} = 0$ and
$\mathcal{A}_{\rho \pi}^{-+} = 0$)
by $2.4$ standard deviations (Fig.~\ref{fig:apm_amp_contour}).
Measurements with a larger data sample
may be needed to observe direct $CP$ violation.

The ratio of the branching fractions
$\mathcal{B}(B^0 \rightarrow \rho^0 \pi^0) / \mathcal{B}(B^0 \rightarrow
\rho^\pm\pi^\mp)$
can be related to our measurement
as
\begin{equation}
 \frac{\mathcal{B}(B^0 \rightarrow \rho^0 \pi^0)}
  {\mathcal{B}(B^0 \rightarrow \rho^\pm\pi^\mp)}
  = 
  \frac{U^+_0}{U^+_+ + U^+_-} \;.
\end{equation}
Our measurement yields
\begin{equation}
 \frac{\mathcal{B}(B^0 \rightarrow \rho^0 \pi^0)}
  {\mathcal{B}(B^0 \rightarrow \rho^\pm\pi^\mp)}
  = 
  0.133 \pm 0.022 \pm 0.023 \;.
\end{equation}
This is consistent with the dedicated quasi-two-body branching fraction measurement
of $B^0 \rightarrow \rho^0 \pi^0$ from Belle~\cite{Dragic:2006yv}:
\begin{equation}
 \frac{\mathcal{B}(B^0 \rightarrow \rho^0 \pi^0)^\mathrm{Belle}}
  {\mathcal{B}(B^0 \rightarrow \rho^\pm\pi^\mp)^\mathrm{WA}}
  = \frac{3.12^{+0.88}_{-0.82}{}^{+0.60}_{-0.76}}{24.0 \pm 2.5}
  = 0.130^{+0.049}_{-0.046}
  \;,
\end{equation}
where we use the world average~\cite{unknown:2006bi}
for the denominator.

We also measure the $CP$ violating parameters
of $B^0 \rightarrow \rho^0 \pi^0$
calculated as
\begin{equation}
 \mathcal{A}_{\rho^0 \pi^0} = -\frac{U^-_0}{U^+_0} \; ,
  \quad
  \mathcal{S}_{\rho^0 \pi^0} = \frac{2 I_0}{U^+_0} \; .
\end{equation}
We obtain
\begin{eqnarray}
 \mathcal{A}_{\rho^0 \pi^0} & = & -0.45 \pm 0.35 \pm 0.32 \;,\\
 \mathcal{S}_{\rho^0 \pi^0} & = & +0.15 \pm 0.57 \pm 0.43 \;,
\end{eqnarray}
with correlation coefficient of $-0.07$.
Our measurement of $\mathcal{A}_{\rho^0 \pi^0}$
is consistent with the previous measurement from Belle~\cite{Dragic:2006yv}.
This is the first measurement of $\mathcal{S}_{\rho^0 \pi^0}$.

\section{\boldmath Constraint on the CKM angle $\phi_2$
 \label{sec:constraint_on_ph2}
}
\begin{figure}[htbp]
 \begin{center}
  \includegraphics[scale=0.75]{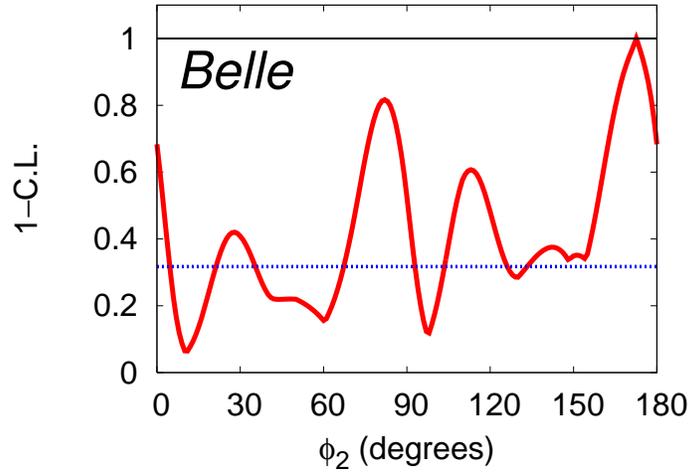}
  \caption{
  $1-\mathrm{C.L.}$ vs. $\phi_2$ obtained by Dalitz analysis.
  The dotted horizontal line corresponds to the $1\sigma$ confidence level.
  \label{fig:phi2_1_cl_plot_dalitz}
  }
 \end{center}
\end{figure}

We constrain the CKM angle $\phi_2$ from our analysis
following the procedure described in
 Ref.~\cite{Snyder:1993mx}.
With three $\bz \rightarrow (\rho\pi)^0$ decay modes,
we have 9 free parameters including $\phi_2$:
\begin{equation}
 \begin{array}{l}
  9 = \mathrm{(6\;complex\;amplitudes=12 \, d.o.f.) + \phi_2}
   \\
  \hspace{8mm}
   \mathrm{
   - (1\;global\;phase)
   - (1\;global\;normalization)
   - (1\;isospin\;relation=2 \, d.o.f.)
   } \;,
 \end{array}
\end{equation}
where we make use of an isospin relation
that relates neutral $B$
 decay processes only~\cite{Lipkin:1991st,Gronau:1991dq}.
Parameterizing the 6 complex amplitudes with 9 free parameters,
we form a $\chi^2$ function using the 26 measurements
of our time-dependent Dalitz plot analysis as constraints.
We first optimize all the 9 parameters to obtain a minimum $\chi^2$,
 $\chi^2_\mathrm{min}$; we then scan $\phi_2$
from $0^\circ$ to $180^\circ$ optimizing the other 8 parameters,
whose resultant minima are defined as $\chi^2(\phi_2)$;
and $\Delta \chi^2(\phi_2)$ is defined as
$\Delta \chi^2(\phi_2) \equiv \chi^2(\phi_2) - \chi^2_\mathrm{min}$.
Performing a Toy MC study
following the procedure described in Ref.~\cite{Charles:2004jd},
we obtain the $\mathrm{1-C.L.}$ plot in Fig.~\ref{fig:phi2_1_cl_plot_dalitz}
from the $\Delta \chi^2(\phi_2)$\footnote{
$\Delta \chi^2(\phi_2)$ is usually expected to follow
a $\chi^2$ distribution with one degree of freedom
and thus the cumulative $\chi^2$ distribution for one degree of freedom
is usually used to convert $\Delta \chi^2(\phi_2)$
into a $\mathrm{1-C.L.}$ plot.
A toy MC study shows, however, that this is not the case for
$B\rightarrow \rho \pi$,
 and an analysis with this assumption yields confidence intervals
with undercoverage.
Thus, we perform a dedicated Toy MC study to obtain the confidence interval.
}.

\begin{figure}[htbp]
 \begin{center}
  \includegraphics[scale=1.2]{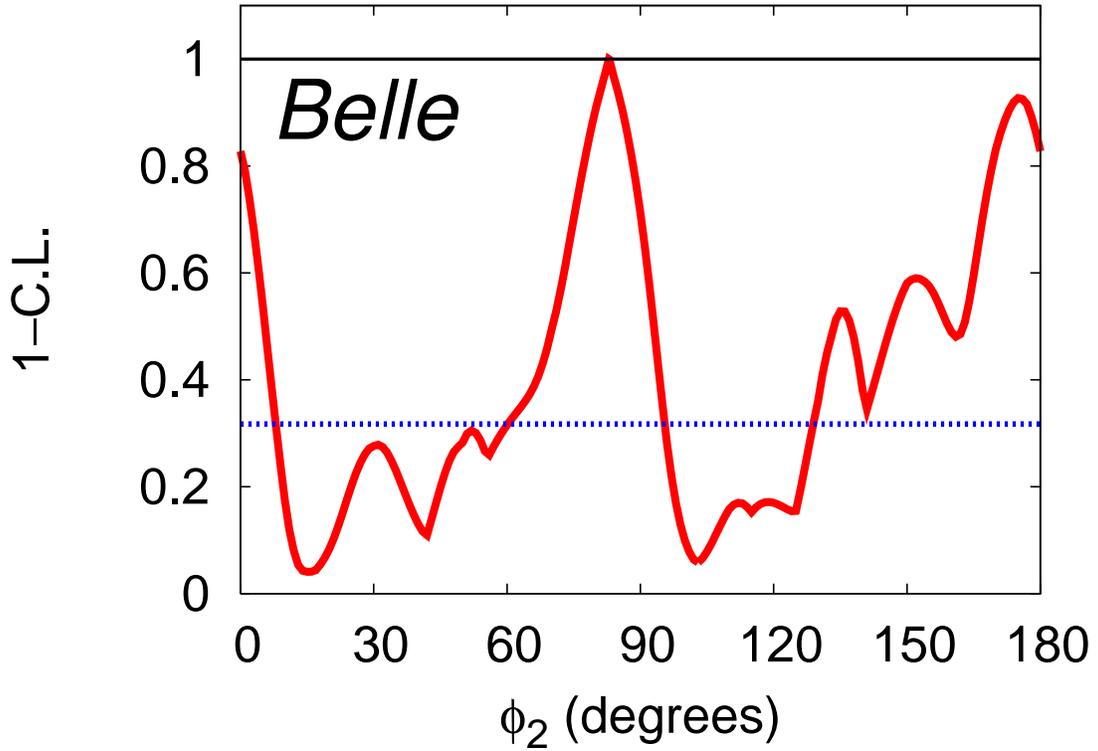}
  \caption{
  $1-\mathrm{C.L.}$ vs. $\phi_2$ obtained by a full
  combined Dalitz and isospin(pentagon) analysis.
  The dotted horizontal line corresponds to the $1\sigma$ confidence level.
  \label{fig:phi2_1_cl_plot}
  }
 \end{center}
\end{figure}

In addition to the 26 observables obtained from our
time-dependent Dalitz plot analysis,
we use the following
world average branching fractions and asymmetries:
$\mathcal{B}(B^0 \rightarrow \rho^\pm \pi^\mp)$,
$\mathcal{B}(B^+ \rightarrow \rho^+ \pi^0)$,
$\mathcal{A}(B^+ \rightarrow \rho^+ \pi^0)$,
$\mathcal{B}(B^+ \rightarrow \rho^0 \pi^+)$,
and
$\mathcal{A}(B^+ \rightarrow \rho^0 \pi^+)$~\cite{unknown:2006bi},
which are not correlated with our 26 observables.
With the 31 measurements above,
we perform a full combined
Dalitz and isospin(pentagon) analysis.
Having 5 related decay modes,
we have 12 free parameters including $\phi_2$:
\begin{equation}
 \begin{array}{l}
  12 =
   \mathrm{(10\;complex\;amplitudes=20 \, d.o.f.) + \phi_2}
   \\
  \hspace{12mm}
   \mathrm{
   - (1\;global\;phase)
   - (4\;isospin\;relations=8 \, d.o.f.)} \;.
 \end{array}
\end{equation}
The detail of $\chi^2$ formation can be found
 in appendix \ref{sec:phi2_chi2_parameterization}.
The $\chi^2_\mathrm{min}$ obtained is 9.8,
which is reasonable for
$\mathrm{31(measurements)}$ $ - $
$\mathrm{12(free\; parameters)} $ $= 19$ degrees of freedom.
Following the same procedure as above,
we obtain the
$\mathrm{1-C.L.}$ plot in the Fig.~\ref{fig:phi2_1_cl_plot}.
We obtain $\phi_2 = (83^{+12}_{-23})^\circ$ as the central value
and $1\sigma$ errors (corresponding to 68.3\% C.L.).
A large CKM-disfavored region
 ($\phi_2<8^\circ$ and $129^\circ<\phi_2$) 
also remains.

\section{Conclusion}
We have performed a full Dalitz plot analysis
of the $B^0 \rightarrow \pi^+\pi^-\pi^0$ decay mode.
The results are consistent with no direct $CP$ violation
and the previous measurement of $B^0 \rightarrow \rho^0\pi^0$
branching fraction and flavor asymmetry.
Combining our analysis and information
from charged $B$ decay modes,
a full Dalitz plot and isospin analysis is performed
to obtain a constraint on $\phi_2$.
We obtain $\phi_2 = (83^{+12}_{-23})^\circ$ as the central value
with $1\sigma$ errors.
However, a large CKM-disfavored region
 ($\phi_2<8^\circ$ and $129^\circ<\phi_2$) also remains.
In principle, with more data we may be able to remove the
additional $\phi_2$ solutions.


\section*{Acknowledgment}

We thank the KEKB group for the excellent operation of the
accelerator, the KEK cryogenics group for the efficient
operation of the solenoid, and the KEK computer group and
the National Institute of Informatics for valuable computing
and Super-SINET network support. We acknowledge support from
the Ministry of Education, Culture, Sports, Science, and
Technology of Japan and the Japan Society for the Promotion
of Science; the Australian Research Council and the
Australian Department of Education, Science and Training;
the National Science Foundation of China and the Knowledge
Innovation Program of the Chinese Academy of Sciencies under
contract No.~10575109 and IHEP-U-503; the Department of
Science and Technology of India; 
the BK21 program of the Ministry of Education of Korea, 
the CHEP SRC program and Basic Research program 
(grant No.~R01-2005-000-10089-0) of the Korea Science and
Engineering Foundation, and the Pure Basic Research Group 
program of the Korea Research Foundation; 
the Polish State Committee for Scientific Research; 
the Ministry of Science and Technology of the Russian
Federation; the Slovenian Research Agency;  the Swiss
National Science Foundation; the National Science Council
and the Ministry of Education of Taiwan; and the U.S.\
Department of Energy.

\appendix

\section{PDFs for time-dependent Dalitz plot analysis
 \label{sec:appendix_pdf_definitions}
}
In this section, we describe the details of the PDF for each component,
which appear in equations
 (\ref{equ:dembc_dalitz_simultaneous-total_pdf})-(\ref{equ:signal_pdf_definition}).

\subsection{Signal PDF
  \label{sec:signal_pdf}
}

\subsubsection{PDF for correctly reconstructed events}
In terms of the event fractions
 for the $l^\mathrm{th}$ flavor tagging region
 ($\mathcal{F}^l_\mathrm{true}$), 
the Dalitz plot dependent efficiency
 ($\epsilon^l$), the $\pi^0$ momentum dependent efficiency correction
taking account of the difference
between data and MC ($\epsilon'$),
 wrong tag fraction ($w_l$), and the wrong tag 
fraction difference between $\bz$ and $\bzb$ ($\dwl$),
the PDF for correctly reconstructed events is given by
\begin{equation}
 \begin{array}{l}
  p_\mathrm{true}^l(\dE, \mbc; m',\theta';\Delta t, q_\mathrm{tag}; p_{\pi^0}) = \\
  \hspace{10mm}
   {\mathcal{F}}^l\cdot p_\mathrm{true}(\Delta E, \mbc; p_{\pi^0})
   \cdot \epsilon^l(m',\theta') \, \epsilon'(p_{\pi^0})
   \cdot|\mathrm{det} \boldsymbol{J}(m',\theta')|\cdot
   p_\mathrm{true}^l( m',\theta';\Delta t, q_\mathrm{tag}),
 \end{array}
\end{equation}
where 
\begin{eqnarray}
 \begin{array}{l}
  p_\mathrm{true}^l( m',\theta';\Delta t, q_\mathrm{tag})  = 
   \frac{e^{-|\Delta t|/\tau_{B^0}}}{4\tau_{B^0}}\cdot
   \Bigl\{(1-q_\mathrm{tag}\dwl)(|A_{3\pi}|^2+|\overline{A}_{3\pi}|^2)
   \\
  \hspace{10mm}+q_\mathrm{tag}(1-2w_l)\cdot
   \left[
    -(|A_{3\pi}|^2-|\overline{A}_{3\pi}|^2)\cos (\dmd \Delta t)
     +2\mathrm{Im}[\frac{q}{p}\overline{A}_{3\pi}A^*_{3\pi}]\sin(\dmd
     \Delta t)
   \right]\Bigr\} \;.
 \end{array}
\end{eqnarray}
For the $\Delta t$ PDF, 
the above equation is convolved with the resolution function~\cite{Tajima:2003bu}.
The $\dE$-$\mbc$ PDF is normalized such that
\begin{equation}
  \int \hspace{-2mm} \int_\mathrm{signal \; region}
   \hspace{-12mm} d\Delta E \; d\mbc \;
   p_\mathrm{true}(\Delta E, \mbc; p_{\pi^0}) = 1
   \quad (\forall p_{\pi^0}) \;,
\end{equation}
since we define the Dalitz plot efficiency for events inside the signal region.
With the PDFs in the $\Delta t$-$q_\mathrm{tag}$ direction
being also normalized to be unity,
the integral inside the signal region, $n_\mathrm{true}$, is,
\begin{equation}
 n_\mathrm{true}
  = \sum_l n^l_\mathrm{true} \;,
\end{equation}
\begin{eqnarray}
 \label{equ:dembc_dalitz_simultaneous-normalization_dalitz}
  n^l_\mathrm{true}
  & \equiv &
  \sum_{q_\mathrm{tag}}
  \int \hspace{-1mm} d \Delta t
  \int \hspace{-2mm} \int_\mathrm{signal \; region}
  \hspace{-15mm} d\Delta E \; d\mbc \;
  \int \hspace{-2mm}
  \int_\mathrm{SDP, \; Veto}
  \hspace{-12mm}
  dm' \; d\theta' \;
  p_\mathrm{true}^l(\dE, \mbc; m',\theta';\Delta t, q_\mathrm{tag}; p_{\pi^0})
  \\
 & = &
  \mathcal{F}^l_\mathrm{true}
  \int \hspace{-2mm}
  \int_\mathrm{SDP, \; Veto}
  \hspace{-12mm}
  dm' \; d\theta' \;
  \epsilon^l_\mathrm{true}(m', \theta') \, \epsilon'(p_{\pi^0})
  \,
  |\mathrm{det} \boldsymbol{J}| \,
  \left( |A_{3\pi}|^2 + |\overline{A}_{3\pi}|^2 \right) \;,
\end{eqnarray}
where the correlation between $p_{\pi^0}$ and $m'$ is properly
taken into account in the integration
of the last line.
The notation
$\int \int_\mathrm{SDP, \; Veto} dm' \; d\theta'$
means integration over the square Dalitz plot
with the vetoed region in the Dalitz plot taken into account.

The $\pi^0$ momentum dependent
 $\dE$-$\mbc$ PDF,
 $p_\mathrm{true}(\dE, \mb; p_{\pi^0})$,
is modeled
using MC-simulated events in
a binned histogram interpolated in the $p_{\pi^0}$ direction,
to which a small correction obtained with
$\bzb\to\rho^-D^{(*)+}$
 is applied
to account for the difference between MC and data.

The Dalitz plot distribution is smeared and distorted by
detection efficiencies and detector resolutions.
We obtain the signal Dalitz plot efficiency from MC
to take the former into account.
We introduce a dependence of the efficiency on the $r$ region,
 $\epsilon^l_\mathrm{true}$,
 since a significanct dependence is observed in MC.
Small corrections, $\epsilon'(p_{\pi^0})$,
 are also applied to the MC-determined efficiency
to account for differences between MC and data.
We use $\bzb\to\rho^-D^{(*)+}$, $\bzb\to\pi^-D^{*-}$, $B^-\to\rho^-D^0$
and $B^-\to\pi^-D^0$ decays to obtain the correction factors.
The detector resolutions are small
compared to the widths of $\rho(770)$ resonances;
this is confirmed by MC to be a negligibly small effect.

\subsubsection{PDF for SCF events}
Approximately 20\% of signal candidates are SCFs,
which are subdivided into  $\sim 4\%$ of NR SCF
and $\sim 16\%$ of CR SCF.
It is therefore important to develop a model that describes
the SCF well.
The time-dependent PDF for SCF events 
is defined as
\begin{equation}
 p_{i}^l(\dE, \mbc; m',\theta';\Delta t, q_\mathrm{tag}) =
  {\mathcal{F}}^l_{i}\cdot p_{i}(\Delta E, \mbc; s_i)
  \cdot 
  p_i(m', \theta'; \Delta t, q_\mathrm{tag}) \;, \quad (i=\mathrm{NR, CR})
\end{equation}
where ${\mathcal{F}}^l_i$ is the event fraction
in each tagging $r$-bin
and
\begin{equation}
 \begin{array}{l}
  p_i(m', \theta'; \Delta t, q_\mathrm{tag})
   =
   \frac{e^{-|\Delta t|/\tau_i}}{4\tau_i}\cdot
   \Bigl\{(1-q_\mathrm{tag}\dwl^{i}) \: p^\mathrm{Life}_{i}(m', \theta')
   \\
  \hspace{20mm}
  +q_\mathrm{tag}(1-2 {w_l}^{i})
  \cdot[-p^{\mathrm{Cos}}_{i}(m', \theta')\cos(\dmd \Delta t)
   +p^{\mathrm{Sin}}_{i}(m', \theta') \sin(\dmd \Delta t)]\Bigr\} \;.
 \end{array}
 \label{equ:dt_pdf_of_SCF}
\end{equation}
The $\dE$-$\mbc$ PDF is normalized inside the signal region as
\begin{equation}
  \int \hspace{-2mm} \int_\mathrm{signal \; region}
   \hspace{-12mm} d\Delta E \; d\mbc \;
   p_{i}(\Delta E, \mbc; s_i) = 1
   \quad (\forall s_i) \;.
\end{equation}
With the PDFs in the $\Delta t$-$q_\mathrm{tag}$ direction
 being normalized to unity
and $\sum_l {\mathcal{F}}^l_i = 1$,
the integral inside the signal region, $n_i$, is
\begin{eqnarray}
 n_i & \equiv & \sum_l \sum_{q_\mathrm{tag}}
  \int \hspace{-1mm} d \Delta t
  \int \hspace{-2mm} \int_\mathrm{signal \; region}
  \hspace{-15mm} d\Delta E \; d\mbc \;
  \int \hspace{-2mm}
  \int_\mathrm{SDP, \; Veto}
  \hspace{-12mm}
  dm' \; d\theta' \;
  p_{i}^l(\dE, \mbc; m',\theta';\Delta t, q_\mathrm{tag})
  \\
 & = &
  \int \hspace{-2mm}
  \int_\mathrm{SDP, \; Veto}
  \hspace{-12mm}
  dm' \; d\theta' \;
  p^\mathrm{Life}_{i}(m', \theta') \;.
\end{eqnarray}

We find that the $\dE$-$\mb$ distribution for SCF has a sizable correlation with
Dalitz plot variables,
but with only one of its two dimensions.
We thus introduce a model
with dependences on the Dalitz plot variable $s_i$.
The variable $s_\mathrm{CR} = s_\pm \equiv \max(s_+, s_-)$ is used,
because the CR SCF can be divided into a $\pi^+$ replaced SCF
and a $\pi^-$ replaced SCF, where $s_-$ ($s_+$) is used for
$\pi^+$ ($\pi^-$) replaced SCF.
Here, we exploit the fact that almost all of the $\pi^+$ ($\pi^-$)
replaced SCF distributes in the region of $s_+ < s_-$ ($s_+ > s_-$).
For the NR SCF, $s_\mathrm{NR} = s_0$.
This parameterization
models the correlation quite well,
with each of the parameters $s_i$
reasonably related to the kinematics of replaced tracks.

Since track ($\pi$) replacement
changes the measured kinematic variables,
the SCF events ``migrate'' in the Dalitz plot
from the correct (or generated) position
to the observed position.
Using MC,
we determine resolution functions
$R_i(m'_\mathrm{obs}, \theta'_\mathrm{obs}; m'_\mathrm{gen},
\theta'_\mathrm{gen})$
to describe this ``migration'' effect,
where $(m'_\mathrm{obs}, \theta'_\mathrm{obs})$
and $(m'_\mathrm{gen}, \theta'_\mathrm{gen})$
are the observed and the generated (correct) 
positions in the Dalitz plot, respectively.
Together with the efficiency function
 $\epsilon_i(m'_\mathrm{gen}, \theta'_\mathrm{gen})$,
which is also obtained with MC,
the Dalitz plot PDF for SCF is described as
\begin{eqnarray}
 p^j_{i}(m', \theta')
  & = & \left[(R_i \cdot \epsilon_i) \otimes p^j_\mathrm{gen}\right](m',
  \theta')
  \\
  & \equiv &
  \int \hspace{-2mm}
  \int_\mathrm{SDP}
  \hspace{-4mm}
  dm'_\mathrm{gen} \; d\theta'_\mathrm{gen} \;
  R_i(m', \theta'; m'_\mathrm{gen}, \theta'_\mathrm{gen})
  \cdot \epsilon_i (m'_\mathrm{gen}, \theta'_\mathrm{gen})
  \cdot p^j_\mathrm{gen} (m'_\mathrm{gen}, \theta'_\mathrm{gen})
  \;,
\end{eqnarray}
\[
 (\; i=\mathrm{CR, NR} \;, \quad j=\mathrm{Life, Cos, Sin}\;)
\]
where
\begin{eqnarray}
 p^\mathrm{Life}_\mathrm{gen} (m'_\mathrm{gen}, \theta'_\mathrm{gen})
  & = &
  |\mathrm{det}\boldsymbol{J}| (|A_{3\pi}|^2 + |\overline{A}_{3\pi}|^2) \;,
  \\
 p^\mathrm{Cos}_\mathrm{gen} (m'_\mathrm{gen}, \theta'_\mathrm{gen})
  & = &
  |\mathrm{det}\boldsymbol{J}| (|A_{3\pi}|^2 - |\overline{A}_{3\pi}|^2) \;,
  \\
 p^\mathrm{Sin}_\mathrm{gen} (m'_\mathrm{gen}, \theta'_\mathrm{gen})
  & = &
  |\mathrm{det}\boldsymbol{J}| 2\mathrm{Im}\left[\frac{q}{p}\overline{A}_{3\pi}A^*_{3\pi}\right] \;.
\end{eqnarray}

For the NR SCF,
the shape of the $\Delta t$ PDF defined in equation
(\ref{equ:dt_pdf_of_SCF})
is exactly the same as correctly reconstructed signal,
i.e., $\tau_\mathrm{NR} = \tau_\bz$,
${w_l}^\mathrm{NR} = w_l$, and $\dwl^\mathrm{NR} = \dwl$,
since the replaced track, $\pi^0$,
is not used for either vertexing or flavor tagging.
On the other hand, for the CR SCF,
the $\Delta t$ PDF is different from correctly reconstructed
signal,
since the replaced $\pi^\pm$ is used for both
vertexing and flavor tagging.
Thus, we use MC-simulated CR SCF events to obtain
 $\tau_\mathrm{CR}$,
 ${w_l}^\mathrm{CR}$, and $\dwl^\mathrm{CR}$,
which are different from those of correctly reconstructed signal events.
In particular,
$\dwl^\mathrm{CR}$
is opposite in sign
for the $\pi^+$ and $\pi^-$ replaced SCFs,
which is due to the fact that the replaced 
$\pi^\pm$ tends to be directly used
for flavor tagging in the slow pion category.

%

\subsection{Continuum PDF}
The PDF for the continuum background is
\begin{equation}
 \begin{array}{l}
  p^l_{qq}(\Delta E, \mbc; m', \theta';\Delta t, q_\mathrm{tag})
  \\
  =
   \mathcal{F}^l_{qq}
   \cdot p_{qq}^l(\dE, \mbc)
   \cdot p_{qq}(m', \theta'; \Delta E, \mbc)
   \cdot \left[ \frac{1+q_\mathrm{tag} A^l(m', \theta')}{2} \right]
   \cdot p_{qq}(\Delta t) \;,
 \end{array}
\end{equation}
where $\mathcal{F}^l_{qq}$ is the event fraction
for each $r$ region obtained in the signal yield fit.
All the terms on the right hand side of the equation are normalized to be unity
so that
\begin{equation}
 \sum_l \sum_{q_\mathrm{tag}}
  \int \hspace{-1mm} d \Delta t
  \int \hspace{-2mm} \int_\mathrm{signal \; region}
  \hspace{-15mm} d\Delta E \; d\mbc \;
  \int \hspace{-2mm}
  \int_\mathrm{SDP, \; Veto}
  \hspace{-12mm}
  dm' \; d\theta' \;
  p^l_{qq}(\Delta E, \mbc; m', \theta';\Delta t, q_\mathrm{tag}) = 1 \;.
\end{equation}

Since the allowed kinematic region is dependent on
$\dE$ and $\mbc$,
the Dalitz plot distribution is dependent on $\dE$ and $\mbc$.
We define a $\dE$-$\mbc$ independent PDF, $p_{qq}(m'_\mathrm{scale},
\theta')$,
where $m'_\mathrm{scale}$ is a re-defined SDP variable
with the kinematic effect taken into account as
\begin{equation}
 m'_\mathrm{scale}
  \equiv
  \frac{1}{\pi}
  \arccos
  \left(
   2 \frac{m_0 - m_0^{\mathrm{min}}}{m_0^{\mathrm{max}} -
   m_0^{\mathrm{min}} + \dE + \Delta \mbc} - 1
  \right) \;,
\end{equation}
where
\begin{equation}
 \Delta \mbc \equiv \mbc - m_{\bz} \;.
\end{equation}
Using the $\dE$-$\mbc$ independent PDF,
$p_{qq}(m', \theta'; \Delta E, \mbc)$
is described as
\begin{equation}
 p_{qq}(m', \theta'; \Delta E, \mbc)
  =
  \frac{1}{N(\dE + \Delta \mbc)} \cdot \frac{\sin(\pi m')}{\sin (\pi
  m'_\mathrm{scale})}
   \cdot
  p_{qq}(m'_\mathrm{scale},\theta')
\end{equation}
for the region,
$m_0^\mathrm{min} < m_0 < \min(m_0^\mathrm{max}, m_0^\mathrm{max}+\dE +
\Delta \mbc)$
($p_{qq}=0$ otherwise),
where $N(\dE + \Delta \mbc)$
and $\sin(\pi m')/\sin (\pi m'_\mathrm{scale})$
 are a normalization factor and the Jacobian
for the parameter transformation of $m'_\mathrm{scale} \rightarrow m'$,
respectively.
We obtain the $p_{qq}(m'_\mathrm{scale},\theta')$ 
distribution from data in part of the sideband region,
 $-0.1\,\mathrm{GeV} < \dE < 0.2\,\mathrm{GeV}$
 and $5.2\,\mathrm{GeV}/c^2 < \mbc < 5.26\,\mathrm{GeV}/c^2$,
where the contribution from $B\bbar$ background is negligible.


Since we find significant flavor asymmetry
depending on the location in the Dalitz plot,
we introduce the following term to take account of it:
\begin{equation}
\frac{1+q_\mathrm{tag} A^l(m', \theta')}{2} \;,
\end{equation}
which is $r$ region dependent.
The asymmetry is anti-symmetric in the direction of $\theta'$,
i.e., $A^l(m', \theta') > 0$ ($A^l(m', \theta') < 0$)
in the region of $\theta'>0.5$ ($\theta < 0.5$),
and the size of the asymmetry is $\sim 20 \%$ at most in
 the best $r$ region.
This effect is due to
the jet-like topology of continuum events;
when an event has a high momentum $\pi^-$ ($\pi^+$) on the $CP$ side,
the highest momentum $\pi$ on the tag side tends to have $+$ ($-$)
charge.
The highest momentum $\pi$ on the tag side with $+$ ($-$) charge
 tags the flavor as $\bz$ ($\bzb$).
Since an event with a high momentum $\pi^-$ ($\pi^+$) resides
in the region $\theta'>0.5$ ($\theta' < 0.5$),
a continuum event in the region $\theta'>0.5$ ($\theta' < 0.5$)
tends to be tagged as $\bz$ ($\bzb$).
We again parameterize the $A^l(m', \theta')$
in a $\dE$-$\mbc$ independent way as
\begin{equation}
 A^l(m', \theta') = A^l(m'_\mathrm{scale}, \theta') \;,
\end{equation}
and model with a two-dimensional polynomial,
whose coefficients are determined by a fit to data in the sideband region.

\subsection{\boldmath $B\bbar$ background PDF}
The treatment of $B\bbar$ background is 
different for $CP$ eigenstate modes
and flavor specific or charged modes.
The PDF for the $CP$ eigenstate modes is
\begin{equation}
 p^l_{BB}(\Delta E, \mbc; m', \theta';\Delta t, q_\mathrm{tag})
  = \mathcal{F}^l_{BB} \cdot p_{BB}(\Delta E, \mbc)
  \cdot p_{BB}(m', \theta')
  \cdot p_{BB}(\Delta t, q_\mathrm{tag}) \;,
\end{equation}
where $p_{BB}(\Delta t, q_\mathrm{tag})$ is a time-dependent $CP$
violation PDF normalized as
\begin{equation}
 \sum_{q_\mathrm{tag}} \int \hspace{-1mm} d\Delta t \;
  p_{BB}(\Delta t, q_\mathrm{tag}) = 1 \;.
\end{equation}
For the flavor specific or charged modes,
the PDF is
\begin{equation}
 p^l_{BB}(\Delta E, \mbc; m', \theta';\Delta t, q_\mathrm{tag})
  = \mathcal{F}^l_{BB} \cdot p_{BB}(\Delta E, \mbc)
  \sum_{q_\mathrm{rec}}
  p_{BB}(m', \theta'; q_\mathrm{rec})
  \cdot p_{BB}(\Delta t, q_\mathrm{tag}, q_\mathrm{rec}) \;,
\end{equation}
where the Dalitz plot PDF $p_{BB}(m', \theta'; q_\mathrm{rec})$
is dependent on the true flavor of the $CP$
(fully reconstructed) side, $q_\mathrm{rec}$,
and the time dependent part is a mixing PDF (lifetime PDF with flavor asymmetry)
for flavor specific (charged) modes.
The $\Delta t$ PDF is normalized as
\begin{equation}
 \sum_{q_\mathrm{tag}} \sum_{q_\mathrm{rec}}
  \int \hspace{-1mm} d\Delta t \;
  p_{BB}(\Delta t, q_\mathrm{tag}, q_\mathrm{rec}) = 1 \;.
\end{equation}

The $\dE$-$\mbc$ PDF and Dalitz plot PDF are obtained mode-by-mode
from MC.
The Dalitz plot PDF of the $CP$ eigenstate modes is assumed to have following
symmetry
\begin{equation}
 p_{BB}(m', \theta')
  = p_{BB}(m', 1-\theta') \;,
\end{equation}
while
that of
 flavor specific and charged modes is assumed to have following symmetry
\begin{equation}
 p_{BB}(m', \theta'; q_\mathrm{rec})
  = p_{BB}(m', 1-\theta'; -q_\mathrm{rec}) \;.
\end{equation}
The total PDF of $B\bbar$ background is a linear combination
of each mode with efficiencies and branching fractions
taken into account.


\section{Correlation matrix of the fit result
 \label{sec:appendix_correlation}
}
Tables
\ref{tab:correl_mtx_26_1}-\ref{tab:correl_mtx_26_3}
show the correlation matrix for the 26 parameters
determined in the time-dependent Dalitz plot analysis,
corresponding to the
total error matrix with statistical and systematic
error matrices combined.
We assume no correlation for the systematic errors.
%
%
\begin{table}[htbp]
  \caption{
  Correlation matrix (1) of the 26 fitted parameters,
  with statistical and systematic errors combined.
  \label{tab:correl_mtx_26_1}
  }
  \begin{tabular*}{14cm}{@{\extracolsep{\fill}}@{\hspace{5mm}}l@{\hspace{5mm}}|cccccccc}
   \hline
   \hline
      & $U^+_-$                  & $U^+_0$                  & $U^{+,\mathrm{Re}}_{+-}$ & $U^{+,\mathrm{Re}}_{+0}$ & $U^{+,\mathrm{Re}}_{-0}$ & $U^{+,\mathrm{Im}}_{+-}$ & $U^{+,\mathrm{Im}}_{+0}$ & $U^{+,\mathrm{Im}}_{-0}$ \\
  \hline
  $U^+_-$                  & $+1.00$ \\
  $U^+_0$                  & $+0.21$ & $+1.00$ \\
  $U^{+,\mathrm{Re}}_{+-}$ & $+0.05$ & $+0.03$ & $+1.00$ \\
  $U^{+,\mathrm{Re}}_{+0}$ & $+0.09$ & $+0.01$ & $+0.01$ & $+1.00$ \\
  $U^{+,\mathrm{Re}}_{-0}$ & $-0.02$ & $-0.09$ & $+0.01$ & $+0.01$ & $+1.00$ \\
  $U^{+,\mathrm{Im}}_{+-}$ & $+0.02$ & $+0.01$ & $+0.03$ & $+0.00$ & $-0.00$ & $+1.00$ \\
  $U^{+,\mathrm{Im}}_{+0}$ & $-0.04$ & $-0.09$ & $-0.00$ & $+0.14$ & $+0.02$ & $-0.00$ & $+1.00$ \\
  $U^{+,\mathrm{Im}}_{-0}$ & $-0.13$ & $-0.10$ & $-0.01$ & $-0.02$ & $+0.02$ & $-0.00$ & $+0.02$ & $+1.00$ \\
  \hline
  $U^-_+$                  & $+0.05$ & $+0.02$ & $-0.00$ & $-0.01$ & $+0.00$ & $-0.02$ & $-0.01$ & $-0.01$ \\
  $U^-_-$                  & $-0.22$ & $-0.07$ & $-0.03$ & $-0.03$ & $-0.03$ & $-0.01$ & $+0.01$ & $+0.02$ \\
  $U^-_0$                  & $+0.05$ & $+0.09$ & $+0.01$ & $+0.00$ & $-0.04$ & $+0.00$ & $-0.05$ & $-0.06$ \\
  $U^{-,\mathrm{Re}}_{+-}$ & $-0.04$ & $-0.02$ & $-0.06$ & $-0.01$ & $-0.00$ & $+0.02$ & $+0.00$ & $+0.01$ \\
  $U^{-,\mathrm{Re}}_{+0}$ & $-0.05$ & $-0.02$ & $-0.00$ & $-0.13$ & $-0.00$ & $+0.00$ & $-0.04$ & $+0.02$ \\
  $U^{-,\mathrm{Re}}_{-0}$ & $-0.02$ & $-0.05$ & $-0.00$ & $-0.00$ & $+0.08$ & $-0.00$ & $+0.02$ & $+0.07$ \\
  $U^{-,\mathrm{Im}}_{+-}$ & $-0.04$ & $-0.02$ & $-0.00$ & $-0.01$ & $-0.00$ & $+0.05$ & $+0.00$ & $+0.01$ \\
  $U^{-,\mathrm{Im}}_{+0}$ & $-0.04$ & $-0.08$ & $-0.01$ & $-0.06$ & $+0.01$ & $-0.00$ & $-0.08$ & $+0.02$ \\
  $U^{-,\mathrm{Im}}_{-0}$ & $-0.02$ & $-0.02$ & $-0.00$ & $-0.00$ & $-0.03$ & $-0.00$ & $+0.00$ & $-0.12$ \\
  \hline
  $I_+$                    & $+0.00$ & $+0.00$ & $-0.01$ & $-0.01$ & $-0.00$ & $-0.02$ & $-0.02$ & $-0.00$ \\
  $I_-$                    & $+0.07$ & $+0.03$ & $-0.01$ & $+0.01$ & $-0.01$ & $+0.03$ & $-0.00$ & $+0.01$ \\
  $I_0$                    & $+0.01$ & $+0.01$ & $+0.00$ & $+0.01$ & $+0.01$ & $+0.00$ & $-0.02$ & $-0.03$ \\
  $I^{\mathrm{Re}}_{+-}$   & $-0.02$ & $-0.00$ & $+0.01$ & $+0.00$ & $+0.00$ & $-0.20$ & $+0.00$ & $+0.00$ \\
  $I^{\mathrm{Re}}_{+0}$   & $+0.00$ & $+0.02$ & $+0.00$ & $-0.12$ & $-0.01$ & $+0.00$ & $-0.01$ & $+0.00$ \\
  $I^{\mathrm{Re}}_{-0}$   & $-0.04$ & $+0.02$ & $-0.01$ & $-0.01$ & $-0.11$ & $+0.00$ & $-0.01$ & $-0.25$ \\
  $I^{\mathrm{Im}}_{+-}$   & $-0.02$ & $-0.01$ & $+0.09$ & $-0.00$ & $-0.00$ & $+0.03$ & $+0.00$ & $+0.00$ \\
  $I^{\mathrm{Im}}_{+0}$   & $+0.01$ & $-0.02$ & $+0.00$ & $+0.04$ & $+0.00$ & $+0.00$ & $+0.07$ & $+0.00$ \\
  $I^{\mathrm{Im}}_{-0}$   & $-0.07$ & $-0.04$ & $-0.01$ & $-0.01$ & $-0.09$ & $-0.01$ & $+0.01$ & $+0.08$ \\
  \hline

  \end{tabular*}
 \vspace{50mm}
\end{table}
\begin{table}[htbp]
  \caption{
  Correlation matrix (2) of the 26 fitted parameters,
  with statistical and systematic errors combined.
  \label{tab:correl_mtx_26_2}
  }
  \begin{tabular*}{15cm}{@{\extracolsep{\fill}}@{\hspace{5mm}}l@{\hspace{5mm}}|ccccccccc}
   \hline
   \hline
      & $U^-_+$                  & $U^-_-$                  & $U^-_0$                  & $U^{-,\mathrm{Re}}_{+-}$ & $U^{-,\mathrm{Re}}_{+0}$ & $U^{-,\mathrm{Re}}_{-0}$ & $U^{-,\mathrm{Im}}_{+-}$ & $U^{-,\mathrm{Im}}_{+0}$ & $U^{-,\mathrm{Im}}_{-0}$ \\
  \hline
  $U^-_+$                  & $+1.00$ \\
  $U^-_-$                  & $-0.06$ & $+1.00$ \\
  $U^-_0$                  & $+0.00$ & $-0.00$ & $+1.00$ \\
  $U^{-,\mathrm{Re}}_{+-}$ & $-0.04$ & $+0.02$ & $-0.00$ & $+1.00$ \\
  $U^{-,\mathrm{Re}}_{+0}$ & $-0.17$ & $+0.03$ & $-0.08$ & $+0.01$ & $+1.00$ \\
  $U^{-,\mathrm{Re}}_{-0}$ & $+0.01$ & $-0.10$ & $-0.18$ & $-0.00$ & $+0.02$ & $+1.00$ \\
  $U^{-,\mathrm{Im}}_{+-}$ & $+0.06$ & $-0.00$ & $-0.00$ & $+0.16$ & $-0.01$ & $+0.00$ & $+1.00$ \\
  $U^{-,\mathrm{Im}}_{+0}$ & $-0.00$ & $+0.01$ & $-0.08$ & $+0.00$ & $+0.03$ & $+0.02$ & $+0.00$ & $+1.00$ \\
  $U^{-,\mathrm{Im}}_{-0}$ & $-0.00$ & $+0.02$ & $+0.01$ & $-0.00$ & $-0.00$ & $+0.02$ & $+0.00$ & $+0.00$ & $+1.00$ \\
  \hline
  $I_+$                    & $-0.02$ & $-0.00$ & $-0.00$ & $+0.02$ & $+0.00$ & $+0.00$ & $-0.02$ & $+0.01$ & $+0.01$ \\
  $I_-$                    & $-0.00$ & $-0.02$ & $+0.01$ & $+0.02$ & $-0.00$ & $-0.04$ & $-0.01$ & $-0.01$ & $-0.08$ \\
  $I_0$                    & $-0.00$ & $-0.01$ & $+0.07$ & $+0.00$ & $+0.01$ & $+0.02$ & $-0.00$ & $-0.04$ & $-0.04$ \\
  $I^{\mathrm{Re}}_{+-}$   & $+0.02$ & $+0.01$ & $-0.00$ & $-0.04$ & $-0.00$ & $-0.00$ & $-0.19$ & $-0.00$ & $+0.00$ \\
  $I^{\mathrm{Re}}_{+0}$   & $-0.00$ & $+0.00$ & $+0.01$ & $-0.00$ & $+0.06$ & $-0.01$ & $+0.00$ & $+0.11$ & $+0.01$ \\
  $I^{\mathrm{Re}}_{-0}$   & $-0.01$ & $+0.07$ & $+0.05$ & $+0.01$ & $-0.00$ & $-0.09$ & $+0.00$ & $-0.01$ & $+0.13$ \\
  $I^{\mathrm{Im}}_{+-}$   & $+0.01$ & $+0.03$ & $-0.00$ & $+0.06$ & $-0.00$ & $-0.00$ & $+0.05$ & $+0.00$ & $+0.00$ \\
  $I^{\mathrm{Im}}_{+0}$   & $+0.02$ & $-0.00$ & $-0.01$ & $-0.00$ & $-0.25$ & $-0.00$ & $+0.00$ & $-0.02$ & $+0.01$ \\
  $I^{\mathrm{Im}}_{-0}$   & $-0.00$ & $+0.03$ & $-0.01$ & $+0.00$ & $+0.00$ & $+0.15$ & $+0.01$ & $+0.02$ & $+0.10$ \\
  \hline

  \end{tabular*}
\end{table}
\begin{table}[htbp]
  \caption{
  Correlation matrix (3) of the 26 fitted parameters,
  with statistical and systematic errors combined.
  \label{tab:correl_mtx_26_3}
  }
  \begin{tabular*}{15cm}{@{\extracolsep{\fill}}@{\hspace{5mm}}l@{\hspace{7mm}}|ccccccccc}
   \hline
   \hline
      & $I_+$                    & $I_-$                    & $I_0$                    & $I^{\mathrm{Re}}_{+-}$   & $I^{\mathrm{Re}}_{+0}$   & $I^{\mathrm{Re}}_{-0}$   & $I^{\mathrm{Im}}_{+-}$   & $I^{\mathrm{Im}}_{+0}$   & $I^{\mathrm{Im}}_{-0}$   \\
  \hline
  $I_+$                    & $+1.00$ \\
  $I_-$                    & $-0.06$ & $+1.00$ \\
  $I_0$                    & $+0.01$ & $+0.02$ & $+1.00$ \\
  $I^{\mathrm{Re}}_{+-}$   & $-0.08$ & $-0.00$ & $-0.00$ & $+1.00$ \\
  $I^{\mathrm{Re}}_{+0}$   & $-0.02$ & $-0.00$ & $-0.10$ & $+0.00$ & $+1.00$ \\
  $I^{\mathrm{Re}}_{-0}$   & $-0.01$ & $+0.14$ & $+0.00$ & $+0.00$ & $+0.00$ & $+1.00$ \\
  $I^{\mathrm{Im}}_{+-}$   & $-0.03$ & $-0.02$ & $-0.00$ & $-0.28$ & $+0.00$ & $+0.00$ & $+1.00$ \\
  $I^{\mathrm{Im}}_{+0}$   & $-0.13$ & $+0.01$ & $-0.12$ & $+0.01$ & $-0.11$ & $-0.00$ & $+0.01$ & $+1.00$ \\
  $I^{\mathrm{Im}}_{-0}$   & $+0.01$ & $-0.12$ & $-0.19$ & $+0.00$ & $+0.03$ & $+0.03$ & $+0.00$ & $+0.03$ & $+1.00$ \\
  \hline

  \end{tabular*}
\end{table}

\section{\boldmath Method of $\phi_2$ constraint
 \label{sec:phi2_chi2_parameterization}
}

\subsection{Formalism}
We define amplitudes as
\begin{eqnarray}
 A^+    & \equiv & A(B^0\rightarrow \rho^+ \pi^-) \;, \\
 A^-    & \equiv & A(B^0\rightarrow \rho^- \pi^+) \;, \\
 A^0    & \equiv & A(B^0\rightarrow \rho^0 \pi^0) \;, \\
 A^{+0} & \equiv & A(B^+\rightarrow \rho^+ \pi^0) \;, \\
 A^{0+} & \equiv & A(B^+\rightarrow \rho^0 \pi^+) \;,
\end{eqnarray}
and
\begin{eqnarray}
 \overline{A}{}^+ & \equiv & \frac{p}{q} A(\overline{B}{}^0\rightarrow \rho^+ \pi^-) \;, \\
 \overline{A}{}^- & \equiv & \frac{p}{q} A(\overline{B}{}^0\rightarrow \rho^- \pi^+) \;, \\
 \overline{A}{}^0 & \equiv & \frac{p}{q} A(\overline{B}{}^0\rightarrow \rho^0 \pi^0) \;, \\
 A^{-0} & \equiv & \frac{p}{q} A(B^-\rightarrow \rho^- \pi^0) \;, \\
 A^{0-} & \equiv & \frac{p}{q} A(B^-\rightarrow \rho^0 \pi^-) \;.
\end{eqnarray}
These amplitudes are obtained from
 1) 26 measurements determined in the time-dependent Dalitz plot
 analysis as well as
 2) branching fractions and asymmetry measurements,
and give a constraint on $\phi_2$.

Equations (\ref{equ:fit_params_first})-(\ref{equ:fit_params_last})
define the relations between the amplitudes for the neutral modes
and the parameters determined in the time-dependent Dalitz plot analysis.
The relations between the branching fractions and asymmetries,
and the amplitudes are
\begin{equation}
 \mathcal{B}(\rho^\pm \pi^\mp) =
  c
  \cdot
  \left(
   |A^+|^2 + |A^-|^2 + |\overline{A}{}^+|^2 + |\overline{A}{}^-|
  \right)
  \cdot 
  \tau_{B^0}
  \label{equ:phi2_constraint_br_pm_definition}
  \;,
\end{equation}
\begin{equation}
 \mathcal{B}(\rho^+ \pi^0) =
  c
  \cdot
  \left(
   |A^{+0}|^2 + |A^{-0}|^2
  \right)
  \cdot 
  \tau_{B^+} \;,
\end{equation}
\begin{equation}
 \mathcal{B}(\rho^0 \pi^+) =
  c
  \cdot
  \left(
   |A^{0+}|^2 + |A^{0-}|^2
  \right)
  \cdot 
  \tau_{B^+} \;,
\end{equation}
\begin{equation}
 \mathcal{A}(\rho^+ \pi^0)
  = \frac{|A^{-0}|^2 - |A^{+0}|^2}{|A^{-0}|^2 + |A^{+0}|^2} \;,
\end{equation}
\begin{equation}
 \mathcal{A}(\rho^0 \pi^+)
  = \frac{|A^{0-}|^2 - |A^{0+}|^2}{|A^{0-}|^2 + |A^{0+}|^2} \;,
\end{equation}
where $c$ is a constant
and the lifetimes $\tau_{B^0}$ and $\tau_{B^+}$ are
introduced to take account of the total width difference
between $B^0$ and $B^+$.
Note that we do not use quasi-two-body parameters
related to neutral modes except for $\mathcal{B}(\rho^\pm \pi^\mp)$,
since they are included in the Dalitz plot parameters.

The amplitudes are expected to follow $SU(2)$ isospin symmetry
to a good approximation \cite{Lipkin:1991st,Gronau:1991dq}
\begin{equation}
 \begin{array}{l}
  A^+ + A^- + 2A^0 = \widetilde{A}{}^+ + \widetilde{A}{}^- + 2\tilde{A}{}^0 \\
  \qquad = \sqrt{2} (A^{+0} + A^{0+}) 
   = \sqrt{2} (\widetilde{A}{}^{-0} + \widetilde{A}{}^{0-}) \;,
 \end{array}
 \label{equ:phi2_constraint_isospin_relation_1}
\end{equation}
\begin{equation}
 A^{+0} - A^{0+} - \sqrt{2} (A^+ - A^-)
  =
  \widetilde{A}^{-0} - \widetilde{A}^{0-} - \sqrt{2} (\widetilde{A}^- - \widetilde{A}^+) \;,
  \label{equ:phi2_constraint_isospin_relation_2}
\end{equation}
where
\begin{equation}
 \widetilde{A}{}^\kappa \equiv e^{-2i\phi_2} \overline{A}{}^\kappa \;,
  \quad
  \widetilde{A}{}^{-0} \equiv e^{-2i\phi_2} A^{-0} \;,
  \quad
  \mathrm{and}
  \quad
  \widetilde{A}{}^{0-} \equiv e^{-2i\phi_2} A^{0-} \;.
  \label{equ:phi2_constraint_tilde_A_def}
\end{equation}
Note that there is an inconsistency
in equation (\ref{equ:phi2_constraint_isospin_relation_2})
 between Ref.~\cite{Lipkin:1991st} and Ref.~\cite{Gronau:1991dq};
 we follow the treatment of Ref.~\cite{Lipkin:1991st}, which we believe is correct.

\subsection{Parameterization}
Here we give two examples of the parameterization
of the amplitudes.
The first example may be more intuitive,
while the second example is well behaved in the fit.
The results are independent of the parameterizations
with respect to the constraint on $\phi_2$.

\subsubsection{Amplitude parameterization}
We can parameterize the amplitudes as follows~\cite{Lipkin:1991st}
\begin{eqnarray}
  A^+ & = & e^{-i\phi_2} T^+ + P^+ \;, \label{equ:phi2_constraint_ampl_param_begin} \\
  A^- & = & e^{-i\phi_2} T^- + P^- \;, \\
  A^0 & = & e^{-i\phi_2} T^0 - \frac{1}{2}(P^+ + P^-) \;, \\
 \sqrt{2} A^{+0} & = & e^{-i\phi_2} T^{+0} + P^+ - P^- \;, \\
 \sqrt{2} A^{0+} & = & e^{-i\phi_2} (T^+ + T^- + 2T^0 - T^{+0}) - P^+ + P^- \;,
\end{eqnarray}
and
\begin{eqnarray}
 \overline{A}{}^+ & = & e^{+i\phi_2} T^- + P^- \;, \\
 \overline{A}{}^- & = & e^{+i\phi_2} T^+ + P^+ \;, \\
 \overline{A}{}^0 & = & e^{+i\phi_2} T^0 - \frac{1}{2}(P^+ + P^-) \;, \\
 \sqrt{2} A^{-0} & = & e^{+i\phi_2} T^{+0} + P^+ - P^- \;, \\
 \sqrt{2} A^{0-} & = & e^{+i\phi_2} (T^+ + T^- + 2T^0 - T^{+0}) - P^+ + P^- \;,
  \label{equ:phi2_constraint_ampl_param_end}
\end{eqnarray}
where the overall phase is fixed with the convention $\mathrm{Im}T^+ = 0$.
Thus, there are 6 complex amplitudes, $T^+, T^-, T^0, P^+, P^-$, and
$T^{+0}$,
corresponding to 11 degrees of freedom;
and $\phi_2$, corresponding to 12 degrees of freedom in total.
This parameterization automatically satisfies the isospin relations
without losing generality,
i.e., the isospin relations are the only assumption here.

\subsubsection{Geometric parameterization}
\begin{figure}[htbp]
 \begin{center}
  \includegraphics[width=0.5\textwidth]{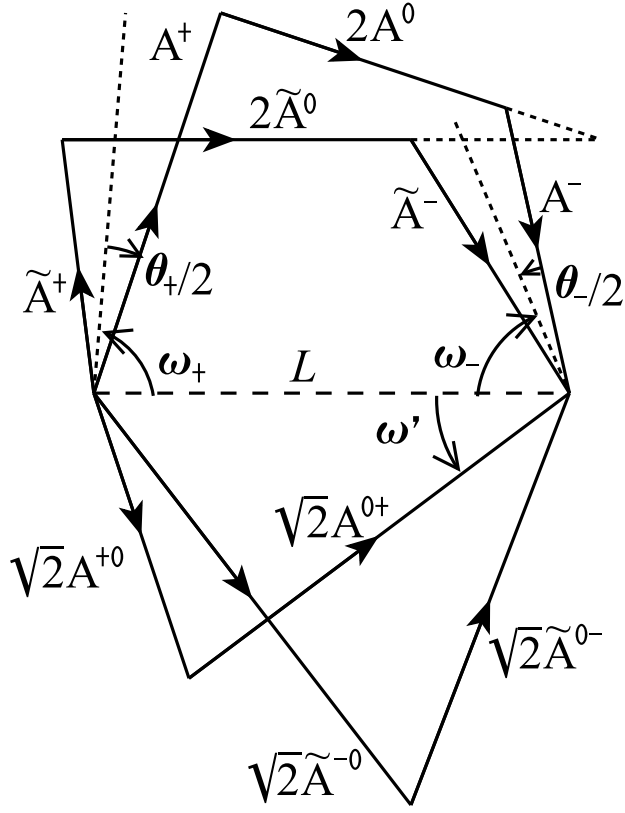}
  \caption{Complex pentagons formed from the $B \rightarrow \rho\pi$ decay
  amplitudes.
  \label{fig:phi2_constraint_geometry}
  }
 \end{center}
\end{figure}

We can parameterize the amplitudes using the
geometric arrangement of Fig.~\ref{fig:phi2_constraint_geometry}
 that satisfies the isospin relation of
 equation (\ref{equ:phi2_constraint_isospin_relation_1}).
This figure is equivalent to Fig.~3 of
Ref.~\cite{Gronau:1991dq},
except that
 the sides corresponding to $B^0 \rightarrow \rho^-\pi^+$
and $B^0 \rightarrow \rho^0\pi^0$
are swapped.
This difference is not physically significant.
We apply this modification only to obtain a better behaved
parameterization;
the parameterization here uses
the angles $\omega_-$ and $\theta_-$
 related to the process $\bz(\bzb) \rightarrow \rho^-\pi^+$,
which are better behaved than those related to
 $\bz(\bzb) \rightarrow \rho^0\pi^0$.

To parameterize the amplitudes, we use $\phi_2$ and
 the following 11 geometric parameters:
\begin{equation}
 \omega_+, \omega_-, \omega', \theta_+, \theta_-, b_+, b_-, b', a_+,
  a_-, L,
\end{equation}
where $b$ and $a$ {\it imply} branching fraction and asymmetry, respectively.
In terms of these parameters,
the amplitudes can be described as follows
\begin{eqnarray}
 A^+ & = & e^{i(\omega_+ + \theta_+/2)} \sqrt{b_+  (1 - a_+) / 2} \;, \\
 \widetilde{A}{}^+ & = & e^{i(\omega_+ - \theta_+/2)} \sqrt{b_+  (1 + a_+) / 2} \;, \\
 A^- & = & e^{i(\omega_- + \theta_-/2)} \sqrt{b_-  (1 - a_-) / 2} \;, \\
 \widetilde{A}{}^- & = & e^{i(\omega_- - \theta_-/2)} \sqrt{b_-  (1 + a_-) / 2} \;, \\
 A^0 & = & (L - A^+ - A^-) / 2 \;, \\
 \widetilde{A}^0 & = & (L - \widetilde{A}{}^+ - \widetilde{A}{}^-)/2 \;, \\
 A^{0+} & = & e^{i\omega'} \sqrt{b'/2} \;, \\
 A^{+0} & = & \frac{L}{\sqrt{2}} - A^{0+} \;, \\
 \widetilde{A}^{0-} & = & \frac{L}{2\sqrt{2}}
  - \left[
     A^{+0} - A^{0+}
     - \sqrt{2} (A^+ - A^-)
     + \sqrt{2} (\widetilde{A}{}^- - \widetilde{A}{}^+)
    \right] / 2 \;,
  \label{equ:phi2_constraint_geom_util_unused_isospin}
  \\
 \widetilde{A}^{-0} & = & \frac{L}{\sqrt{2}} - \widetilde{A}{}^{0-} \;.
\end{eqnarray}
Equation (\ref{equ:phi2_constraint_geom_util_unused_isospin})
exploits the isospin relation of equation
 (\ref{equ:phi2_constraint_isospin_relation_2}),
which Fig.~\ref{fig:phi2_constraint_geometry} does not incorporate geometrically.
The phase $\phi_2$ enters when the $\widetilde{A}$'s are converted
into $\overline{A}$'s with equation (\ref{equ:phi2_constraint_tilde_A_def}).
When we perform the analysis only with the time-dependent Dalitz plot
observables and without the information from charged decay modes,
we remove the parameters $\omega'$ and $b'$ from the fit
and fix $L$ to be a constant.

This geometric parameterization has 
a substantial advantage
in terms of required computational resources,
 compared to the parameterization
based on the $T$ and $P$ amplitudes described in the
previous section.
In the procedure to constrain $\phi_2$,
the minimum $\chi^2$ has to be calculated for each
value of $\phi_2$.
To avoid local minima, initial values of the parameters
for the minimization have to be scanned
and this inflates the computing time,
which increases exponentially with the number of parameters.
However, the number of parameters to be scanned
decreases in the geometric parameterization.
Among 11 parameters except for $\phi_2$, five of them,
$b_+, b_-, b', a_+$, and $a_-$,
are related to the branching fractions and asymmetries.
Since in most cases they do not have multiple solutions,
we do not have to scan the initial values of them.
In addition,
the optimum initial value for $L$
can also be determined using other parameters
and $b_0$,
the nominal branching fraction of $B^0 \rightarrow
\rho^0\pi^0$,
from the following relation
\begin{equation}
 b_0 = \left|L - e^{i\omega_+} \sqrt{b_+/2}
	- e^{i\omega_-} \sqrt{b_-/2}\right|^2 \;,
\end{equation}
up to a two-fold ambiguity.
Here $b_0$ is calculated using the input parameters as
\begin{equation}
 b_0 = 
  \frac{U^+_0}{U^+_+ + U^+_-} \cdot
  \frac{\mathcal{B}(\rho^\pm \pi^\mp)}{c \cdot \tau_{B^0}} \;,
\end{equation}
based on equations
 (\ref{equ:fit_params_first})
 and (\ref{equ:phi2_constraint_br_pm_definition}).
The explicit solution for the optimal initial value of $L$ is
\begin{equation}
  L = \mathrm{Re}\gamma \pm \sqrt{b_0 - \left(\mathrm{Im} \gamma
  \right)^2} \quad
  \left(
   \quad
    \gamma \equiv 
    e^{i\omega_+} \sqrt{b_+/2}
    + e^{i\omega_-} \sqrt{b_-/2}
   \quad
  \right) \;.
\end{equation}
When $b_0 - \left(\mathrm{Im} \gamma \right)^2 < 0$,
there is no real-valued solution and
$L = \mathrm{Re}\gamma$ is the optimum initial value.
With the optimum values calculated above,
the initial value of $L$ does not have to be scanned,
except for the two fold ambiguity.
Consequently, the number of parameters
to be scanned in this parameterization is only five,
corresponding to
$\omega_+, \omega_-, \omega', \theta_+$, and $\theta_-$,
while all of 11 or maybe 10 parameters
have to be scanned in the $T$ and $P$ amplitude parameterization.
This leads to a substantial reduction
of the computational resources required.


\bibliographystyle{apsrev.bst}
\bibliography{cites}

\begin{thebibliography}{26}
\expandafter\ifx\csname natexlab\endcsname\relax\def\natexlab#1{#1}\fi
\expandafter\ifx\csname bibnamefont\endcsname\relax
  \def\bibnamefont#1{#1}\fi
\expandafter\ifx\csname bibfnamefont\endcsname\relax
  \def\bibfnamefont#1{#1}\fi
\expandafter\ifx\csname citenamefont\endcsname\relax
  \def\citenamefont#1{#1}\fi
\expandafter\ifx\csname url\endcsname\relax
  \def\url#1{\texttt{#1}}\fi
\expandafter\ifx\csname urlprefix\endcsname\relax\def\urlprefix{URL }\fi
\providecommand{\bibinfo}[2]{#2}
\providecommand{\eprint}[2][]{\url{#2}}

\bibitem[{\citenamefont{Kobayashi and Maskawa}(1973)}]{Kobayashi:1973fv}
\bibinfo{author}{\bibfnamefont{M.}~\bibnamefont{Kobayashi}} \bibnamefont{and}
  \bibinfo{author}{\bibfnamefont{T.}~\bibnamefont{Maskawa}},
  \bibinfo{journal}{Prog. Theor. Phys.} \textbf{\bibinfo{volume}{49}},
  \bibinfo{pages}{652} (\bibinfo{year}{1973}).

\bibitem[{\citenamefont{Carter and Sanda}(1980)}]{Carter:1980hr}
\bibinfo{author}{\bibfnamefont{A.~B.} \bibnamefont{Carter}} \bibnamefont{and}
  \bibinfo{author}{\bibfnamefont{A.~I.} \bibnamefont{Sanda}},
  \bibinfo{journal}{Phys. Rev. Lett.} \textbf{\bibinfo{volume}{45}},
  \bibinfo{pages}{952} (\bibinfo{year}{1980}).

\bibitem[{\citenamefont{Carter and Sanda}(1981)}]{Carter:1980tk}
\bibinfo{author}{\bibfnamefont{A.~B.} \bibnamefont{Carter}} \bibnamefont{and}
  \bibinfo{author}{\bibfnamefont{A.~I.} \bibnamefont{Sanda}},
  \bibinfo{journal}{Phys. Rev.} \textbf{\bibinfo{volume}{D23}},
  \bibinfo{pages}{1567} (\bibinfo{year}{1981}).

\bibitem[{\citenamefont{Bigi and Sanda}(1981)}]{Bigi:1981qs}
\bibinfo{author}{\bibfnamefont{I.~I.~Y.} \bibnamefont{Bigi}} \bibnamefont{and}
  \bibinfo{author}{\bibfnamefont{A.~I.} \bibnamefont{Sanda}},
  \bibinfo{journal}{Nucl. Phys.} \textbf{\bibinfo{volume}{B193}},
  \bibinfo{pages}{85} (\bibinfo{year}{1981}).

\bibitem[{Cha()}]{ChargeConjugate}
\bibinfo{note}{Throughout this paper, the inclusion of the charge conjugate
  decay mode is implied unless otherwise stated.}

\bibitem[{\citenamefont{Snyder and Quinn}(1993)}]{Snyder:1993mx}
\bibinfo{author}{\bibfnamefont{A.~E.} \bibnamefont{Snyder}} \bibnamefont{and}
  \bibinfo{author}{\bibfnamefont{H.~R.} \bibnamefont{Quinn}},
  \bibinfo{journal}{Phys. Rev.} \textbf{\bibinfo{volume}{D48}},
  \bibinfo{pages}{2139} (\bibinfo{year}{1993}).

\bibitem[{\citenamefont{Lipkin et~al.}(1991)\citenamefont{Lipkin, Nir, Quinn,
  and Snyder}}]{Lipkin:1991st}
\bibinfo{author}{\bibfnamefont{H.~J.} \bibnamefont{Lipkin}},
  \bibinfo{author}{\bibfnamefont{Y.}~\bibnamefont{Nir}},
  \bibinfo{author}{\bibfnamefont{H.~R.} \bibnamefont{Quinn}}, \bibnamefont{and}
  \bibinfo{author}{\bibfnamefont{A.}~\bibnamefont{Snyder}},
  \bibinfo{journal}{Phys. Rev.} \textbf{\bibinfo{volume}{D44}},
  \bibinfo{pages}{1454} (\bibinfo{year}{1991}).

\bibitem[{\citenamefont{Gronau}(1991)}]{Gronau:1991dq}
\bibinfo{author}{\bibfnamefont{M.}~\bibnamefont{Gronau}},
  \bibinfo{journal}{Phys. Lett.} \textbf{\bibinfo{volume}{B265}},
  \bibinfo{pages}{389} (\bibinfo{year}{1991}).

\bibitem[{\citenamefont{Kurokawa and Kikutani}(2003)}]{KEKB}
\bibinfo{author}{\bibfnamefont{S.}~\bibnamefont{Kurokawa}} \bibnamefont{and}
  \bibinfo{author}{\bibfnamefont{E.}~\bibnamefont{Kikutani}},
  \bibinfo{journal}{Nucl. Instrum. Meth.} \textbf{\bibinfo{volume}{A499}},
  \bibinfo{pages}{1} (\bibinfo{year}{2003}), \bibinfo{note}{and other papers
  included in this volume.}

\bibitem[{\citenamefont{Abashian et~al.}(2002)}]{Belle}
\bibinfo{author}{\bibfnamefont{A.}~\bibnamefont{Abashian}} \bibnamefont{et~al.}
  (\bibinfo{collaboration}{Belle Collaboration}), \bibinfo{journal}{Nucl.
  Instrum. Meth.} \textbf{\bibinfo{volume}{A479}}, \bibinfo{pages}{117}
  (\bibinfo{year}{2002}).

\bibitem[{\citenamefont{Ushiroda}(2003)}]{Ushiroda:2003bu}
\bibinfo{author}{\bibfnamefont{Y.}~\bibnamefont{Ushiroda}}
  (\bibinfo{collaboration}{Belle SVD2 Group}), \bibinfo{journal}{Nucl. Instrum.
  Meth.} \textbf{\bibinfo{volume}{A511}}, \bibinfo{pages}{6}
  (\bibinfo{year}{2003}).

\bibitem[{\citenamefont{Gounaris and Sakurai}(1968)}]{Gounaris:1968mw}
\bibinfo{author}{\bibfnamefont{G.~J.} \bibnamefont{Gounaris}} \bibnamefont{and}
  \bibinfo{author}{\bibfnamefont{J.~J.} \bibnamefont{Sakurai}},
  \bibinfo{journal}{Phys. Rev. Lett.} \textbf{\bibinfo{volume}{21}},
  \bibinfo{pages}{244} (\bibinfo{year}{1968}).

\bibitem[{\citenamefont{Aubert et~al.}(2004)}]{Aubert:2004iu}
\bibinfo{author}{\bibfnamefont{B.}~\bibnamefont{Aubert}} \bibnamefont{et~al.}
  (\bibinfo{collaboration}{BaBar Collaboration}) (\bibinfo{year}{2004}),
  \eprint{hep-ex/0408099}.

\bibitem[{\citenamefont{Tajima et~al.}(2004)}]{Tajima:2003bu}
\bibinfo{author}{\bibfnamefont{H.}~\bibnamefont{Tajima}} \bibnamefont{et~al.},
  \bibinfo{journal}{Nucl. Instrum. Meth.} \textbf{\bibinfo{volume}{A533}},
  \bibinfo{pages}{370} (\bibinfo{year}{2004}), \eprint{[hep-ex/0301026]}.

\bibitem[{\citenamefont{Kakuno et~al.}(2004)}]{Kakuno:2004cf}
\bibinfo{author}{\bibfnamefont{H.}~\bibnamefont{Kakuno}} \bibnamefont{et~al.},
  \bibinfo{journal}{Nucl. Instrum. Meth.} \textbf{\bibinfo{volume}{A533}},
  \bibinfo{pages}{516} (\bibinfo{year}{2004}), \eprint{[hep-ex/0403022]}.

\bibitem[{\citenamefont{Abe et~al.}(2005)}]{Abe:2004mz}
\bibinfo{author}{\bibfnamefont{K.}~\bibnamefont{Abe}} \bibnamefont{et~al.}
  (\bibinfo{collaboration}{Belle Collaboration}), \bibinfo{journal}{Phys. Rev.}
  \textbf{\bibinfo{volume}{D71}}, \bibinfo{pages}{072003}
  (\bibinfo{year}{2005}), \eprint{[hep-ex/0408111]}.

\bibitem[{\citenamefont{Chen et~al.}(2005)}]{Chen:2005dr}
\bibinfo{author}{\bibfnamefont{K.~F.} \bibnamefont{Chen}} \bibnamefont{et~al.}
  (\bibinfo{collaboration}{Belle Collaboration}), \bibinfo{journal}{Phys. Rev.}
  \textbf{\bibinfo{volume}{D72}}, \bibinfo{pages}{012004}
  (\bibinfo{year}{2005}), \eprint{[hep-ex/0504023]}.

\bibitem[{\citenamefont{Albrecht et~al.}(1990)}]{Albrecht:1990am}
\bibinfo{author}{\bibfnamefont{H.}~\bibnamefont{Albrecht}} \bibnamefont{et~al.}
  (\bibinfo{collaboration}{ARGUS Collaboration}), \bibinfo{journal}{Phys.
  Lett.} \textbf{\bibinfo{volume}{B241}}, \bibinfo{pages}{278}
  (\bibinfo{year}{1990}).

\bibitem[{\citenamefont{Eidelman et~al.}(2004)}]{Eidelman:2004wy}
\bibinfo{author}{\bibfnamefont{S.}~\bibnamefont{Eidelman}} \bibnamefont{et~al.}
  (\bibinfo{collaboration}{Particle Data Group}), \bibinfo{journal}{Phys.
  Lett.} \textbf{\bibinfo{volume}{B592}}, \bibinfo{pages}{1}
  (\bibinfo{year}{2004}), \bibinfo{note}{and 2005 partial update
  (http://pdg.lbl.gov).}

\bibitem[{\citenamefont{{Heavy Flavor Averaging Group
  (HFAG)}}(2006)}]{unknown:2006bi}
\bibinfo{author}{\bibnamefont{{Heavy Flavor Averaging Group (HFAG)}}}
  (\bibinfo{year}{2006}), \bibinfo{note}{hep-ex/0603003; and online update of
  Winter 2006 (http://www.slac.stanford.edu/xorg/hfag).}

\bibitem[{\citenamefont{Ablikim et~al.}(2004)}]{Ablikim:2004qn}
\bibinfo{author}{\bibfnamefont{M.}~\bibnamefont{Ablikim}} \bibnamefont{et~al.}
  (\bibinfo{collaboration}{BES Collaboration}), \bibinfo{journal}{Phys. Lett.}
  \textbf{\bibinfo{volume}{B598}}, \bibinfo{pages}{149} (\bibinfo{year}{2004}),
  \eprint{[hep-ex/0406038]}.

\bibitem[{\citenamefont{Muramatsu et~al.}(2002)}]{Muramatsu:2002jp}
\bibinfo{author}{\bibfnamefont{H.}~\bibnamefont{Muramatsu}}
  \bibnamefont{et~al.} (\bibinfo{collaboration}{CLEO Collaboration}),
  \bibinfo{journal}{Phys. Rev. Lett.} \textbf{\bibinfo{volume}{89}},
  \bibinfo{pages}{251802} (\bibinfo{year}{2002}), \eprint{[hep-ex/0207067]}.

\bibitem[{\citenamefont{Aitala et~al.}(2001)}]{Aitala:2000xu}
\bibinfo{author}{\bibfnamefont{E.~M.} \bibnamefont{Aitala}}
  \bibnamefont{et~al.} (\bibinfo{collaboration}{E791 Collaboration}),
  \bibinfo{journal}{Phys. Rev. Lett.} \textbf{\bibinfo{volume}{86}},
  \bibinfo{pages}{770} (\bibinfo{year}{2001}), \eprint{[hep-ex/0007028]}.

\bibitem[{\citenamefont{Long et~al.}(2003)\citenamefont{Long, Baak, Cahn, and
  Kirkby}}]{Long:2003wq}
\bibinfo{author}{\bibfnamefont{O.}~\bibnamefont{Long}},
  \bibinfo{author}{\bibfnamefont{M.}~\bibnamefont{Baak}},
  \bibinfo{author}{\bibfnamefont{R.~N.} \bibnamefont{Cahn}}, \bibnamefont{and}
  \bibinfo{author}{\bibfnamefont{D.}~\bibnamefont{Kirkby}},
  \bibinfo{journal}{Phys. Rev.} \textbf{\bibinfo{volume}{D68}},
  \bibinfo{pages}{034010} (\bibinfo{year}{2003}), \eprint{[hep-ex/0303030]}.

\bibitem[{\citenamefont{Dragic et~al.}(2006)}]{Dragic:2006yv}
\bibinfo{author}{\bibfnamefont{J.}~\bibnamefont{Dragic}} \bibnamefont{et~al.}
  (\bibinfo{collaboration}{Belle Collaboration}), \bibinfo{journal}{Phys. Rev.}
  \textbf{\bibinfo{volume}{D73}}, \bibinfo{pages}{111105}
  (\bibinfo{year}{2006}).

\bibitem[{\citenamefont{Charles et~al.}(2005)}]{Charles:2004jd}
\bibinfo{author}{\bibfnamefont{J.}~\bibnamefont{Charles}} \bibnamefont{et~al.}
  (\bibinfo{collaboration}{CKMfitter Group}), \bibinfo{journal}{Eur. Phys. J.}
  \textbf{\bibinfo{volume}{C41}}, \bibinfo{pages}{1} (\bibinfo{year}{2005}),
  \eprint{[hep-ph/0406184]}.

\end{thebibliography}

\end{document}